\documentclass[extra]{gji}
\newcommand{\abit}{0.920}

\usepackage{mathrsfs,subfig}
\usepackage{multicol}
\usepackage[dvips]{graphicx}
\usepackage{subeqn,url,amstext}
\usepackage{times,bbold}
\newcommand{\plth}{\pl_\theta}
\newcommand{\bgamma}{\mbox{\boldmath$\gamma$}}
\newcommand{\gammaS}{\gamma_{\thetaS}}
\newcommand{\gammaL}{\gamma_{\thetaL}}
\newcommand{\gth}{\gamma_\theta}
\newcommand{\gthp}{\gamma_{\theta'}}

\newcommand{\gSthS}{(\gamma_S)_{\thetaS}}
\newcommand{\gSthSu}{(\gamma_S)_{\thetaSu}}
\newcommand{\gSthpS}{(\gamma_S)_{\thetapS}}
\newcommand{\gtk}{\gth\ofk}
\newcommand{\gtlk}{{\gth}\unL\ofk}
\newcommand{\gtsk}{{\gth}\unS\ofk}

\newcommand{\gtpk}{\gthp\ofk}
\newcommand{\mth}{m_\theta}
\newcommand{\mthp}{m_{\theta'}}

\newcommand{\mthS}{m_{\thetaS}}
\newcommand{\mthSu}{m_{\thetaSu}}
\newcommand{\mthSp}{m_{\thetapS}}
\newcommand{\mthL}{m_{\thetaL}}
\newcommand{\unS}{_{\hsomm S}}
\newcommand{\unL}{_{\hsomm L}}
\newcommand{\unth}{_{\theta}}
\newcommand{\unthL}{_{\thetaL}}
\newcommand{\unthS}{_{\thetaS}}
\newcommand{\unthp}{_{\theta'}}
\newcommand{\Ab}{\mbf{A}}
\newcommand{\Abinv}{\mbf{A}^{-1}}
\newcommand{\Abth}{\Ab\unth}
\newcommand{\AbD}{\Ab_D}
\newcommand{\Abft}{\Ab_\ft}
\newcommand{\Abr}{\Ab_r}
\newcommand{\AbthL}{\Ab\unthL}
\newcommand{\AbthS}{\Ab\unthS}
\newcommand{\Abthp}{\Ab\unthp}

\newcommand{\Fdd}{\mcF_{\!DD}}

\newcommand{\Fff}{\mcF_{\!\ft\!\ft}}
\newcommand{\Frr}{\mcF_{\!rr}}

\newcommand{\Fdf}{\mcF_{\!D\!\ft}}

\newcommand{\Ffr}{\mcF_{\!\ft r}}

\newcommand{\Fdr}{\mcF_{\!Dr}}

\newcommand{\Fthth }{\mcF_{\theta\theta}}
\newcommand{\Fthsths}{\mcF_{\thetaS\thetaS}}
\newcommand{\Fthlthl}{\mcF_{\thetaL\thetaL}}
\newcommand{\Jthth }{\mcJ_{\theta\theta}}
\newcommand{\Hththp}{   F_{\theta\theta'}}
\newcommand{\Fththp}{\mcF_{\theta\theta'}}
\newcommand{\Jththp}{\mcJ_{\theta\theta'}}
\newcommand{\mcL}{{\mathcal{L}}} 
 
\newcommand{\mcP}{{\mathcal{P}}} 
\newcommand{\mcLw}{{\tilde{\mcL}}}
\newcommand{\mcLk}{\mcL_\kb}
\newcommand{\mcLS}{\mcL\unS}

\newcommand{\Soom}{\Soo^{-1}}
\newcommand{\cSoom}{\cSoo^{-1}}
\newcommand{\Soo}{S_{11}}

\newcommand{\cSij}{\mcS_{ij}}
\newcommand{\cSoo}{\mcS_{11}}
\newcommand{\cSook}{\cSoo\ofk}
\newcommand{\cStt}{\mcS_{22}}
\newcommand{\cSttk}{\cStt\ofk}
\newcommand{\cSot}{\mcS_{12}}
\newcommand{\cSotk}{\cSot\ofk}
\newcommand{\cSto}{\mcS_{21}}
\newcommand{\cStok}{\cSto\ofk}
\newcommand{\cSoij}{\mcS_{\circ ij}}
\newcommand{\cSooo}{\mcS_{\circ 11}}
\newcommand{\cSott}{\mcS_{\circ 22}}

\newcommand{\cSoto}{\mcS_{\circ 21}}
\newcommand{\qform}[1]{\Hbo\Hrm{#1}\hsom\Hbo}
\newcommand{\hform}[1]{\cSoo^{-1}\Hbo\Hrm\hsomm{#1}\hsom\Hbo}
\newcommand{\lform}[1]{\Lbo\Trm{#1}\hsom\Lbo}
\newcommand{\thetaL}{\theta\unL}
\newcommand{\thetaLu}{\theta^{}\unL}
\newcommand{\thetapL}{\theta'\unL}
\newcommand{\thetaS}{\theta\unS}
\newcommand{\thetaSu}{\theta^{}\unS}
\newcommand{\thetapS}{\theta'\unS}
\newcommand{\thetaw}{\tilde{\theta}}
\newcommand{\btheta}{\mbox{\boldmath$\theta$}}
\newcommand{\oft}{(\btheta)}
\newcommand{\hbt}{\mbox{\boldmath$\hat{\theta}$}}
\newcommand{\bthetaw}{\mbox{\boldmath$\tilde{\theta}$}}
\newcommand{\bthetawt}{\mbox{\boldmath$\tilde{\theta}$}{}\Tit}
\newcommand{\oftw}{(\bthetaw)}
\newcommand{\hbtw}{\mbox{\boldmath$\hat{\tilde{\theta}}$}}
\newcommand{\hbtwt}{\mbox{\boldmath$\hat{\tilde{\theta}}$}{}\Tit}

\newcommand{\hbtwc}{\mbox{\boldmath$\hat{\tilde{\theta}}$}_u}
\newcommand{\oftwhc}{(\hbtwc)}

\newcommand{\oftwz}{([\bthetawt\,\,\,0]\Tit)}
\newcommand{\ofts}{(\btheta\unS)}
\newcommand{\oftr}{(\btruth)}
\newcommand{\oftrw}{(\btruthw)}

\newcommand{\ofth}{(\hbt)}
\newcommand{\bthetaL}{\mbox{\boldmath{$\theta$}}\unL}
\newcommand{\bthetaS}{\mbox{\boldmath{$\theta$}}\unS}

\newcommand{\bthetafull}{[D\,\,\,\ft\,\,\,\st\,\,\,\nu\,\,\,\rho\,]\Tit}
\newcommand{\bthetafullr}{[D\,\,\,\ft\,\,\,r\,\,\,\st\,\,\,\nu\,\,\,\rho\,]\Tit}
\newcommand{\bthetaSfull}{[\st\,\,\nu\,\,\,\rho\,]\Tit}

\newcommand{\bthetaLfullr}{[D\,\,\,\ft\,\,\,r]\Tit}
\newcommand{\xb}{\mbf{x}}
\newcommand{\yb}{\mbf{y}}
\newcommand{\xbp}{\mbf{x'}}

\newcommand{\kb}{\mbf{k}} 
\newcommand{\kbp}{\mbf{k'}} 
 
\newcommand{\dbk}{\,d\kb}
\newcommand{\dbx}{\,d\xb}
\newcommand{\dtbk}{\,d\kb\,d\kb'}

\newcommand{\dkkp}{\delta({\kb,\kb'})}

\newcommand{\ofk}{(\kb)}
\newcommand{\ofsk}{(k)}

\newcommand{\ofx }{(\xb)}
\newcommand{\ofkp}{(\kb')}
\newcommand{\ofxp}{(\xb')}
\newcommand{\ofkpp}{(\kb'')}
\newcommand{\ofkppp}{(\kb''')}
\newcommand{\normu}{\sum_{\kb}}
\newcommand{\twnorml}{\frac{2}{K}\sum_{\kb}}
\newcommand{\norml}{\frac{1}{K}\sum_{\kb}}
\newcommand{\sumkp}{\sum_{\kb'}}
\newcommand{\normlt}{\frac{1}{K^2}\sum_{\kb}}
\newcommand{\normltu}{\frac{1}{K^2}\sum_{\kb^{}}}
\newcommand{\tnorml}{\frac{1}{2K}\sum_{\kb}}
\newcommand{\xik}{\xi\ofsk}

\newcommand{\xit }{\xi^2}
\newcommand{\xitk}{\xi^2\ofsk}

\newcommand{\chik }{\chi\ofsk}

\newcommand{\chit }{\chi^2}

\newcommand{\phik}{\phi\ofsk}

\newcommand{\phit}{\phi^2}

\newcommand{\phixi}{\phi\hsom\xi}

\newcommand{\inprob}{\stackrel{\mcP}{\longrightarrow}}
\newcommand{\inlaw}{\stackrel{\mcL}{\longrightarrow}}
\newcommand{\omat}{\left(\!\!\barray}
\newcommand{\omats}{\left[\!\!\barray}
\newcommand{\cmat}{\earray\!\!\right)}
\newcommand{\cmats}{\earray\!\!\right]}
\newcommand{\tr}{{\mathrm{tr}}}
\newcommand{\Hrm}{^{\sst{\mathbf{H}}}}
\newcommand{\Hbfp}{'^{\sst{\mathbf{H}}}}

\newcommand{\Trm}{^{\sst{\mathrm{T}}}}
\newcommand{\Tit}{^{\sst{T}}}
\newcommand{\Trminv}{^{\sst{-\mathrm{T}}}}
\newcommand{\Tcal}{^{\sst{\mathcal{T}}}}
\newcommand{\Tcalinv}{^{\sst{-\mathcal{T}}}}
\newcommand{\sst}{\scriptstyle} 
\newcommand{\half}{{\textstyle\frac{1}{2}}}
\newcommand{\shalf}{^{1/2}}

\newcommand{\bzero}{\mbf{0}} 
\newcommand{\rmo}{^{+}}
\newcommand{\rmt}{^{-}}
\newcommand{\barray}{\begin{array}}
\newcommand{\earray}{\end{array}}
\newcommand{\bnabla}{\mbox{\boldmath$\nabla$}}
\newcommand{\blt}{\mbf{\Lambda}_\theta}

\newcommand{\ssts}{\scriptscriptstyle}
\newcommand{\nnr}{\nonumber}
\newcommand{\Dj}{\Delta_j}
\newcommand{\Do}{\Delta_1}
\newcommand{\Dou}{\Do^{}}
\newcommand{\Dt}{\Delta_2}
\newcommand{\Dtu}{\Dt^{}}
\newcommand{\hth}{\hat{\theta}}
\newcommand{\tl}{\theta\unL}
\newcommand{\ts}{\theta\unS}
\newcommand{\pl}{\partial}

\newcommand{\lth}{\mcL(\btheta)}

\newcommand{\Ho}{\mcH_1}
\newcommand{\Ht}{\mcH_2}
\newcommand{\Hj}{\mcH_j}
\newcommand{\bmcH}{\mbox{\boldmath$\mathcal{H}$}}
\newcommand{\bmcHH}{\mbox{\boldmath$\mathcal{H}^H$}}
\newcommand{\Hoo}{\mcH_{11}}
\newcommand{\Htt}{\mcH_{22}}
\newcommand{\Hot}{\mcH_{12}}
\newcommand{\Hto}{\mcH_{21}}
\newcommand{\Hij}{\mcH_{ij}}
\newcommand{\Hno}{\mcH_{\circ 1}}
\newcommand{\Hnt}{\mcH_{\circ 2}}
\newcommand{\Hnj}{\mcH_{\circ j}}
\newcommand{\bmcHo}{\bmcH_\circ}
\newcommand{\bmcHot}{\mbox{\boldmath$\mathcal{H}_\circ^T$}}
\newcommand{\bmcHoH}{\mbox{\boldmath$\mathcal{H}_\circ^H$}}
\newcommand{\nHno}{H_{\circ 1}}
\newcommand{\nHnt}{H_{\circ 2}}
\newcommand{\dcHo}{\omats{c}d\Hno\ofk \\
d\Hnt\ofk \cmats}

\newcommand{\Hovec}{\omats{c} \nHno\ofk \\
\hsom \nHnt\ofk\cmats}

\newcommand{\dcH}{\omats{c}d\Ho\ofk \\
d\Ht\ofk\cmats}

\newcommand{\btruth}{\btheta_0}
\newcommand{\btruthw}{\bthetaw_0}
\newcommand{\btruthwt}{\bthetawt_0}
\newcommand{\truth}{\theta_0}
\newcommand{\Gpn}{\mcG_{\circ 2}}
\newcommand{\Gpf}{\mcG_{\circ\circ}}
\newcommand{\Hpf}{\mcH_{\circ\circ}}
\newcommand{\Gpns}{\mcG^*_{\circ 2}}
\newcommand{\nGpn}{G_{\circ 2}}
\newcommand{\Qo}{\mcQ_\circ}
\newcommand{\Qon}{\mcQ_1}
\newcommand{\Qt}{\mcQ_2}
\newcommand{\Qf}{\mcQ_f}
\newcommand{\ekzt}{e^{kz_2}}
\newcommand{\gto}{\gamma^2_\circ}
\newcommand{\gtf}{\gamma^2_f}
\newcommand{\hsps}{\hspace{0.75em}}
\newcommand{\hsem}{\hspace{1em}}

\newcommand{\hsom}{\hspace{0.1em}}
\newcommand{\hsomm}{\hspace{-0.1em}}
\newcommand{\hsum}{\hspace{0.05em}}
\newcommand{\hsumm}{\hspace{-0.05em}}
\newcommand{\dekf}{1+\frac{Dk^4}{g\Dt}}
\newcommand{\dekp}{1+\frac{Dk^4}{g\Do}}

\newcommand{\ft}{{f^2}}
\newcommand{\fto}{{f^2_0}}
\newcommand{\hft}{\widehat{\ft}}
\newcommand{\hst}{\widehat{\sigma^2}}
\newcommand{\st}{{\sigma^2}}
\newcommand{\sto}{{\sigma^2_0}}
\newcommand{\noin}{\noindent} 
\newcommand{\be}{\begin{equation}} 
\newcommand{\ee}{\end{equation}} 
\newcommand{\ber}{\begin{eqnarray}} 
\newcommand{\eer}{\end{eqnarray}} 
\newcommand{\mbf}{\mathbf} 
\newcommand{\mcH}{{\mathcal{H}}} 
\newcommand{\mcZ}{{\mathcal{Z}}} 
\newcommand{\bmcF}{\mbox{\boldmath$\mathcal{F}$}}
\newcommand{\thetax}{\theta_{\!\times}}
\newcommand{\bthetax}{\btheta_{\!\times}}
\newcommand{\hbtx}{\hbt_{\!\times}}
\newcommand{\hbtxt}{\hbtx\Tit}
\newcommand{\bthetaxt}{\btheta_{\!\times}\Tit}
\newcommand{\bmcFx}{\bmcF_{\!\times}}

\newcommand{\bmcFo}{\bmcF_{\!\circ}}

\newcommand{\bmcFxo}{\bmcF_{\!\times\!\circ}}
\newcommand{\bmcFox}{\bmcF_{\!\circ\!\times}}
\newcommand{\bmcFxt}{\bmcFx\Tcal}

\newcommand{\bmcFw}{\mbox{\boldmath$\tilde{\mathcal{F}}$}}

\newcommand{\bmcS}{\mbox{\boldmath$\mcS$}}
\newcommand{\bmcC}{\mbox{\boldmath$\mcC$}}
\newcommand{\mcSo}{{\mcS_{\circ}}}
\newcommand{\bmcSo}{\mbox{\boldmath$\mcS_{\circ}$}}

\newcommand{\Sbar}{\bar\mcS}
\newcommand{\Lbar}{\bar\mcL}
\newcommand{\Lbark}{\bar\mcLk}
\newcommand{\bbmcSo}{\mbox{\boldmath$\Sbar_{\circ}$}}
\newcommand{\bmcSoinv}{\mbox{\boldmath$\mcS_{\circ}^{-1}$}}
\newcommand{\bbmcSoinv}{\mbox{\boldmath$\Sbar_{\circ}^{-1}$}}
\newcommand{\bmcM}{\mbf{M}}
\newcommand{\bmcMD}{\bmcM_D}
\newcommand{\bmcMDk}{\bmcM_D\ofsk}
\newcommand{\bmcJ}{\mbox{\boldmath$\mathcal{J}$}}

\newcommand{\bmcZo}{\mbox{\boldmath$\mathcal{Z}_{\circ}$}}
\newcommand{\bmcZoH}{\mbox{\boldmath$\mathcal{Z}_{\circ}^H$}}
\newcommand{\mcJ}{\mathcal{J}}
\newcommand{\mcF}{{\mathcal{F}}}
\newcommand{\mcS}{{\mathcal{S}}} 
\newcommand{\mcO}{{\mathcal{O}}} 
 
\newcommand{\mcC}{{\mathcal{C}}} 
\newcommand{\mcQ}{{\mathcal{Q}}} 
\newcommand{\mcG}{{\mathcal{G}}} 
\newcommand{\mcI}{{\mathcal{I}}} 
\newcommand{\mcNC}{{\mathcal{N}^C}}
\newcommand{\mcN}{\mathcal{N}}
\newcommand{\Tb}{\mbf{T}} 
\newcommand{\Tbo}{\mbf{T}_\circ} 
\newcommand{\dTb}{\mbox{\boldmath$\Delta$}\mbf{T}}

\newcommand{\Pb}{\mbf{P}} 
\newcommand{\Ib}{\mbf{I}} 
\newcommand{\Zb}{\mbf{Z}} 
\newcommand{\Zbo}{\mbf{Z}_\circ} 
\newcommand{\Zbho}{\mbf{\hat{Z}}_0} 
\newcommand{\Fb}{\mbf{F}} 
\newcommand{\Fbw}{\mbf{\tilde{F}}}

\newcommand{\Lbo}{\mbf{L}_\circ} 
\newcommand{\Lboinv}{\Lbo^{-1}} 
 
\newcommand{\bLbo}{\mbf{\bar{L}}_\circ} 
\newcommand{\bLboinv}{\bLbo^{-1}}

\newcommand{\Tbinv}{\Tb^{-1}}
\newcommand{\dTbinv}{\dTb^{-1}}
\newcommand{\Tboinv}{\Tbo^{-1}}

\newcommand{\Vb}{\mbf{V}}

\newcommand{\Hb}{\mbf{H}} 
\newcommand{\Hbo}{\mbf{H}_\circ} 
\newcommand{\Hbok}{\mbf{H}_\circ\hsomm\ofk} 
\newcommand{\Hbokp}{\mbf{H}_\circ\hsomm\ofkp}

\newcommand{\var}{\mbox{var}} 
 
\newcommand{\cov}{\mbox{cov}} 
\newcommand{\fracd}[2]{\frac{\displaystyle{#1}}{\displaystyle{#2}}}
\newcommand{\for}{\quad\mbox{for}\quad} 
\newcommand{\foral}{\quad\mbox{for all}\hspace{0.5em}} 
\newcommand{\also}{\quad\mbox{and}\quad} 
\newcommand{\where}{\quad\mbox{where}\quad} 
 
\newcommand{\intnyq}{\int\!\!\!\!\int}
\newcommand{\intnyqt}{\intnyq\!\!\!\!\intnyq}
\newcommand{\pHbo}{p_{\Hbo}}
\newcommand{\pHbok}{p_{\Hbok}}
\newcommand{\pHbokp}{p_{\Hbokp}}
\newcommand{\NCOI}{\mcNC\!\!\left(\bzero,\Ib\right)}
\newcommand{\NCOSbo}{\mcNC\!\!\left(\bzero,\bbmcSo\right)}
\newcommand{\NOh}{\mcN\!\left(\bzero,{\textstyle\frac{1}{2}}\Ib\right)}
\begin{document}
\onecolumn
\title[Maximum-likelihood estimation of flexural rigidity]
{Maximum-likelihood estimation of lithospheric flexural rigidity,
 initial-loading fraction, and load correlation, under isotropy}   
\author[Simons and Olhede]{
Frederik J.~Simons$^{1,2}$ and Sofia C.~Olhede$^3$ \\
$^1$Department of Geosciences, Princeton University,
Princeton, NJ 08544, USA\\
$^2$Associated Faculty, Program in Applied \& Computational
Mathematics, Princeton University, Princeton, NJ 08544, USA\\  
$^3$Department of Statistical Science, University College London,
London WC1E 6BT, UK\\
E-mail:  fjsimons@alum.mit.edu, s.olhede@ucl.ac.uk}
\maketitle
\begin{summary}
  Topography and gravity are geophysical fields whose joint
  statistical structure derives from interface-loading processes
  modulated by the underlying mechanics of isostatic and flexural
  compensation in the shallow lithosphere. Under this dual
  statistical-mechanistic viewpoint an estimation problem can be
  formulated where the knowns are topography and gravity and the
  principal unknown the elastic flexural rigidity of the lithosphere.
  In the guise of an equivalent ``effective elastic thickness'', this
  important, geographically varying, structural parameter has been the
  subject of many interpretative studies, but precisely how well it is
  known or how best it can be found from the data, abundant
  nonetheless, has remained contentious and unresolved throughout the
  last few decades of dedicated study. The popular methods whereby
  admittance or coherence, both spectral measures of the relation
  between gravity and topography, are inverted for the flexural
  rigidity, have revealed themselves to have insufficient power to
  independently constrain both it and the additional unknown
  initial-loading fraction and load-correlation factors, respectively.
  Solving this extremely ill-posed inversion problem leads to
  non-uniqueness and is further complicated by practical
  considerations such as the choice of regularizing data tapers to
  render the analysis sufficiently selective both in the spatial and
  spectral domains. Here, we rewrite the problem in a form amenable to
  maximum-likelihood estimation theory, which we show yields unbiased,
  minimum-variance estimates of flexural rigidity, initial-loading
  fraction and load correlation, each of those separably resolved with
  little \textit{a posteriori} correlation between their estimates. We
  are also able to separately characterize the isotropic spectral
  shape of the initial-loading processes. Our procedure is well-posed
  and computationally tractable for the two-interface case. The
  resulting algorithm is validated by extensive simulations whose
  behavior is well matched by an analytical theory with numerous tests
  for its applicability to real-world data examples.
\end{summary}

\section{I~N~T~R~O~D~U~C~T~I~O~N{\hsps}A~N~D{\hsps}M~O~T~I~V~A~T~I~O~N} 
\label{introduction}

With a remarkable series of papers, all entitled \textit{Experimental
  Isostasy}, Dorman and Lewis heralded in an era of Fourier-based
estimation in geophysics, using gravity and topography to study
isostasy ``experimentally'', that is, without first assuming a
particular mechanistic model such as Airy or Pratt compensation
\cite[]{Dorman+70,Lewis+70a,Lewis+70b,Dorman+72}. All three papers
remain essential reading for us today. 

The first in the series introduced the basic point of view by which
Earth is regarded as a linear time-invariant system and the
unknown ``isostatic response'' is the transfer function:
\begin{quote}
  \textit{The linear system here is the earth: The input is the
    topography, or more precisely, the stress due to the topography
    across some imaginary surface, say sea level, and the output is
    the gravity field due to the resulting compensation.} 
 \cite[][p.~3360.]{Dorman+70}
\end{quote}
In keeping with classical systems identification practice, or in their
words, \textit{through the fruits of linear mathematics, in 
  particular, harmonic analysis and the convolution theorem} 
 \cite[][p.~3358]{Dorman+70}, the
recovery of the impulse response practically suggested itself: 
\begin{quote}
\textit{If the earth is linear in its response to the crustal loading of the
topography, the response of the earth's gravity field to this loading
can be represented as the two-dimensional convolution of the topography
with the earth's isostatic response function. [...] Through
transformation into the frequency domain, the convolution becomes
multiplication, and one is led directly to the result that the
isostatic response function is equal to the inverse transform of the
quotient of the transforms of the Bouguer gravity anomaly and the
topography.} \cite[][p.~3357.]{Dorman+70}
\end{quote}
% The estimate was formed by averaging, in practice, over angles in
% polar coordinates or over orders in spherical harmonic space.
% This averaging is a first stabilization of the noise.
Contingent upon establishing the validity of the linear assumption in
interpreting the data, subsequently, the isostatic response function
was to be ``inverted'', i.e. by computing the \textit{density changes
  at depth that would be required to fit the experimentally determined
  response function} \cite[][p.~3361]{Dorman+70}.
% Under the assumption of strict locality this is the inverse
% Laplace transform of the isostatic response function in the planar
% case or the inverse Mellin transform of the spherical harmonic
% expansion coefficients of the isostatic response function in the
% spherical case.
However, due to various forms of measurement, geological or
modeling ``noise'', \textit{[t]he problems involved in computing the
  inverse [...] of an   experimentally determined function are
  formidable} \cite[][p.~3361]{Dorman+70}, even when strictly local
compensation is assumed and the solution is, in principle, unique.   

The second paper \cite[]{Lewis+70a,Lewis+70b} was devoted to
discussing the numerous geophysical and numerical strategies by which
the least-squares inversion of the experimentally derived response can
be accomplished at all. Broadly speaking, these involve any or all of
(a)~modification of the data, e.g. by windowing prior to Fourier
transformation, (b)~modification of the recovered response, e.g. by
averaging, smoothing, or limiting the frequency interval of interest,
(c)~conditioning of the unknown density profile, e.g. by series
expansion or imposing hard bounds, and (d)~stabilizing the inversion,
e.g. by iteration, frequency weighting, or the addition of minimum
$\ell_1$ norm constraints on the density profile. As a result, many
possible local density profiles can be found that ``explain'', in the
$\ell_2$ sense, the observed response curves, and an appeal has to be
made to independent outside information, e.g. from seismology and
geodynamics, to make the final selection. Regardless of the ultimate
outcome of this exercise in deciding over which depth the compensating
mass anomalies occur, the modeling procedure allows for the 
computation of the so-called ``isostatic anomaly''. The latter is
thereby defined as that portion of the variation in the observed
terrestrial gravity field that cannot be explained by the difference
in measurement position on or above the reference geoid (which leads
to the free-air anomaly), nor of the anomalous mass contained in
the topography above the reference geoid (hence the Bouguer anomaly)
--- but, most importantly, also not by the assumption of a linear
isostatic compensation mechanism, at whichever depth or however
regionally this is being accommodated
\cite[][]{Lambeck88,Blakely95,Watts2001,Turcotte+2002,Hofmann+2006}.

In their third and final paper \cite[]{Dorman+72} the authors employed
\cite{Backus+70} theory to obtain and interpret the result of the
inversion of isostatic response functions by way of depth-averaging
kernels rather than solving for particular profiles, which had shown
considerable non-uniqueness and possibly unphysical oscillations. But
even admitting that only localized averages of the anomalous density
structure could be considered known, the authors concluded that the
available data called for the compensation of terrestrial topography
by density variations down to at least 400~km depth, i.e.  involving
not only Earth's crust but also its mantle.

If in these papers the main objective was to make isostatic anomaly
maps and to recover local density variations at depth to explain the
cause of isostasy where possible, to do the latter reliably arguments
needed to be made that \textit{involve the strength of the crust and
  upper mantle} \cite[][p.~3371]{Lewis+70a}. In practice, this led the
authors to decide that \textit{the constitution of the earth is such
  that it is at least able to support mass anomalies of wavelengths
  equal to the depth at which they occur} \cite[][p.~3383]{Lewis+70a}.
This \textit{contradictio in terminis} (it is no longer a strictly
local point of view) was the very one that led \cite{VeningMeinesz31}
to argue against the hypotheses of Airy and Pratt: strength implies
lateral transfer of stress which is incompatible with the tenets of
local isostasy \cite[]{Lambeck88,Watts2001}.

Following a similar line of reasoning in replacing local by regional
compensation mechanisms, \cite{McKenzie+76} and \cite{Banks+77}
presented a new theoretical framework by which the observed
admittance, indeed the ratio of Fourier-domain gravity anomalies to
topography \cite[]{Karner82}, could be interpreted in terms of a
regional compensation mechanism that involves flexure of a thin
(compared to the wavelength of the deformation) elastic plate (a
``lithosphere'' defined in its response to long-term, as opposed to
seismic stresses) overlying an inviscid mantle (an ``asthenosphere'',
again referring to its behavior over long time scales). No longer was
the local density structure the driving objective of the inversion of
the isostatic response curve, but rather the thickness over which the
density anomalies could plausibly occur, assuming a certain limiting
mantle density. This subversion of the question how to best explain
gravity and topography data became the now dominant quest for the
determination of the flexural rigidity or strength, $D$, of the
lithosphere thus defined. The theory of plates and shells
\cite[]{Timoshenko+59} could then be applied to translate~$D$ into the
``effective'' elastic plate thickness, $T_e$, upon the further
assumption of a Young's modulus and Poisson's ratio. A tripartite
study entitled \textit{An analysis of isostasy in the world's oceans}
\cite[]{Watts78,Cochran79,Detrick+79} went around the globe
characterizing~$T_e$ in a plate-tectonic context. Subsequent additions
to the theory involved a few changes to the physics of how deformation
was treated, e.g. by considering that the isostatic response may be
anisotropic \cite[]{Stephenson+80}, taking into account non-linear
elasticity and finite-amplitude topography \cite[]{Ribe82},
visco-elasticity and erosional feedbacks \cite[]{Stephenson84}, and
updating the force balance to include also lateral, tectonic, stresses
\cite[]{Stephenson+85a}. None of these considerations changed the
basic premise. With the methodology for effective elastic thickness
determination firmly established, the way was paved for its
rheological interpretation \cite[e.g.][]{McNutt+82,McNutt84,Burov+95}.

A first hint that not all was well in the community came when 
transfer function theory was applied to measure the
strength of the continents. \cite{McNutt+78} concluded from
admittance analysis that, on the whole, Australia (an old
continent) might not have any strength, and would thus be in
complete local isostatic equilibrium. On the contrary,
\cite{Zuber+89} concluded on the basis of coherence analysis
that the Australian continental effective elastic thickness well
exceeded 100~km.  This apparent contradiction was found despite the
observed admittance and coherence being merely different ``summaries''
of gravity and topography: spectral ratios that both estimate the
underlying isostatic transfer function. At least part of the
discrepancy could be ascribed to the treatment of subsurface loads in
the formulation of the forward model \cite[]{Forsyth85}. With
\cite{Bechtel+90}, and numerous others after them, these authors led
the next decade in which a ``thick'' (greater than 100~km) continental
lithosphere was espoused.  Then, \cite{McKenzie+97} started a decade
of making effective arguments %advocacy
for ``thin'' continents (no more than 25~km), a controversial position
with many ramifications \cite[]{Jackson+2002,Burov+2006} that was
hotly contested and remains so today
\cite[]{Banks+2001,Swain+2003b,McKenzie2003,McKenzie2010}.  

Three developments happened on the way to the current state,
with sound arguments made on both sides of the debate.
Inverting coherence between Bouguer gravity and topography yielded
thicker lithospheres than working with the admittance between the
free-air gravity and the topography.  
There was discussion over the
treatment of ``buried loads'' and how to solve for the
subsurface-to-surface loading ratio. Finally, there were arguments
over the best way by which to form spectral estimates of either
admittance or coherence. Among others, \cite{Perez+2004},
\cite{Perez+2005} and \cite{Kirby+2009} % attempted diplomacy
provided some reconciliation by making
estimates of effective elastic thickness that were based on both
free-air admittance and Bouguer coherence, respectively. They argued
the equivalence of the results when either method was applied in a
``consistent'' formulation, taking into account the finite window size
of any patch of available data. Still, large differences remained,
experiments on synthetic data showed significant bias and large
variance, and a clear consensus failed to arise. \cite{Macario+95},
\cite{McKenzie2003} and \cite{Kirby+2009} investigated the effect of
the statistical correlation between surface and subsurface loads. For
their part, \cite{Diament85}, \cite{Lowry+94},
\cite{Simons+2000,Simons+2003a}, \cite{Ojeda+2002},
\cite{Kirby+2004,Kirby+2008a,Kirby+2008b} and \cite{Audet+2007}
focused on the spectral estimation of admittance and coherence via
maximum-entropy, multitaper and wavelet-based methods, and identified
the spectral bias, leakage and variance inherent in those.  Much as
the controversy involved the geological consequences of a
thick versus a thin lithosphere, with only gravity and topography as
the primary observations and no significant divergence in viewing the
physics of the problem, that is, of elastic flexure in a
multilayered system, over time the arguments evolved into a debate
that was mostly about spectral
analysis. Least-squares fitting of admittance and coherence functions,
however determined, had become synonymous with the process of
elastic-thickness determination.  

The appropriateness of using least squares is not something that can
be taken for granted but rather needs to be carefully assessed, as
was pointed out early on in this context by \cite{Dorman+72}, \cite{Banks+77},
\cite{Stephenson+80} and \cite{Ribe82}, which, however, also focused
on other issues that have since received more attention. 
Admittance and coherence are ``statistics'': functions of the data
with non-Gaussian distributions even if the data themselves are Gaussian
\cite[]{Munk+66,Carter+73,Walden90b,Thomson+91,Touzi+96,Touzi+99}. Estimators
for flexural rigidity based on any given method have their own
distributions, though not necessarily ones with a tractable form. Without
knowledge of the joint properties of admittance- and coherence-based
estimators it is impossible to assess the relative merits of any
method for a given data set or true parameter regime; with current
state-of-the-art understanding it is not even clear if the two methods
are statistically inconsistent.  

At this juncture this paper aims for a return to the basics, by asking
the question: ``What information does the relation between
  gravity and topography contain about the (isotropic) strength of the
  elastic lithosphere?'' and by formulating an answer that returns the
full statistical distribution of the estimates derived from such data.
As such, it should %  encompass all prior work
provide a framework for the interpretation of the early work on which
we build: as others before us we
are merely using the measurable ingredients of gravity, topography and
the flexure equations. However, as we shall see, we do not need to consider 
this a two-step process by which first the transfer function needs to
be estimated non-parametrically and then the inversion for structural
parameters performed with the estimated transfer function as ``data''.
This approach amounts to a loss of most of the degrees of freedom in
the data, replacing them with spectral ratios estimated at a much
smaller set of wavenumbers, and with much of the important information
on the flexural rigidity compromised due to lack of resolution at the
low wavenumbers. Rather, we can treat it as an optimization problem that
uses everything we know about gravity and topography available as data
to directly construct a maximum-likelihood solution for the
lithospheric parameters of interest. These are returned together with
comprehensive knowledge of their uncertainties and dependencies, and
with a statistical apparatus to evaluate how well they explain the
data; the analysis of the residuals then informing us where the
modeling assumptions were likely violated. By the principle of
functional invariance the maximum-likelihood solution for elastic
thickness and loading ratio also returns the maximum-likelihood
estimates of the coherence and admittance themselves, which can then
be compared to those obtained by other methods. Admittance may be
superior to coherence, or vice versa, in particular scenarios, but
only maximum-likelihood, by definition, produces solutions that are
preferred globally for all parameter regimes
\cite[]{Pawitan2001,Severini2001,Young+2005}. Finally, we note that
understanding the likelihood is also a key component of fully Bayesian
solution approaches \cite[e.g.][]{Mosegaard+95,Kaipio+2005}.  

\section{B~A~S~I~C{\hsps} F~R~A~M~E~W~O~R~K}
\label{basics}

Despite their singular focus on deriving density profiles to
reconstruct the portion of the Bouguer gravity field that is linearly
related to the topography and thereby ``explain'' the isostatic
compensation of surface topography to first order, even when the
strength of the lithosphere had to be effectively prescribed, Dorman
and Lewis' \textit{Experimental Isostasy}~1, 2 and~3 contained
virtually all of the elements of the analysis of gravity and
topography by which the problem could be turned around to the, in the
words of \cite{Lambeck88} ``vexing'', question ``What is the
flexural strength of the lithosphere''? The elements applicable to
the analysis were the expressions for admittance and coherence between
topography and the Bouguer, free-air, and isostatic residual gravity
anomalies, the averaging or smoothing required to statistically stabilize the
estimate of the transfer function that is the intermediary between the
data and the model obtained by inversion for the unknown parameters
(if not the density distribution, then the mechanical properties of the
plates), the notion of correlated and uncorrelated noise of various
descriptions: indeed all of the ingredients that will form the
vernacular of our present contribution. In this section we redefine all
primary quantities of interest in a manner suitable for the
statistical development of the problem.

We treat Earth locally as a Cartesian system. Our chosen
coordinate system has $\mbf{x}=(x_1,x_2)$ in the horizontal plane and
defines~$\mbf{\hat{z}}$ pointing up: depths in Earth are
negative. A density contrast located at interface~$j$ is found at
depth~$z_j\le0$, and is denoted 
\be\label{dj}
\Dj=\rho_j-\rho_{j-1}.
\ee
Two layers is the minimum required to capture the full complexity of
the general problem which may, of course, contain any number of
layers. In a simple two-layer system, the first interface, at $z_1=0$,
is the surface of the solid Earth, and $\rho_0$ is the density of the
air (or water) overlying it. The density of the crust is  $\rho_1$,
and the second interface, at $z_2\le0$,  separates the crust from the
mantle with density $\rho_2$. 

For now we use the term ``topography'' very generally to describe any
departure from flatness at any surface or subsurface interface. By
``gravity'' we mean the ``anomaly'' or ``disturbance''; both are
differences in gravitational acceleration with respect to a certain
reference model. These departures in elevation and acceleration are
all small: we consider topography to be a small height perturbation of
a constant-depth interface, and neglect higher-order finite-amplitude
effects on the gravity. We always assume that the ``loads'', the
stresses exerted by the topography, occur at the density interfaces
and not anywhere else. If not in the space domain, $\xb$, we will work
almost exclusively in the Fourier domain, using the wave vector~$\kb$
or wavenumber (spatial frequency) $k=\|\kb\|$. We only distinguish
between both domains when we need to, and then only by their
argument. All of this corresponds to standard
practice~\cite[]{Watts2001}.  

\begin{figure}\centering
\includegraphics[width=\abit\textwidth]{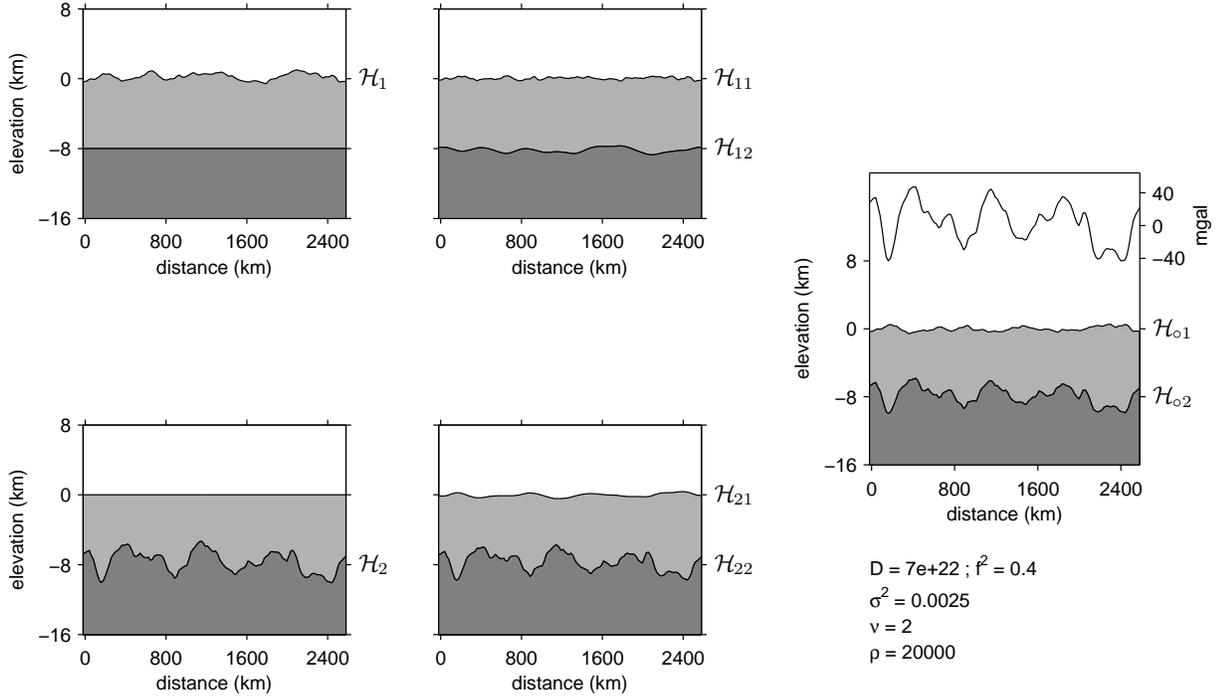}
\caption{\label{oh0}Synthetic data representing the standard model,
  identifying the initial, $\Hj$, equilibrium, $\Hij$, and final
  topographies, $\Hnj$, emplaced on a lithosphere with flexural
  rigidity~$D$. The initial loads were generated from the Mat\'ern
  spectral class with parameters $\st$, $\rho$ and $\nu$; they were
  not correlated, $r=0$, and the spectral proportionality was $\ft$.
  Also shown, by the black line, is the Bouguer gravity anomaly,
  $\Gpn$. The density contrasts used were $\Do=2670$~kg$\hsom$m$^{-3}$
  and $\Dt= 630$~kg$\hsom$m$^{-3}$, respectively. All symbols and
  processes are clarified in the text. They will furthermore be
  identified and briefly explained in Table~\ref{tablelitho}.}
\end{figure}

Looking ahead we draw the readers' attention to Fig.~\ref{oh0},
which contains a graphical representation of the
problem. Fig.~\ref{oh0} is, in fact, the result of a data simulation
with realistic input parameters. Many of the details of its construction
remain to be introduced and many of the symbols remain to be
clarified. What is important here is that we seek to build an
understanding of how, from the observations of gravity and topography,
we can invert for the flexural rigidity of the lithosphere in this
two-layer case. The observables (rightmost single panel) are the sum
of the flexural responses (middle panels) of two initial
interface-loading processes (leftmost panels)  which have occurred in
unknown proportions and with unknown correlations between them. 

\subsection{Spatial and spectral representation, theory and
  observation}
\label{Cramer}

Writing~$\mcH$ and~$\mcG$ without argument we will be referring quite
generically to the random processes  ``topography'' and ``gravity''
respectively, though when we consider either physical quantity
explicitly in the spatial or spectral domain we will distinguish them
accordingly as 
\be\label{generic0}
\mcH\ofx\quad\mbox{or}\quad\mcG\ofx\qquad \mbox{(in space)},\also
d\mcH\ofk\quad\mbox{or}\quad d\mcG\ofk\qquad \mbox{(in spectral space)},
\ee
where they depend on spatial position~$\xb$ or on wave vector~$\kb$, respectively.
In doing so we use to the \cite{Cramer42} spectral
representation under which~$d\mcH\ofk$ and~$d\mcG\ofk$ are
well-defined orthogonal increment
processes~\cite[]{Brillinger74,Percival+93}, in the sense that at any
point in space we may write
\be\label{fourierfirst}
\mcH\ofx
=\intnyq e^{i\kb\cdot\xb}\,d\mcH\ofk\also
\mcG\ofx
=\intnyq e^{i\kb\cdot\xb}\,d\mcG\ofk
.
\ee
We make the assumption of stationarity such that for every
point~$\xb$ under consideration all equations of the type~(\ref{fourierfirst})
are statistically equivalent.
% Expectation, covariance only depend on separation not location. 
 We further assume that both processes will be either
strictly bandlimited or else decaying very fast with increasing
wavenumber~$k=\|\kb\|$ such that we may restrict all integrations over
spectral space to the Nyquist
plane~$\kb\in[-\pi,\pi]\times[-\pi,\pi]$. While this is certainly a
geologically reasonable assumption we would at any rate be without
recourse in the face of the broadband bias and aliasing
that would arise unavoidably if it were violated. For simplicity~$\xb$ maps out a
rectangle that can be sampled on an~$M\times N\approx 2K$ grid given by
\be\label{xgrid}
\xb=\big\{
%(m\,\Delta x_1, n\,\Delta x_2)
(m, n)
\,: \quad m=0,\dots,M-1\
\,; \quad n=0,\dots,N-1\
\big\}
.
\ee
In the non-rarified world of geophysical data analysis we will not be
dealing with stochastic processes directly, rather with particular
realizations thereof. These are our gravity and topography
data, observed on finite domains, to which we continue to
refer as~$\mcH\ofx$ and~$\mcG\ofx$. The modified Fourier transform of
these measurements, obtained after sampling and windowing with a
certain function~$w_K(\xb)$, is
\be
\label{fourierH} H\ofk=\sum_\xb
w_K\ofx \mcH\ofx e^{-i\kb\cdot\xb} = \sum_\xb w_K\ofx\Bigg(\,\,\intnyq
e^{i\kb'\cdot\xb}\,d\mcH\ofkp\Bigg) e^{-i\kb\cdot\xb}= \intnyq
W_K(\kb-\kb')\,d\mcH\ofkp .  
\ee 
In this expression $W_K\ofk$ is the unmodified Fourier transform of
the energy-normalized applied window,
\be
\label{Wdefine} W_K\ofk=\sum_\xb w_K\ofx e^{-i\kb\cdot\xb} .  
\ee
The spectral density or spectral covariance of continuous stationary
processes is defined as the ensemble average (denoted by angular
brackets)  
\be
\label{specdens} \langle d\mcH\ofk \hsom d\mcH^*\hsomm\ofkp\rangle=
\mcS_{\mcH\mcH}\ofk\dtbk\dkkp , 
\ee
whereby we denote complex conjugation with an asterisk and $\dkkp$ is
the Dirac delta function. There can be no covariance between non-equal
wavenumbers if the spatial covariance matrix is to be dependent on
spatial separation and not location, as from eqs~(\ref{fourierfirst})
and~(\ref{specdens}) 
\be\label{specdensx} 
\langle \mcH\ofx \mcH^*\hsomm\ofxp\rangle=
\intnyqt e^{i\kb\cdot\xb} e^{-i\kbp\cdot\xbp} \langle d\mcH\ofk
d\mcH^*\hsomm\ofkp\rangle =\intnyq e^{i\kb\cdot(\xb-\xbp)}
\mcS_{\mcH\mcH}\ofk\dbk =\mcC_{\mcH\mcH}(\xb-\xbp) .  
\ee 
In contrast to eq.~(\ref{specdens}), as follows readily from
eqs~(\ref{fourierH}) and~(\ref{specdens}), the covariance between the
modified Fourier coefficients of the finite sample is 
\be\label{specblur} \langle H\ofk
H^*\hsomm\ofkp\rangle= \intnyqt
W^{}_K(\kb-\kb'')W_K^*(\kb'-\kb''')\,\langle d\mcH\ofkpp
d\mcH\ofkppp\rangle= \intnyq
W^{}_K(\kb-\kb'')W_K^*(\kb'-\kb'')\hsom\mcS_{\mcH\mcH}(\kb'')\dbk'' .
\ee 
Eqs~(\ref{fourierH}) and~(\ref{specblur}) show that the theoretical
fields~$d\mcH\ofk$ and their spectral densities~$\mcS_{\mcH\mcH}\ofk$
are out of reach of observation from spatially finite sample sets.
Spectrally we are always observing a version of the ``truth'' that is
``blurred'' by the observation window. Even if, or rather, especially
when the windowing is implicit and only consists of transforming a
certain rectangle of data, this effect will be felt. For example,
whereas the true spectral density is obtained by Fourier
transformation of the covariance at all lags, denoted by the summed
infinite series
\be\label{infinite} \mcS_{\mcH\mcH}\ofk=\sum_{-\infty}^{+\infty}
e^{-i\kb\cdot\yb} \mcC_{\mcH\mcH}(\yb)
=\intnyq\sum_{-\infty}^{+\infty} e^{-i(\kb-\kbp)\cdot\yb}
\mcS_{\mcH\mcH}\ofkp\dbk' =\intnyq\delta(\kb,\kbp)\,
\mcS_{\mcH\mcH}\ofkp\dbk' , 
\ee
a blurred spectral density is what we obtain after observing only a
finite set, denoted by the summed finite series 
\ber\label{finitea}
\Sbar_{\mcH\mcH}\ofk&=& \sum_\yb e^{-i\kb\cdot\yb}
\mcC_{\mcH\mcH}(\yb)
=\frac{1}{K}\sum_\xb\sum_\xbp e^{-i\kb\cdot\xb}e^{i\kb\cdot\xbp}
\intnyq e^{i\kbp\cdot\xb} e^{-i\kbp\cdot\xbp}
\mcS_{\mcH\mcH}\ofkp\dbk'
\\
&=&
\intnyq 
\frac{1}{K}\sum_\xb\sum_\xbp 
e^{-i(\kb-\kbp)\cdot\xb}
e^{i(\kb-\kbp)\cdot\xbp} 
\mcS_{\mcH\mcH}\ofkp\dbk'
=\intnyq \left|F_K(\kb-\kbp)\right|^2 \mcS_{\mcH\mcH}\ofkp\dbk'
,\label{finiteb}
\eer
with~$|F_K|^2$ denoting Fej\'er's kernel~\cite[]{Percival+93}. 
The design of suitable windowing functions \cite[in this geophysical
context, see, e.g.,][]{Simons+2000,Simons+2003a,Simons+2011a}, is
driven by the desire to mold what we can calculate from the
observations into estimators of these ``truths'' that are as ``good''
as possible, e.g. in the minimum mean-squared error sense; we will
keep the windows or tapers~$w_K(\xb)$ and the convolution
kernels~$W_K\ofk$ generically in all of the formulation. For the
gravity observable, whose spectral density is denoted
$\mcS_{\mcG\mcG}$, we find the modified Fourier coefficients and the
spectral covariance, respectively, as
\be\label{fourierG}
G\ofk=
\intnyq W_K(\kb-\kb')\,d\mcG\ofkp
\also
\langle G\ofk G^*\hsomm\ofkp\rangle= \intnyq
W_K(\kb-\kb'')W^*_K\hsomm(\kb'-\kb'')\hsom\mcS_{\mcG\mcG}(\kb'')\dbk''
.
\ee
Finally, we will need to sample $H\ofk$, $W_K\ofk$, and $G\ofk$ on a
grid of wavenumbers. Exploiting the Hermitian symmetry that applies in
the case of real-valued physical quantities, for an~$M\times N$ data
set we select the half-plane consisting of the~$K=M\times (\lfloor
N/2\rfloor+1)$ wave 
vectors %that satisfy
\be\label{kgrid}
\kb=
\left\{
\left(\frac{2\pi}{M}\left[-\left\lfloor\frac{M}{2}\right\rfloor+m\right],\,
\frac{2\pi}{N} n\right)
\quad:\quad
m=0,\dots,M-1\,;\quad
n=0,\dots,\left\lfloor\frac{N}{2}\right\rfloor
\right\}
.
\ee
The quantities $H\ofk$, $W_K\ofk$, and $G\ofk$ are complex except at the
dc wave vectors $(0,0)$ and the Nyquist wave vectors $(0,\pi)$,
$(-\pi,0)$ and $(-\pi,\pi)$ if they exist in eq.~(\ref{kgrid}), which
depends on the parity of~$M$ and~$N$. 

\subsection{Topography}\label{sectopo}

As mentioned before, we apply the term ``topography'', $\mcH$,
generically to any small perturbation of the Cartesian  
reference surface, which is assumed to be flat. Specifically, we need
to distinguish between what we shall call `initial', `equilibrium' and
`final' topographies, respectively. In the classic multilayer 
loading scenario reviewed by, e.g., \cite{McKenzie2003} and
\cite{Simons+2003a}, as the $j$th~interface gets loaded by
an initial topography, the singly-indexed quantity~$\Hj$, a
configuration results in which each of the interfaces expresses this
loading by assuming an equilibrium topography, which is identified as
the double-indexed quantity~$\Hij$. The first subscript refers to
the interface on which the initial loading occurs; the second to the
interface that reflects this process. The state of this equilibrium is
governed by the laws of elasticity, as we will see in the next
section. All of these equilibrium configurations combine into what we
shall call the final topography on the $j$th~interface,
namely $\Hnj$, where the~$\circ$ is meant to evoke the
summation over all of the interfaces that have generated
initial-loading contributions. 

Thus, in a two-layer scenario, what in common parlance is called
``the'' topography, i.e. the final, observable height of mountains and the
depth of valleys expressed with respect to a certain neutral reference
level, will be called~$\Hno$, and this then will be the sum of the two
unobservable components $\Hoo$ and $\Hto$. In other words, the final
``surface'' topography is
\be\label{H01}
\Hno=\Hoo+\Hto
.
\ee
Likewise, the final ``subsurface'' topography, $\Hnt$, is
given by the sum of two 
unobservable components $\Hot$ and $\Htt$, totaling
\be\label{H02}
\Hnt=\Hot+\Htt
.
\ee
This last quantity,~$\Hnt$, is not directly observable but can be
calculated from the Bouguer gravity anomaly, as we describe
below. Both $\Hoo$ and $\Hot$ refer to the same geological
loading process occurring on the first 
interface but being expressed  on the first and second interfaces,
respectively. In a similar way, $\Hto$ and $\Htt$ refer to the
process loading the second interface which thereby produces topography
on the first and second interfaces, respectively.  

While postponing the discussion on the mechanics to the next section
it is perhaps intuitive that a positive height perturbation at one
interface creates a negative deflection at another interface:
``mountains'' have ``roots'', as has been known since the days of
Airy~\cite[]{Watts2001}. The initial-loading topography, then, is given by
the difference between these two equilibrium components. At the first
and second interfaces, respectively, we will have  for the initial
topographies at the surface and subsurface, respectively,
\ber
\Ho&=&\Hoo-\Hot\label{H1},\\
\Ht&=&\Htt-\Hto\label{H2}
.
\eer

The sum of all of the equilibrium topographies, at all of the
interfaces in this system and thus requiring two subscripts $\circ$,
is given by 
\be\label{H00}
\Hpf=\Hoo+\Hot+\Hto
+\Htt
,
\ee
which is a quantity that we can only access through the free-air
gravity anomaly that it generates, as we shall see. 

\subsection{Flexure}\label{secflex}

Mechanical equilibrium exists between $\Hoo$ and $\Hot$ on the
one hand, and $\Hto$ and $\Htt$ on the other. The equilibrium
refers to the balance between hydrostatic driving and restoring
stresses, which depend on the density contrasts, and the stresses
resulting from the elastic strength of the lithosphere. Introducing
the  flexural rigidity~$D$, in units of N$\hsom$m, we obtain the biharmonic
flexural or plate equation~\cite[]{Banks+77,Turcotte+82} as follows on
the first 
(surface) interface:  
\begin{subequations}
\label{biharm0}
\be
\left(\nabla^4+\frac{g\Dt}{D}\right)\Hot\ofx
=-\frac{g\Do}{D}\Hoo\ofx
,
\label{biharm1}
\ee
and at the second (subsurface) level, we have 
\be
\left(\nabla^4+\frac{g\Do}{D}\right)\Hto\ofx
=-\frac{g\Dt}{D}\Htt\ofx
.\label{biharm2}
\ee
\end{subequations}
The mechanical constant~$D$ is the objective of our study:
geologically, this yields to what is commonly referred to as the
``integrated strength'' of the lithosphere, which can be usefully
interpreted under certain assumptions as an equivalent or
``effective'' elastic thickness. This quantity, $T_e$, in units of m,
relates to~$D$ by a simple scaling involving the Young's modulus~$E$
and Poisson's ratio, $\nu$, as is well
known~\cite[e.g.][]{Ranalli95,Watts2001,Kennett+2008}. Here we follow
these authors and simply define
\be\label{tedef}
D=\frac{ET_e^3}{12(1-\nu^2)}
.
\ee

Much has been written about what $T_e$ really ``means'' in a
geological
context~\cite[]{Lowry+94,Burov+95,Lowry+95,McKenzie+97,Burov+2006}.
This discussion remains outside of the scope of this study. Moreover,
eqs~(\ref{biharm0}) are the only governing equations that we
shall consider in this problem. It is not
exact~\cite[e.g.][]{McKenzie+76,Ribe82}, it is not
complete~\cite[e.g.][]{Turcotte+82}, and it may not even be right
\cite[e.g.][]{Karner82,Stephenson+85a,McKenzie2010}. For that matter,
a single, isotropic $D$ may be an oversimplification
\cite[]{Stephenson+80,Lowry+95,Simons+2000,Simons+2003a,Audet+2004a,Swain+2003b,Kirby+2006}.
However, the neglect of higher-order terms, additional tectonic terms
in the force balance, time-dependent visco-elastic effects and elastic
anisotropy remain amply justified on geological grounds. It should be
clear, however, that any consideration of such additional complexity
will amount to a change in the governing equations~(\ref{biharm0}),
which we reserve for further study. 

At the surface, eq.~(\ref{biharm1}) is solved in the Fourier domain as
\be\label{H12}\label{H12xidef}
d\Hot\ofk  =-d\Hoo\ofk  \Dou\Dt^{-1}\xi^{-1}\ofsk
,
\ee
where we have defined the dimensionless wavenumber-dependent transfer
function baptized by~\cite{Forsyth85} 
\be
\xik=\dekf
\label{xidef}
.
\ee
At the subsurface, eq.~(\ref{biharm2}) has the solution
\be\label{H21}\label{H21phidef}
d\Hto\ofk  =-d\Htt\ofk  \Do^{-1}\Dtu\hsom\phi^{-1}\ofsk
,
\ee
with the dimensionless filter function
\be
\phik=\dekp
.\label{phidef}
\ee
All of the physics of the problem is contained in the equations in
this section. As a final note we draw attention to the
assumption that the interfaces at which topography is generated and
those on which the resulting deformation is expressed coincide: this
is the first of the important simplifications introduced
by~\cite{Forsyth85}. This assumption, though not universally
made~\cite[e.g.][]{McNutt83,Banks+2001}, is broadly held to be
valid. Finding~$D$ in this context is the estimation problem with
which we shall concern ourselves.

\subsection{Gravity}\label{secgrav}

Every perturbation from flatness by topography generates a
corresponding effect on the gravitational acceleration when compared
to the reference state. We relate the gravity anomaly to the
disturbing topography by the density perturbation~$\Dj$ and
account for the exponential decay of the gravity field from the
depth~$z_j\le 0$ where it was generated. The ``free-air''
gravitational anomaly~\cite[]{Hofmann+2006} from the topographic
perturbation at the $j$th~interface that results from the
$i$th~loading process is given in the spectral domain by  
\be\label{fanom}
d\mcG_{ij}\ofk  =2\pi G\Dj d\Hij\ofk  e^{kz_j}
,
\ee
where~$G$ is the universal gravitational constant, in
m$^3$$\hsom$kg$^{-1}$s$^{-2}$, not to be confused with the gravity anomaly
itself. Once again this equation is inexact in 
assuming local Cartesian geometry~\cite[]{Turcotte+82,McKenzie2003}
and neglecting higher-order finite-amplitude
effects~\cite[]{Parker72,Wieczorek+98}, but for our purposes, this
``infinite-slab approximation'' will be good enough. The observable
free-air anomaly is the sum of all contributions of the
kind~(\ref{fanom}), thus in the two-layer case
\be\label{fanomall}
d\Gpf\ofk  =d\mcG_{11}\ofk  +d\mcG_{12}\ofk  +d\mcG_{21}\ofk  
+d\mcG_{22}\ofk   
.
\ee
The Bouguer gravity anomaly is derived from the free-air anomaly by
assuming a non-laterally varying density contrast across the surface
interface. It thus removes the gravitational effect from the
observable surface topography~\cite[]{Blakely95}, and is given by
\ber
d\Gpn\ofk  &=&d\mcG_{12}\ofk  +d\mcG_{22}\ofk\\
&=&2\pi G\Dt\hsom\ekzt d\Hnt\ofk\label{todown}\\
&=&2\pi G\Dt\hsom\ekzt
\left[d\Hot\ofk  +d\Htt\ofk  \right]\\
&=&-2\pi G\Dt\hsom\ekzt
\left[d\Hoo\ofk  \Dou\Dt^{-1}\xi^{-1}\ofsk
-d\Htt\ofk  \right]\label{bga}
.
\eer
In this reduction, we have used eqs~(\ref{fanom})--(\ref{fanomall}),
(\ref{H02}) and~(\ref{H12}). For simplicity we shall write the Bouguer
anomaly as
\be\label{Gpnchidef}
\label{banomaly}
d\Gpn\ofk=\chik\hsom d\Hnt\ofk
,
\ee
defining one more function, which acts like a harmonic ``upward continuation''
operator~\cite[]{Blakely95},
\be\label{chidef}
\chik=2\pi G \Dt\hsom\ekzt
.
\ee

At this point we remark that topography and gravity, in one form or
another, are the only measurable geophysical quantities to help us
constrain the value of~$D$. The Bouguer anomaly~$\Gpn$ is usually
computed from the free-air anomaly~$\Gpf$ and the
topography~$\Hno$, assuming a density contrast~$\Do$. Any estimation
problem that deals with any combination of these variables should thus
yield results that are equivalent to within the error in the estimate
\cite[]{Tarantola2005}, though whether the free-air or the Bouguer
gravity anomaly is used as the primary quantity in the estimation
process could have an effect on the properties of the solution
depending on the manner by which it is found --- a paradox that this
paper will eliminate.

\subsection{Observables, deconvolution, and loading} 
\label{observables}

We are now in a position to return to writing explicit forms for the
theoretical observables from whose particular realizations (the data),
ultimately, we desire to estimate the flexural rigidity~$D$. These are
the final ``surface'' topography,  given by combining eqs~(\ref{H01})
and~(\ref{H21phidef}) as  
\be\label{gho1}
d\Hno\ofk=d\Hoo\ofk-d\Htt\ofk\hsom\Do^{-1}\Dtu\hsom \phi^{-1}\ofsk.
\ee
By analogy we shall write for the final ``subsurface'' topography that
which we can obtain by ``downward continuation'' \cite[]{Blakely95} of
the Bouguer gravity anomaly. From eq.~(\ref{Gpnchidef}), or combining
eqs~(\ref{H02}) and~(\ref{H12xidef}) this quantity is then
\be
\label{gho2}
d\Hnt\ofk=\chi^{-1}\ofsk\hsom d\Gpn\ofk=
-d\Hoo\ofk\hsom\Dou\Dt^{-1}\xi^{-1}\ofsk
+d\Htt\ofk 
.
\ee
The dependence on the parameter of interest, the flexural
rigidity~$D$, is non-linear through the ``lithospheric filters''
$\phi$~and~$\xi$. While both~$\Hno$ and $\Hnt$ can thus be
``observed'' (or at least calculated from observations) we are for the
moment taciturn about the  complexity caused by the potentially
unstable inversion of the parameter~$\chi$ \cite[see
also][]{Kirby+2011}. We return to this issue in
Section~\ref{modeltest}.  

Combining eqs~(\ref{H1})--(\ref{H2}) with
eqs~(\ref{H12xidef})--(\ref{H21phidef}) and then substituting the
results in eqs~(\ref{gho1})--(\ref{gho2}) yields the equations that
relate the observed topographies on either interface with the applied
loads. Without changing from the expressions first derived by
\cite{Forsyth85} these have come to be called the
``load-deconvolution'' equations
\cite[]{Lowry+94,Banks+2001,Swain+2003a,Perez+2004,Kirby+2008a,Kirby+2008b}.
They are usually expressed in matrix form as
\be\label{loadcon}
\dcHo=
\omats{cc}
\fracd{\Dt\hsom\xik}{\Do+\Dt\hsom\xik} &
\fracd{-\Dt}{\Do\phik+\Dt} \\
\fracd{-\Do}{\Do+\Dt\hsom\xik} &
\fracd{\Do\phik}{\Do\phik+\Dt}
\cmats
\omats{c}
d\Ho\ofk \\
d\Ht\ofk 
\cmats,
\ee
with the inverse relationships given by
\be\label{loaddecon}
\omats{c}
d\Ho\ofk \\
d\Ht\ofk 
\cmats
=
\frac{1}{\phik\xik-1}
\omats{cc}
\fracd{\phik[\Do+\Dt\hsom\xik]}{\Dt} &
\fracd{\Do+\Dt\hsom\xik}{\Do} \\
\fracd{\Do\phik+\Dt}{\Dt} &
\fracd{\xik[\Do\phik+\Dt]}{\Do}
\cmats
\dcHo.
\ee
It should be noted that when~$D=0$, in the absence of any lithospheric
flexural strength, thus in the case of complete Airy isostasy,
$\phixi=1$ at all wavenumbers, and no such solutions exist. In that
case the problem of reconstructing the initial loads has become
completely degenerate. 

Armed with these solutions we can solve for the equilibrium
loads. Combining eqs~(\ref{H1})--(\ref{H2}) with
eqs~(\ref{H12xidef})--(\ref{H21phidef}) returns usable forms for
$\Hoo$ and $\Htt$, and substituting the results back into
eqs~(\ref{H12xidef})--(\ref{H21phidef}) returns $\Hot$ and
$\Hto$, all in terms of the initial loads~$\Ho$ and
$\Ht$, as
\begin{subequations}
\ber
d\Hoo\ofk&=&d\Ho\ofk\fracd{\Dt\hsom\xik}{\Do+\Dt\hsom\xik}\also
d\Htt\ofk\hspace{0.8em}=\hspace{0.8em}
d\Ht\ofk\fracd{\Do\phik}{\Do\phik+\Dt}\\ 
d\Hot\ofk&=&d\Ho\ofk\fracd{-\Do}{\Do+\Dt\hsom\xik}\also
d\Hto\ofk\hspace{0.8em}=\hspace{0.8em}
d\Ht\ofk\fracd{-\Dt}{\Do\phik+\Dt}
.
\eer
\end{subequations}

To complete this section we formulate the initial-loading stresses,
in kg$\hsom$m$^{-1}$s$^{-2}$, at each interface as 
\ber
\mcI_1&=&g\Do\Ho,\label{I1}\\
\mcI_2&=&g\Dt\Ht.\label{I2}
\eer

All variables that we have introduced up to this point are listed in
Table~\ref{tablelitho}, to which we further refer for units and short
descriptions. We are now also in the position of further interpreting
Fig.~\ref{oh0}, once again drawing the readers' attention to the heart
of the problem, which is the estimation of the single parameter, the
flexural rigidity $D$, which is responsible for generating, from the
initial loads (left), the equilibrium topographies (middle) whose
summed effects (right) we observe in the form of ``the'' topography
and the (Bouguer) gravity anomaly. 

\begin{table*}
\centering
\begin{tabular}{cclcc}\hline 
 & unit & description & character & eq.\\\hline
$D$ & N$\hsom$m & flexural rigidity of the lithosphere & \textbf{estimated} &
  (\ref{biharm0}) \\[-1em] 
& & & \\
$g$ & m$\hsom$s$^{-2}$ & reference gravitational acceleration & assumed & (\ref{biharm0})\\
$G$ & m$^3$$\hsom$kg$^{-1}$s$^{-2}$ & universal gravitational constant &
  assumed & (\ref{biharm0}) \\
$z_j$ &  m & location of $j$th interface & assumed & (\ref{dj})\\ 
$\Delta_j$ & kg$\hsom$m$^{-3}$ & density contrast across $j$th interface & assumed
& (\ref{dj})\\
[-0.75em] 
& & & \\
$\Ho$ & m & initial topography applied by loading of interface
1 & calculable & (\ref{H1}) \\ 
$\Ht$ & m & initial  topography applied by loading of
interface 2 & calculable & (\ref{H2}) \\ 
$\mcI_{1}$ & Pa & initial load applied at interface 1 & calculable & (\ref{I1}) \\
$\mcI_{2}$ & Pa & initial load applied at interface 2 & calculable & (\ref{I2}) \\[-0.5em]
& & & \\
$\Hoo$ & m & equilibrium topography of interface 1 produced by loading at
interface 1  & calculable   & (\ref{H12xidef})\\
$\Hto$ &  m & equilibrium  topography of interface 1 produced by loading at
 interface 2 & calculable  & (\ref{H21phidef})\\
$\Hno$ &  m & final topography of interface 1 resulting from all
interface loading & \textit{measured} & (\ref{H01}) \\[-0.75em]
& & & \\
$\Hot$ &  m & equilibrium topography of interface 2 produced by loading at
interface 1  & calculable  & (\ref{H12xidef})  \\
$\Htt$ &  m & equilibrium topography of interface 2 produced by loading at
 interface 2 & calculable  & (\ref{H21phidef})\\
$\Hnt$ &  m & final topography of interface 2 resulting from all
interface loading & \textit{calculated} & (\ref{H02}) \\[-0.75em]
& & & \\
$\xi$ &  & filter relating topographies on both interfaces resulting
from loading at interface 1  & calculated
& (\ref{xidef})\\
$\phi$ & & filter relating topographies on both interfaces resulting
from loading at interface 2  & calculated & (\ref{phidef})\\
$\chi$ &  & filter by which final topography on interface 2 maps into
 Bouguer anomaly  & calculated& (\ref{chidef})\\[-0.75em]
& & & \\
$\Hpf$& m & sum of all topographic expressions
of all loading processes & calculable & (\ref{H00})\\
$\Gpf$ & m$\hsom$s$^{-2}$ & free-air gravitational anomaly due to all loading and flexure &
\textit{measured} & (\ref{fanomall})\\  
$\Gpn$ & m$\hsom$s$^{-2}$ & Bouguer gravitational anomaly due to all loading and flexure &
\textit{calculated} & 
(\ref{bga})\\[-0.75em]
& & & \\
$\Qo$ & m$^2$s$^{-2}$ & complex admittance of Bouguer anomaly and topography &
estimable & (\ref{Qfull})\\
$\gto$ &  & real coherence-squared of Bouguer anomaly and
topography & 
estimable & (\ref{Qfull2})\\
\hline
$\cSij$ & m$^2$ & (cross-)spectral density between initial topographies at interfaces~$i$ and~$j$ & \textbf{estimated} &(\ref{specdensij}) \\
$\cSoij$ & m$^2$ & (cross-)spectral density between final topographies at interfaces~$i$ and~$j$ & estimable &(\ref{specdensoij}) \\
%% $\cSoo$ & m$^2$ & spectral density of initial topography at interface 1 & estimable &(\ref{specdensij}) \\
%% $\cStt$ & m$^2$ & spectral density of initial topography at interface 2 & estimable & (\ref{specdensij})\\
%% $\cSot$ & m$^2$ & cross-spectral density between initial topographies at interface 1 and 2 & estimable & (\ref{specdensij})\\
$r$ & &  correlation coefficient between initial loading at interface 1 and 2   & \textbf{estimated} &(\ref{corrdef})\\
$\ft$ &  &  ratio of the spectral densities of the initial loads at interface 1 and 2
  & \textbf{estimated} & (\ref{ftdef})\\[-0.75em]
& & & \\
$\Qf$ & s$^{-2}$ &  Bouguer/topography admittance for uncorrelated proportional loading at both interfaces & estimable
 & (\ref{Qf})\\
$\Qon$ & s$^{-2}$ &  Bouguer/topography admittance for loading only at interface 1  & estimable
 & (\ref{Q1})\\
$\Qt$ & s$^{-2}$ &  Bouguer/topography admittance for loading only at interface 2  & estimable
 & (\ref{Q2})\\[-0.75em]
& & & \\
$\gtf$ &  &  Bouguer/topography coherence for uncorrelated proportional loading at both interfaces& estimable
 & (\ref{gtf})\\
\hline\end{tabular}
\caption{\label{tablelitho}Subset of symbols used in this paper, their units and physical
description, their role in our estimation process, and relevant equation numbers.} 
\end{table*}

\subsection{Admittance and coherence}

Modeled after eq.~(\ref{specdens}),  the
Fourier-domain relation between the theoretical observable 
quantities that are the surface topography~$\Hno$ and the Bouguer
gravity anomaly~$\Gpn$ is encapsulated by the complex-valued
theoretical Bouguer admittance, which we define as
\be\label{Qfull}
\Qo\ofk  =\frac{\langle d\Gpn^{}\ofk  \hsom d\Hno^*\hsomm\ofk  \rangle} 
{\langle d\Hno^{}\ofk  \hsom d\Hno^*\hsomm\ofk  \rangle}
  =\chi\ofsk\frac{\langle d\Hnt^{}\ofk  \hsom d\Hno^*\hsomm\ofk  \rangle} 
{\langle d\Hno^{}\ofk  \hsom d\Hno^*\hsomm\ofk  \rangle}
.\ee
A quantity whose expression eliminates the dependence on the
location of the first interface contained in the term~$\chi$ of
eq.~(\ref{Gpnchidef}) is the real-valued Bouguer
coherence-squared, the Cauchy-Schwarz bounded quantity
% <|a||b|>\le\sqrt(<a,a><b,b>)
\be
\gto\ofk  =
\frac{|\langle d\Gpn^{}\ofk \hsom d\Hno^*\hsomm\ofk  \rangle|^2} 
{\langle d\Hno^{}\ofk  \hsom d\Hno^*\hsomm\ofk  \rangle
\langle d\Gpn^{}\ofk  \hsom d\Gpns\ofk  \rangle}
=
\frac{|\langle d\Hnt^{}\ofk \hsom d\Hno^*\hsomm\ofk  \rangle|^2} 
{\langle d\Hno^{}\ofk  \hsom d\Hno^*\hsomm\ofk  \rangle
\langle d\Hnt^{}\ofk  \hsom d\Hnt^*\hsomm\ofk  \rangle}
,\label{Qfull2}\qquad
0\le\gto\ofk \le 1.
\ee
As illustrated by eqs~(\ref{specblur})--(\ref{fourierG}), similarly,
the values of either ratio when calculated using actual
observations~$\nHno$ and~$\nGpn$ or~$\nHnt$, with or without explicit
windowing, will be estimators for eqs~(\ref{Qfull}) and~(\ref{Qfull2}),
but will never manage to recover more than a blurred version of the
true cross-power spectral density ratios that they are, and with an
estimation variance that will depend on how the required averaging is
implemented~\cite[]{Thomson82,Percival+93}. Despite the various
attempts by many
authors~\cite[]{Diament85,Lowry+94,Simons+2000,Simons+2003a,Kirby+2004,Kirby+2011,Audet+2007,Simons+2011a}
to design optimal data treatment, wavelet or (multi-)windowing
procedures, with the common goal to minimize the combined effect of
such bias or leakage and estimation variance, in the end this may result in a
well-defined (non-parametric) estimate for coherence and admittance,
but the actual quantity of interest, the flexural rigidity, $D$, still
has to be determined from that. As we wrote in the
Introduction, understanding the statistics of the estimators for~$D$
derived from estimates of coherence or admittance depends on fully
characterizing their distributional properties: a daunting task that,
to our knowledge, has never been successfully attempted. Without this,
however, we will never know which method is to be preferred under
which circumstance. Moreover, we will never be able to properly
characterize the standard errors of the estimates except by exhaustive
trial and error~\cite[see,
  e.g.,][]{Perez+2004,Crosby2007,Kalnins+2009} from data that are 
synthetically generated. This is no trivial
task~\cite[]{Macario+95,Ojeda+2002,Kirby+2008a,Kirby+2008b,Kirby+2009};
we return to this issue later. 

We have hereby reached the essence of this paper: our goal is to
estimate flexural rigidity~$D$ from observed topography~$\Hno$ and
gravity~$\Gpn$; estimates based on inversions of estimated admittance
and coherence have led to widely different results, a general lack of
understanding of their statistics, and thus a failure to be able to
judge their interpretation. We must thus abandon doing this via the
intermediary of admittance~$\Qo$ and coherence~$\gto$, and rather
focus on directly constructing the best possible estimator for~$D$
from the data. This realization is not unlike that made in the last
decade by the seismological community, where the inversion of (group
velocity? phase velocity?) surface-wave dispersion curves or
individual-phase travel-time measurements has made way for
``full-waveform inversion'' in its many guises
\cite[e.g.][]{Tromp+2005,Tape+2007}. There too, the model is called to
explain the data that are actually being collected by the instrument,
and not via an additional layer of measurement whose statistics must
remain incompletely understood, or modeled with too great a precision.
In cosmology, the power-spectral density of the
cosmic microwave background radiation \cite[]{Dahlen+2008} is but a
step towards the determination of the cosmological parameters of
interest \cite[e.g.][]{Jungman+96,Knox95, Oh+99}. 

\section{T~H~E{\hsps}S~T~A~N~D~A~R~D{\hsps}M~O~D~E~L}
\label{standardmodel}

The essential elements of a geophysical and statistical nature as they
had been broadly understood by the late 1970s were reintroduced in the
previous section in a consistent framework. In this section we discuss
the important innovations and simplifications brought to the problem
by \cite{Forsyth85}. In a nutshell, in his seminal paper,
\cite{Forsyth85} made a series of model assumptions that resulted in
palatable expressions for the admittance and the coherence as defined
in eqs~(\ref{Qfull}) and~(\ref{Qfull2}), neither of which would
otherwise be of much utility in actually ``solving'' the problem of
flexural rigidity estimation from gravity and topography. The first
two of these were already contained in eq.~(\ref{biharm0}): loading
and compensation occur discretely at one and the same set of
interfaces, and the constant describing the mechanical behavior of the
system is a scalar parameter that does not depend on wavenumber nor
direction. The first assumption might be open for debate, and indeed
alternatives have been considered in the
literature~\cite[e.g.][]{Banks+77,Banks+2001}, but reconsidering it
would not fundamentally alter the nature of the problem. The second: isotropy of
the lithosphere, which is certainly only a null hypothesis  \cite[see,
e.g.][and many observational studies that work on the premise that it
must indeed be
rejected]{Stephenson+80,Bechtel89,Simons+2000,Simons+2003a,Swain+2003b,Kirby+2006}, 
does require a treatment that is to be revisited but presently falls
outside the scope of this work.  To facilitate the subsequent
treatment we restate the equations of Section~\ref{observables} in
matrix form. 

\subsection{Flexure of an isotropic lithosphere, revisited}

We shall consider the primary stochastic variables to be the
initial-loading topographies~$\Ho$ and~$\Ht$, respectively, and
describe their joint properties, and their relation to the theoretical
observable final topographies $\Hno$ and $\Hnt$ by defining the spectral
increment process vectors
\be\label{vecdH}
d\bmcH\ofk=\dcH
\also
d\bmcHo\ofk=\dcHo
.
\ee
Subsequently, we express the process by which lithospheric flexure
maps one into the other in the shorthand notation
\be\label{Mdef}
d\bmcHo\ofk=\bmcMDk\, d\bmcH\ofk\also
d\bmcH\ofk=\bmcM^{-1}_D\ofsk\, d\bmcHo\ofk,
\ee
where the real-valued entries of the non-symmetric lithospheric
matrices $\bmcMDk$ and $\bmcM_D^{-1}\ofk$ can be read off
eqs~(\ref{loadcon})--(\ref{loaddecon}) and the functional dependence
on the scalar constant flexural rigidity~$D$ is implied by the 
subscript. We now define the (cross-)spectral densities between the
individual entries in the initial-topography vector~$d\bmcH\ofk$ as in
eq.~(\ref{specdens}) by writing  
\begin{subequations}\label{specdensij}
\be
\langle d\mcH^{}_i\ofk \hsom d\mcH^*_j\ofkp\rangle=
\cSij\ofk\dtbk\dkkp
,
\ee
and form the spectral matrix~$\bmcS\ofk$ from
these elements using the Hermitian transpose as 
\be\label{cSdefine}
\langle
d\bmcH\ofk\hsom d\bmcHH\!\ofkp
\rangle=
\bmcS\ofk\dtbk\dkkp=
\omats{cc}
\cSook  & \cSotk \\
\cStok  & \cSttk 
\cmats\!\dtbk\dkkp
.
\ee
\end{subequations}
Lithospheric flexure transforms the spectral matrix of the initial
topographies, $\bmcS\ofk$, to that of the final topographies,
$\bmcSo\ofk$, defined as
\be\label{specdensoij}\label{cSodefdef}
\langle
d\bmcHo\ofk \hsom d\bmcHoH\ofkp
\rangle=\bmcSo\ofk\dtbk\dkkp
\also
\langle d\mcH^{}_{\circ i}\ofk \hsom d\mcH^*_{\circ j}\ofkp\rangle= 
\cSoij\ofk\dtbk\dkkp
,
\ee 
via the mapping implied by eqs~(\ref{Mdef})
through~(\ref{specdensoij}). We specify  
\be\label{cSodefine}
\bmcSo\ofk=
\bmcMDk\hsom\bmcS\ofk\hsom\bmcM_D\Trm\ofsk
.
\ee
We can now see that the theoretical admittance and coherence of
eqs~(\ref{Qfull})--(\ref{Qfull2}) can equivalently be written as  
\be\label{Qfull3}
\Qo\ofk=\chi\ofsk\frac{\cSoto\ofk}{\cSooo\ofk}
\also
\gto\ofk =
\frac{|\cSoto\ofk|^2}{\cSooo\ofk\hsom\cSott\ofk}
,
\ee
which explains why so many authors before us have focused on admittance and
coherence calculations as a spectral estimation problem.

To be valid spectral matrices of real-valued bivariate fields, the
complex-valued $\bmcS\ofk$ and $\bmcSo\ofk$ only need to possess
Hermitian symmetry, that is, invariance under the conjugate transpose,
and be positive-definite, that is, have non-negative real eigenvalues.
The spectral variances of the initial and final topographies at the
individual interfaces, $\cSook\ge 0$ and $\cSttk \ge 0$, both
arbitrarily depend on~$\kb$, but without dependence between $\kb\ne
\kbp$.  The only additional requirements are that
$\cSotk=\mcS^*_{21}\ofk$ and $|\cSotk|^2\le\cSook\cSttk$. The general
form of~$\bmcS\ofk$ as a stationary random process can be rewritten
with the aid of a coherency or spectral correlation coefficient,
$r\ofk$, which expresses the relation between the components of
surface and subsurface initial topography as
\be\label{corrdef}
r\ofk=\frac{\cSotk}{\sqrt{\cSook}\sqrt{\cSttk}}
,\where
|r\ofk|\le 1
\foral\kb
.
\ee
This correlation coefficient is in general complex-valued as the two
fields may be spatially slipped versions of one another. The
representation  
\be
\bmcS\ofk=
\omats{cc}
\cSook  & r\ofk\hsom\sqrt{\cSoo\ofk}\sqrt{\cStt\ofk} \\
r^*\hsomm\ofk\sqrt{\cSoo\ofk}\sqrt{\cStt\ofk}  & \cStt\ofk
\cmats,\foral\kb,
\ee
is simply a most complete description of a bivariate random spectral
process~\cite[]{Christakos92}.  

Should we make the additional assumption of joint isotropy for all of
the loads, the spectral matrices would both be real and symmetric,
$\bmcS\ofk=\bmcS\ofsk$ and $\bmcSo\ofk=\bmcSo\ofsk$. In keeping with
the notation from eq.~(\ref{specdensx}), we would require a spatial
covariance matrix to only depend on distance, not direction. With
$\theta$ the angle between $\kb$ and $\xb-\xbp$ we would have the
real-valued 
\be\label{Ciso}
\bmcC(\xb-\xbp)= \intnyq e^{i\kb\cdot(\xb-\xbp)}
\bmcS\ofk\dbk= \intnyq e^{ik \|\xb-\xbp\|\cos\theta} \bmcS\ofsk
\,d\theta \, k\,dk= 2\pi\int J_0(k \|\xb-\xbp\|)\, \bmcS\ofsk \, k\,dk
=\bmcC(\|\xb-\xbp\|)
, \ee
with $J_0$ the real-valued zeroth-order Bessel function of the first
kind. With $\bmcS$ real, the spectral variances and covariances
between top and bottom loading components would all be real-valued and
so would the correlation coefficient $r\ofk=r\ofsk$. It is important
to note that the isotropy of the fields individually does not imply
their joint isotropy. Two such fields can be spatially slipped
versions of one another, but with slippage in a particular direction
the fields may remain marginally isotropic but their joint structure
will not.

\subsection{Correlation between the initial loads}

Statistically, eqs~(\ref{specdensij}) and~(\ref{corrdef}) imply that
the initial-loading topographies on the two interfaces are related
spectrally as 
\be\label{Hmult}
d\Ht\ofk=
r\ofk\frac{\sqrt{\cSttk}}{\sqrt{\cSook}}d\Ho\ofk
+d\Ho^\perp\ofk=
p\ofk d\Ho\ofk
+d\Ho^\perp\ofk
,
\ee
whereby~$\Ho^\perp\ofx$,  the zero-mean orthogonal complement
to~$\Ho\ofx$, is uncorrelated with it at all lags. The interpretation
of what should cause a possible ``correlation'' between the
initial-loading topographies must be
geological~\cite[]{McGovern+2002,McKenzie2003,Belleguic+2005,Wieczorek2007,Kirby+2009}. 
Erosion 
\cite[e.g.][]{Stephenson84,Aharonson+2001a} is typically amenable to 
the description articulated by eq.~(\ref{Hmult}), though much work
remains to be done in this area to make it apply to the most general
of settings. Under isotropy of the loading, the implication is that
the initial subsurface loading~$\Ht\ofx$ can be generated from the
initial surface loading~$\Ho\ofx$ by a radially symmetric convolution
operator~$p\ofx$,  
\be\label{Hconv}
\Ht\ofx=\intnyq p(\xb-\xb')\Ho\ofxp\dbx'+\Ho^\perp\ofx.
\ee

By selecting the initial loads~$\Hj$ as the primary variables of the
flexural estimation problem, and not the equilibrium~$\Hij$ or
final loads~$\Hnj$,  we now have the correlation~$r$ between
the initial loads to consider in the subsequent
treatment. Geologically, this puts us in a bit of a quandary, since if
eq.~(\ref{Hmult}) holds, this can only mean that one loading process
``follows the other in time'', ``reacting to it''. However, the temporal
dimension has not entered our discussion at all, and if it did, it
would certainly make sense to choose the correlation between the
equilibrium load on one and the initial load on the
other interface as the one that matters. The linear relationship~(\ref{Mdef})
between the loads renders these two viewpoints mathematically
equivalent. Our definition of eq.~(\ref{corrdef}) is chosen to be
mathematically convenient because it is most in line with the choices
to be made in the next section. 

\cite{Forsyth85} deemed correlations between surface
and subsurface loads to be potentially important but he did not make the
determination of the correlation coefficient~(\ref{corrdef}) part of
the estimation procedure for the flexural rigidity~$D$, which was
instead predicated on the assumption, his third by our count, that
$r\ofk=0$. He did recommend computing the correlation coefficient
between the initial loads via eq.~(\ref{Mdef}), after the inversion
for~$D$, and using the results to aid with the interpretation
\cite[see, e.g.,][]{Zuber+89}. Studies by \cite{Macario+95},
\cite{Crosby2007}, \cite{Wieczorek2007} and \cite{Kirby+2009} have
since shed more light on how to do this more quantitatively, but to
our knowledge no-one has actually attempted to determine the
best-fitting correlation coefficient as part of an inversion for
flexural rigidity.   

\subsection{Proportionality between the initial loads}

\cite{Forsyth85} introduced the `loading fraction' as
the subsurface-to-surface ratio of the power spectral densities of the
initial-loading stresses~$\mcI_2$ and~$\mcI_1$, and thus from
eqs~(\ref{I1})--(\ref{I2}) and~(\ref{specdensij}) we can write 
\be\label{ftdef}
\ft\ofk  =\frac{\langle d\mcI^{}_2\ofk\hsom d\mcI^*_2\ofk  \rangle}
{\langle d\mcI_1^{}\ofk  \hsom d\mcI^*_1\ofk  \rangle}
=\frac{\Dt^2\cSttk}{\Do^2\cSook},\qquad f\ge 0
.
\ee
This definition is fairly consistently applied in the
literature~\cite[e.g.][]{Banks+2001}, though \cite{McKenzie2003} has
preferred to parameterize by the fraction each of the loads
contributes to the total, which is handy for situations with multiple
interfaces \cite[see][]{Kirby+2009} and subsurface-only loading.
Eq.~(\ref{ftdef}) is a statement of proportionality of the power
spectral densities of the initial loads, $\mcS_{2}$ and~$\mcS_{1}$.
With this constraint, which we identify as his fourth assumption,
\cite{Forsyth85} was able to factor~$\mcS_{11}$ out of the spectral
matrix $\bmcS$ in eq.~(\ref{specdensij}), which as we recall from the
previous section, by his third assumption had no off-diagonal terms,
to arrive at simplified expressions for~$\bmcSo$ of
eq.~(\ref{specdensoij}), which acquires off-diagonal terms through
eq.~(\ref{cSodefine}), and ultimately for the admittance~$\Qo$ and
coherence~$\gto$ in eq.~(\ref{Qfull3}). We revisit these quantities in
the next section but conclude with the general form of the initial-loading
spectral matrix that is implied by the definition of proportionality,
which is
\be
\bmcS\ofk=\cSook
\omats{cc}
1 & r\ofk f\ofk\Dou\Dt^{-1} \\
r^*\hsomm\ofk f\ofk\Dou\Dt^{-1}  & \ft\ofk\Do^2\Dt^{-2}
\cmats.
\ee

With what we have obtained so far: flexural isotropy of the
lithosphere, $\bmcMDk$, correlation of the initial-loading processes,
$r\ofk$, and proportionality of the initial-loading processes,
$\ft\ofk$, the spectral matrix (\ref{cSodefine}) of the final
topographies --- those we measure --- is given by  
\be\label{S0def} 
\bmcSo\ofk=
\cSook\hsom\Tbo\ofk=
\cSook
\hsom[\Tb\ofk+\dTb\ofk],
\ee
where we have defined the auxiliary matrices
\be\label{Tdef}
\Tb\ofk=
\omat{cc}
\xit+\ft\ofk\Do^2\Dt^{-2}
&
-\Dou\Dt^{-1}\xi-\ft\ofk\Do^3\Dt^{-3}\phi
\\
-\Dou\Dt^{-1}\xi-\ft\ofk\Do^3\Dt^{-3}\phi&
\Do^2\Dt^{-2}+\ft\ofk\Do^4\Dt^{-4}\phit
\cmat
\left(\frac{\Dt}{\Do+\Dt\hsom\xi}\right)^2
,
\ee
\be\label{dTdef}
\dTb\ofk=
r\ofk f\ofk
\omat{cc}
-2\Dou\Dt^{-1}\xi & \Do^2\Dt^{-2}[\phixi+1]\\  
\Do^2 \Dt^{-2} [\phixi+1] & -2\Do^3 \Dt^{-3} \phi  
\cmat
\left(\frac{\Dt}{\Do+\Dt\hsom\xi}\right)^2
.
\ee
We define both~$\Tb$ and~$\dTb$ so that we can easily revert to a
model of zero correlation,  in which case $\dTb=\bzero$. Note that we
are silent about the dependence on wavenumber by using 
the shorthand notation $\xi$ and $\phi$ for the lithospheric
filters~(\ref{xidef}) and~(\ref{phidef}), but have kept the full
forms of the correlation coefficient~$r\ofk$ and the loading
ratio~$\ft\ofk$ to stress that they are in general functions of the
wave vector as defined by eqs~(\ref{corrdef}) and~(\ref{ftdef}).
In general~$r$ will be complex and of magnitude smaller than
or equal to unity, and $\ft$ (and $f$) will be real and positive.

\subsection{Admittance and coherence for proportional and correlated
  initial loads}\label{admicoh}

Via eqs~(\ref{S0def})--(\ref{dTdef}) we have explicit access to the
(cross-)spectral densities between the individual elements in the
final-topography vector~$d\bmcHo$, as required to evaluate
eq.~(\ref{specdensoij}).  We shall now consider those for the special
case where both $r\ofk=r$ and $\ft\ofk=\ft$ are constants, no longer
varying with the wave vector. Then, following eq.~(\ref{Qfull3}), we
obtain simple expressions for the admittance and coherence that we
shall further specialize to a few end-member cases for comparison with
those treated in the prior literature. We hereby complete
Table~\ref{tablelitho} to which we again refer for a summary of the
relevant notation.

The Bouguer-topography admittance, for correlated and proportional
initial loads with constant correlation~$r$ and proportion~$\ft$, is
\be\label{Qo} %SCO Checked Aug 24 2011
\Qo\ofsk=-2\pi G\Do\ekzt
\frac{\xi+\ft\Do^2\Dt^{-2}\phi-rf\Dou \Dt^{-1} [\phixi+1]}
{\xit+\ft\Do^2\Dt^{-2}-2rf\Dou\Dt^{-1}\xi}
.
\ee
Spectrally, this is a function of wavenumber, $k$, only, since the power
spectra of the loading topographies, which both may vary (similarly,
because of their proportionality)  with the wave vector~$\kb$ have
been factored out. This admittance can be complex-valued since the load
correlation may be, unless the power spectra of the loading
topographies are isotropic. At $k=0$ the admittance yields the
density contrast~$\Delta_1$.

Assuming that the loads are uncorrelated but proportional simplifies
the Bouguer-topography admittance to the familiar expression
\be\label{Qf} 
\Qf\ofsk =-2\pi G\Do\ekzt
\fracd{\xi+\ft \Do^2\Dt^{-2}\phi}
{\xit+\ft  \Do^2\Dt^{-2}}
.
\ee
In scenarios where only top or only bottom loading is present,
we get the original expressions~\cite[]{Turcotte+82,Forsyth85}
\be\label{Q1}
\Qon\ofsk =-2\pi G\Do\xi^{-1}\ekzt
,
\ee
\be\label{Q2}
\Qt\ofsk =-2\pi G\Do\phi\hsom\ekzt
,
\ee
where, as expected and easily verified, 
\be\label{limQf}
\lim_{f=0} \Qf\rightarrow
\Qon\also\lim_{f=\infty} \Qf\rightarrow \Qt.
\ee

The Bouguer-topography coherence, for correlated and proportional
initial loads with constant correlation~$r$ and proportion~$\ft$, is
\be\label{gto}
\gto\ofsk=
\fracd{\left(\xi+ \ft\Do^2\Dt^{-2}\phi -rf\Do\Dt^{-1}[\phixi+1] \right)^2}
{\left(\xit+\ft\Do^2\Dt^{-2} -2rf\Do\Dt^{-1}\xi\right)
\left(1+\ft\Do^2\Dt^{-2}\phit -2rf\Do\Dt^{-1}\phi\right)
},
\ee
which, as the admittance, is a function of wavenumber~$k$ regardless
of the power spectral densities of the loading topographies. Unlike
the admittance it has lost the dependence on the depth to the second
interface, $z_2$, and it is always real, $0\le \gto\le 1$.

When the initial loads are uncorrelated but proportional the
Bouguer-topography coherence is, as according to \cite{Forsyth85},
simply 
\be\label{gtf}
\gtf\ofsk=
\fracd{\left(\xi+ \ft\Do^2\Dt^{-2}\phi \right)^2}
{\left(\xit+\ft\Do^2\Dt^{-2}  \right)
\left(1+\ft\Do^2\Dt^{-2}\phit  \right)}
.
\ee
This expression was solved by  \cite{Simons+2003a} for the
wavenumber at which $\gto=1/2$, the diagnostic~\cite[]{Simons+2002a}
\be
k_{\ssts{1/2}}=\left(\frac{g}{2Df}
\left[\Dt-f(\Do+\Dt)+\ft\Do+\sqrt\beta\right]\right)^{1/4}
,
\label{k4}
\ee
\noin where
$
\beta=
\Dt^2+
2f
\left(\Dt^2-\Do\Dt\right)
+
\ft\left(\Do^2+\Dt^2+4\Do\Dt\right)
-2f^3
\left(\Do\Dt-\Do^2
\right)
+f^4\Do^2
.
$
In the paper by \cite{Simons+2003a} eq.~(\ref{k4}) appears with a typo
in the leading term, which was briefly the cause of some confusion in
the literature \cite[]{Kirby+2008a,Kirby+2008b}. 

\begin{figure}\centering
\rotatebox{-0}{\includegraphics[width=\textwidth]{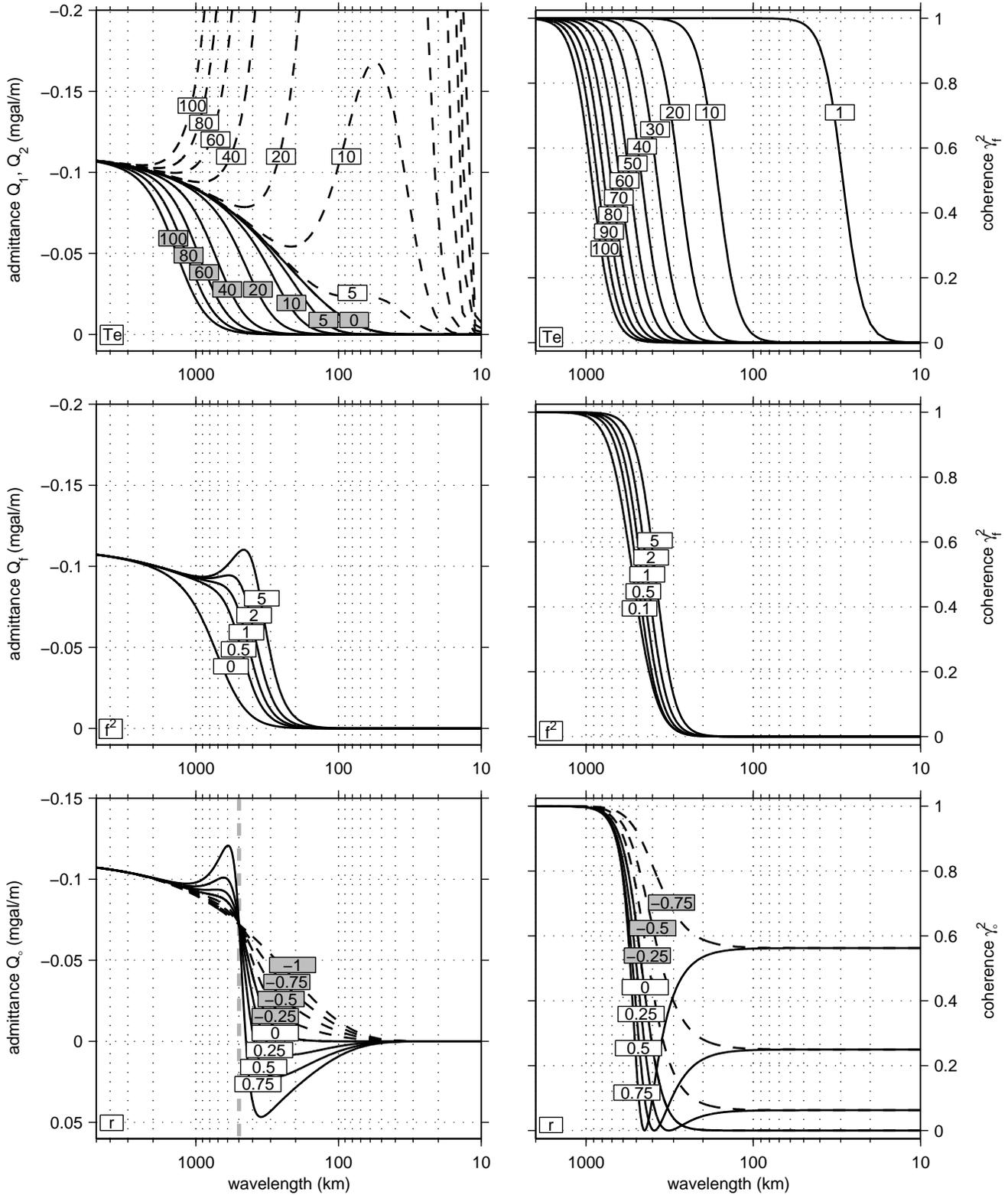}}
\caption{\label{FF1} Expected values of the admittance and coherence
  between Bouguer gravity anomalies and topography in two-interface
  models, derived in Section~\ref{admicoh}. All models have
  identical density structures, $z_1=0$~km, $z_2=35$~km,
  $\Do=2670$~kg$\hsom$m$^{-3}$ and $\Dt= 630$~kg$\hsom$m$^{-3}$,
  Young's and Poisson moduli $E=1.4\times 10^{11}$~Pa and $\nu=0.25$.
  (\textit{Left column}) Admittance curves for top-only ($\ft=0$) and
  bottom-only ($\ft=\infty$) loading as a function of the effective
  elastic thickness, $T_e$ (\textit{top left}); for mixed-loading
  models at constant $T_e=40$~km with varying loading fractions $\ft$,
  but without load correlation (\textit{middle left}); and for models
  at constant $T_e=40$~km and $\ft=1$ but with various
  load-correlation coefficients~$r$ (\textit{bottom left}), as
  indicated in the legend. (\textit{Right column}) Coherence curves
  for a fixed-loading scenario at constant $\ft=1$ but with various
  values for $T_e$ (\textit{top right}); for constant $T_e=40$~km and
  varying values of~$\ft$, without correlation (\textit{middle right});
  and at constant $\ft=1$ and $T_e=40$~km but for varying load
  correlation~$r$ (\textit{bottom right}), as annotated.  }
\end{figure}

Fig.~\ref{FF1} displays the individual effects that varying flexural
rigidity, loading fraction and load correlation have on the expected
admittance and coherence curves. Regardless of the fact that much of
the literature to this date has been concerned with the estimation of
the admittance and coherence from the available data, and regardless
of the justifiably large amount of attention devoted to the role of
windowing and tapering to render these estimates spatially selective
and spectrally free from excessive leakage; regardless, in summary, of
any practicality to the actual methodology by which admittance and
coherence are being estimated and how the behavior of their estimates
affects the behavior of the estimated parameter of interest, the
flexural rigidity, $D$, we show these curves to gain an appreciation
of the complexity of the task at hand. No matter how well we may be
able to recover the ``true'' admittance and coherence behavior, the
issue remains that they need to be interpreted --- inverted --- for a
model that ultimately needs, or can, return an estimate for $D$ but
also of the initial-loading fraction, $\ft$, and also of the
correlation coefficient, $r$. Each of these have distinct
sensitivities but overlapping effects on the predicted behavior of the
measurements: selecting one end-member model (top-loading or
bottom-loading only, for example, or disregarding the very possibility
of load correlation, or imposing a certain non-vanishing value on the
loading fraction or load-correlation coefficient) remains but one
choice open to alternatives, and constraining all three is a task
that, thus far, nobody has successfully attempted. Fig.~\ref{FF1}
serves as a visual reminder of the limitations of admittance- and
coherence-based estimation. However much information these statistical
summaries of the gravity and topography data contain, it is not easily
accessible for navigation in the three-dimensional space of $D$, $\ft$
and~$r$.   

\subsection{Load correlation, proportionality and the standard model}

The expressions in the previous section show how difficult it is to
extract the model parameters~$D$, $\ft$ and~$r$ individually from
admittance or coherence. \cite{Forsyth85} argued that coherence
depends on~$\ft$ much more weakly than admittance, but what is
important for the estimation problem is how the three parameters of
interest vary together functionally: whether they occur in terms by
themselves or as products, in which variations of powers, and so
on. The geometry of the objective functions used to estimate the
triplet of parameters, together with the distribution of any random
quantities the objective functions contain, determine the properties
of the estimators. We return to the question of identifiability after
we have presented the new maximum-likelihood estimation method. For
that matter, \cite{Forsyth85} suggested ignoring the load correlation,
setting $r=0$, and finding an estimate for the flexural rigidity~$D$
using a constant initial guess for the loading fraction~$\ft$ and the
coherence modeled as $\gtf$ in eq.~(\ref{gtf}), and then using
eqs~(\ref{loaddecon}), (\ref{I1})--(\ref{I2}) and~(\ref{ftdef}) to
compute a wavenumber-dependent estimate of~$\ft$, which can then be
plugged back into eq.~(\ref{gtf}) as a variable, and iterating this
procedure to convergence. However, this allows for as many degrees of
freedom as there are ``data'', thereby running the risk that an
ill-fitting~$D$ can be reconciled with the data by adjustment with a
very variable~$\ft$. It is unclear in this context what
``ill-fitting'' or ``very variable'' should mean, and thus it is hard
to think of objective criteria to accomplish this. \cite{McKenzie2003}
showed misfit surfaces for the (free-air) admittance for varying~$D$
and varying~$\ft$ held constant over all wavenumbers. These figures
show prominent trade-offs, suggesting a profound lack of
identifiability of~$D$ and~$\ft$ with such a method.

Even more importantly, \cite{McKenzie2003} emphasized the possibility
of non-zero correlations between the initial loads, deeming 
those prevalent in many areas of low-lying topography, on old portions
of the continents: precisely where the discrepancy between estimates
for elastic thickness derived from different methods has been leading
to so much controversy. As an alternative to the \cite{Forsyth85}
method, \cite{McKenzie+97} suggested estimating~$D$ and~$\ft$ from the
free-air admittance in the wavenumber regime where surface topography
and free-air gravity are most coherent. The rationale for this
procedure is that there might be loading scenarios resulting in
gravity anomalies but not (much) topography, a situation not accounted
for in the \cite{Forsyth85} model that can, however, be described by
initial-load correlation. \cite{Kirby+2009}, most recently, discussed
the differences between both approaches, only to conclude that neither
estimates the complete triplet $(D,\ft,r)$ of parameters (rigidity,
proportionality, correlation) without shortcuts. Once again the
statistical understanding required to evaluate whether either of these
techniques results in ``good'' estimators is lacking.  

That the cause of ``internal loads without topographic expression''
can indeed be attributed to correlation in the
sense of~(\ref{corrdef}) can be readily demonstrated by considering what
it takes for the final, observable, surface topography~$\Hno$ to
vanish exactly. Solving eq.~(\ref{loadcon}) or eq.~(\ref{Mdef}) and
using eqs~(\ref{xidef}) and (\ref{phidef}) returns the conditions that
the first and second initial topographies are related to each other as
\be\label{vanish}
d\Ht\ofk=\xi(k)\hsom d\Ho\ofk,
\ee
which, using eqs~(\ref{specdensij}), (\ref{ftdef}) and~(\ref{corrdef}),
implies the following equivalent relations between them:
\be\label{vanish2}
\cSttk=\xik\hsom\cSotk=\xik\hsom\cStok=\xitk\hsom\cSook,
\qquad
\ft\ofsk=\Do^{-2}\Dt^2\hsom\xitk,
\qquad
r=1.
\ee
This set of equations together with our model very strongly constrain
both fields. Thus, as noted by \cite{McKenzie2003} and others after him
\cite[]{Crosby2007,Wieczorek2007,Kirby+2009}, a situation of internal
loading that results in no net final topography may arise when the
initial-loading topographies are perfectly correlated,
balancing one another according to
eqs~(\ref{vanish})--(\ref{vanish2}).  We can find a more complete
condition for this scenario by equating eqs~(\ref{vanish})
and~(\ref{Hmult}), which returns an expression for the orthogonal
complement~$d\Ho^\perp$; when this is required to vanish
non-trivially we obtain the seemingly more general condition 
\be\label{rfcon}
r\ofk f\ofk=\frac{\Dt}{\Do}\xik
,\qquad 0\le r\ofk\le 1
.
\ee
Requiring that the final surface topography have a
vanishing variance~$\cSooo$, substituting
eqs~(\ref{S0def})--(\ref{dTdef}) into eq.~(\ref{cSodefdef}), we
need to satisfy
\be\label{rfcon2}
r\ofk f\ofk=\frac{\xitk+\ft\ofk\Do^2\Dt^{-2}}{2\Dou\Dt^{-1}\xik}
,\qquad 0\le r\ofk\le 1
.
\ee
The correlation coefficients in eqs~(\ref{rfcon})--(\ref{rfcon2}) must
be real-valued since all of the other quantities involved are. Both
eq.~(\ref{rfcon}) and eq.~(\ref{rfcon2}) should be equivalent, and
together they imply eq.~(\ref{vanish2}). We are thus left to conclude
that for the observable surface topography to vanish, the correlation
between initial surface and subsurface loading must be
perfect and positive,  $r=1$.  Solving the quadratic
equation~(\ref{rfcon2}) for~$f$ yields real-valued results only when
$|r|^2-1\ge 0$, thus $r=1$ for positive but non-constant~$f$, as expected. 

The above considerations have put perhaps unusually strong constraints
on the spectral forms of the final topography~$\Hno\ofk$
or~$\cSooo\ofk$. From eq.~(\ref{fourierfirst}) we learn that in doing
so, the spatial-domain observables~$\Hno\ofx$ can never be non-zero.
On the other hand, an observed $\Hno\ofx$ could be zero over a
restricted patch without its Fourier transform or its spectral density
vanishing exactly everywhere. Alternatively, it can be very nearly
zero, and this may also practically hamper approaches based on
admittance or coherence which contain (estimates of) the
term~$\cSooo\ofk$ in the denominator (see eq.~\ref{Qfull3}). When the
observed topography becomes small, higher-order neglected terms may
become prominent. Furthermore, there may be mixtures of loads with and
without topographic expression \cite[]{McKenzie2003}. Speaking quite
generally, there will be areas with some correlation between the
initial loads, and we should take this into account in the estimation.
Either one of the load correlation or load fraction may vary with
wavenumber. What emerges from this discussion is that the isotropic
flexural rigidity~$D$, the initial-load correlation~$r\ofk$, and the
initial-load proportionality~$\ft\ofk$ should all be part of the
``standard model'' of flexural studies. The last two concepts were
introduced by \cite{Forsyth85}, even though he did not further discuss
the case of non-zero correlation.

As we wrote in the first paragraph in this section, Forsyth's first
assumption was that the depth of compensation and the depth of loading
in fact coincide. He writes that the assumption of collocation of
these hypothetical interfaces and their precise location at depth in
Earth may well be the largest contributor to uncertainty in the
estimates for flexural strength, but also that there may be 
\textit{a priori}, e.g. seismological, information to help constrain the
depth~$z_2$. Thus, much like the density contrasts~$\Delta_1$ and
$\Delta_2$, we will not include the depth to the second interface
$z_2$ as a quantity to be estimated directly. Rather, we will consider
them known inputs to our own estimation procedure and evaluate their
suitability after the fact by an analysis of the likelihood functions
and of the distribution of the residuals.

\section{M~A~X~I~M~U~M~-~L~I~K~E~L~I~H~O~O~D{\hsps}T~H~E~O~R~Y}\label{secmle}

Measurements of ``gravity'' and ``topography'', which we consider free
from observational noise, can be interpreted as undulations, 
$\Hno$ and~$\Hnt$, of the surface and one  subsurface density
interface, with density contrasts, $\Do$  and $\Dt$, located at depths
$z_1=0$ at $z_2$ in Earth, respectively. Geology 
and ``tectonics'' produce initial topographic loads, $\Ho$ and $\Ht$,
on these previously undisturbed interfaces. These are treated as a
zero-mean bivariate, stationary, random process vector, $d\bmcH$,
fully and  most generally described by a spectral matrix,
$\bmcS\ofk$, under the assumption that the higher-order moments
of~$\bmcH\ofx$ are not too prominent~\cite[]{Brillinger74}. For this
paper we assume isotropy of the loading process,
$\bmcS=\bmcS\ofsk$. The lithosphere is modeled as a coupled set of
differential equations, whose action is described by the 
spectral-domain matrix~$\bmcM_D$, which depends on a single,
scalar parameter of interest, the flexural rigidity~$D$. Since our
observations have experienced the linear mapping
$d\bmcHo=\bmcM_D\,d\bmcH$, their spectral matrix is
$\bmcSo\ofsk=\bmcMDk\hsom\bmcS\ofsk\hsom\bmcM_D\Trm\ofsk$, and the
objective is to recover~$D$, we are led to study~$\bmcSo(k)$. This
includes its off-diagonal terms, which depend on the correlation
coefficient of the loads at either interface, $-1\le r\ofsk\le1$,
recall $r\ofsk\in\mathbb{R}$, and, under the assumption of
proportionality of the initial-loading spectra, on a loading fraction,
$\ft\ofsk$. As part of the estimation we will thus also recover
information about the loading process~$\bmcS$.  

All previous studies in the geophysical context of lithospheric
thickness determination have first estimated admittance and coherence,
ratios of certain elements of~$\bmcSo$ whose estimators have joint
distributions that have not been studied. These were then used in
inversion for estimates of~$D$ whose statistics have remained unknown.
In the remainder of this paper we construct a maximum-likelihood
estimator \textit{sensu} \cite{Whittle53}, directly from the data
``gravity'' and ``topography'', and the ``known'' parameters $\Do$,
$\Dt$, and~$z_2$. The unknowns are $D$, $r$ and~$\ft$, and, as we
shall see shortly, three more parameters by which we guarantee
isotropy of the loading process $\bmcS$ through a commonly utilized
functional form. That this is more ambitious than the original
objectives by \cite{Forsyth85} and the modifications by
\cite{McKenzie2003} is because the reduction of the data to admittance
or coherence obliterates information that we are able to recover in
some measure. We study the properties of the new estimators and derive
the distributions of the residuals. When the procedure is applied to
actual data, these should tell us where to adjust the assumptions used
in designing the model.

\subsection{Choice of spectral parameterization, $\st,\nu,\rho$}

In the above we have seen that the primary descriptor of what causes
the observed behavior is the spectral matrix~$\bmcS\ofsk$ from which
the initial interface-loading topographies are being generated. After
the assumption of spectral proportionality of the loading at the two
interfaces, the expressions for admittance and coherence no longer
contain any information about this particular quantity, though of
course the deviations of the observed admittance and coherence from
the models discussed in Section~\ref{admicoh} still might. However,
this information is no longer in an easily accessible
form. Furthermore, coherence and admittance are typically estimated
non-parametrically: the infinitely many, or rather, $2K=M\times N$
dimensions of the data are reduced to a small number of wavenumbers at
which they are being estimated, thus there is a loss of
$\mcO(K)$ degrees of freedom. At the low frequencies, most
tapering methods experience a further reduction in resolution, which
is detrimental especially in estimating the value of thick
lithospheres from relatively small data grids, as is well appreciated
in the geophysical literature. 

Here, we will simply parameterize the initial loading using a ``red''
model, thereby avoiding such a loss. We may consult
\cite{Goff+88,Goff+89b}, \cite{Carpentier+2007} or
\cite{Gneiting+2010} for such models.  Here we do, however, make the
very strong assumption of isotropy. This is unlikely to be satisfied
in real-world situations, as spectral-domain anisotropy is part and
parcel of all geological
processes~\cite[]{Goff+91,Carpentier+2009a,Carpentier+2009b,Goff+2010}.
Relaxing the isotropic loading assumption introduces considerable
extra complications. Our reluctance to handle anisotropic loading
situations stems from the fact that their estimation might be confused
statistically with a possible anisotropy in the lithospheric response:
we can thus not easily study one without studying the other.

At this point we collect the parameters that we wish to estimate into
a vector. To begin with, the ``lithospheric'' parameters, flexural
rigidity~$D$, loading ratio~$\ft$ and load correlation~$r$ are  
\be\label{thetaL}
\bthetaL=\bthetaLfullr
.
\ee
We denote a generic element of this vector as~$\tl$. For the spectrum of
the initial-loading topographies we choose the isotropic
Mat\'ern spectral class, which has legitimacy in geophysical
circles \cite[]{Goff+88,Stein99,Guttorp+2006}. We specify
\be\label{materndef} 
\cSoo\ofsk=\frac{\st\nu^{\nu+1} 4^\nu}{\pi  (\pi\rho)^{2\nu}}
\left(\frac{4\nu}{\pi^2\rho^2}+k^2\right)^{-\nu-1},
\ee
whose parameters we collect in the set
\be\label{thetaS}
\bthetaS=\bthetaSfull
,
\ee
with generic element~$\ts$. The third parameter, $\rho$, is distinct
from the mass density, as will be clear from the context.
The full set of parameters that we  wish to
estimate problem is contained in the vector
\be\label{btheta}
\btheta=[\bthetaL\Tit\,\,\,\bthetaS\Tit]\Tit=
\bthetafullr
,
\ee
whose general element we denote by~$\theta$. For future reference we
define the parameter vector that omits all consideration of the
correlation as 
\be\label{bthetawi}
\bthetaw=\bthetafull
.
\ee

\begin{figure}\centering
\rotatebox{-90}{\includegraphics[height=0.95\textwidth]{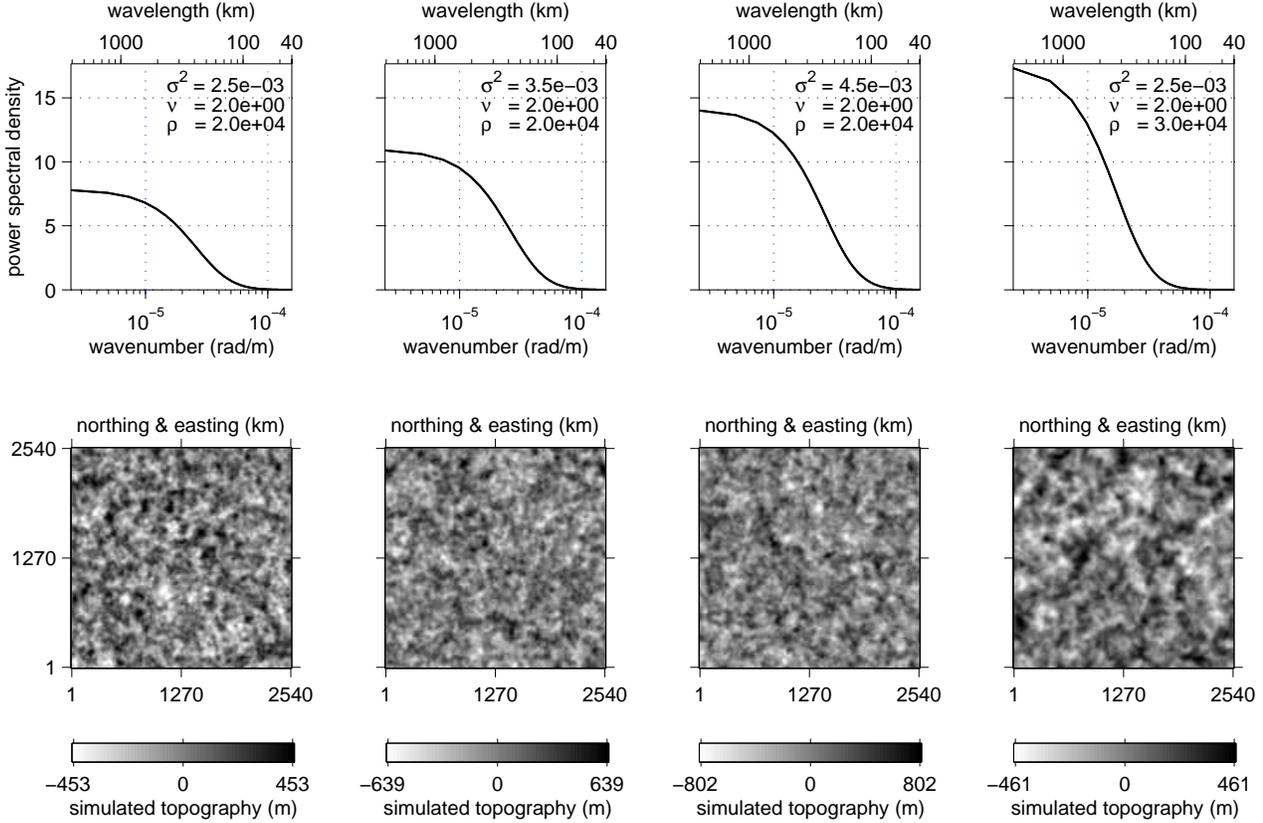}}  
\caption{\label{oh2}Synthetic ``topographies''  generated from the Mat\'ern
spectral class with parameters $\st$, $\nu$ and $\rho$ as indicated in
the legends. \textit{(Top row)} Power spectral densities as given by
eq.~(\ref{materndef}). \textit{(Bottom row)} Spatial realizations of
Gaussian random processes with the power spectral densities as shown
in the top row.}
\end{figure}

Fig.~\ref{oh2} shows a number of realizations of isotropic
Mat\'ern processes with different spectral parameters. As can be seen
the parameters~$\st$ (``variance'') and~$\rho$ (``range'') impart an
overall sense of scale to the distribution while~$\nu$
(``differentiability'') affects its shape \cite[]{Stein99,Paciorek2007}.  

\subsection{The observation vectors, $d\bmcH$ and $\Hb$}

% Say something generic about how we observe one spatial-domain
% realization of some process that we surmise exists in spectral form.  
In Section~\ref{basics} we introduced the standard statistical point
of view on stationary processes~\cite[]{Brillinger74,Percival+93}. We
specified how this applies to a finite set of geophysical observations
that can be defined in a two-layer system, which we revealed to be the
various types of ``topography'' and ``gravity'', and which are mapped
into one another by the differential equations describing
``flexure''. Subsequently, we introduced the matrix formalism that
describes the connections between the various geophysical observables
and the initial driving forces that produce them, which we used
extensively in Section~\ref{standardmodel} to discuss the standard
approach of determining the unknown parameters of the flexural
differential equation and the relative importance and correlation of
the loading processes acting across either layer interface, which are
of geophysical interest~\cite[e.g.][]{Forsyth85,McKenzie2003}. To
address the problem of how to properly estimate these unknowns
and their distribution, we now return to the statistical formalism
espoused in Section~\ref{Cramer} in order to clarify how the
``theorized'' geophysical observables, i.e. the spectral processes
describing the various kinds of topography~$d\mcH\ofk$ and gravity
anomalies~$d\mcG\ofk$ are being shaped into the ``actual''
observations. Those are the windowed Fourier transforms~$H\ofk$
and~$G\ofk$ of particular realizations of topography and gravity as we
can calculate from finite spatial data sets~$\mcH\ofx$ and~$\mcG\ofx$
measured in nature. In the spectral domain we continue to
distinguish by the choice of font the theory (calligraphic) from what
we can actually calculate (italicized). In the spatial
domain, there is no need to define anything but~$\mcH\ofx$
or~$\mcG\ofx$. 

\subsubsection{In theory: infinite length and continuous}
\label{theory}

We recall that the spectral matrix~$\bmcSo\ofk$, given  by
eq.~(\ref{S0def}), of the vector of final, observable,
topographies~$d\bmcHo\ofk$ defined in
eqs~(\ref{vecdH})--(\ref{cSodefine}),  is separable in the 
sought-after parameter vectors~$\bthetaS$ and~$\bthetaL$ by the
factoring of the spectral density~$\cSoo\ofsk$ of the initial-loading
topographies, 
\be\label{split} 
\bmcSo\ofk=\cSoo\ofsk\Tbo\ofk=
\cSoo\ofsk
\left[\Tb\ofk+\dTb\ofk\right]. 
\ee
In writing eq.~(\ref{split}) we emphasize the wavenumber-only
dependence of the ``spectral'' matrix~$\cSoo\ofsk$, which is
isotropic, but keep the full wavevector dependence of the
``lithospheric'' matrices~$\Tb\ofk$ and~$\dTb\ofk$ to make sure they
have the same dimensions as the data. However, in the case of
isotropic loading both~$\Tb\ofk$ and~$\dTb\ofk$ will also only depend
on wavenumber, and they will both be real. We thus rewrite
eqs~(\ref{Tdef})--(\ref{dTdef}) with the dependencies $\phik$,
$\xik$, $r\ofsk$ and $\ft\ofsk$ implicit in this sense, 
\be\label{Tdefagain}
\Tb\ofk=
\omat{cc}
\xit+\ft\Do^2\Dt^{-2}
&
-\Dou\Dt^{-1}\xi-\ft\Do^3\Dt^{-3}\phi
\\
-\Dou\Dt^{-1}\xi-\ft\Do^3\Dt^{-3}\phi&
\Do^2\Dt^{-2}+\ft\Do^4\Dt^{-4}\phit
\cmat
\left(\frac{\Dt}{\Do+\Dt\hsom\xi}\right)^2
,
\ee
\be\label{dTdefagain}
\dTb\ofk=
r f
\omat{cc}
-2\Do\Dt^{-1}\xi & \Do^2\Dt^{-2}[\phixi+1]\\  
\Do^2 \Dt^{-2} [\phixi+1] & -2\Do^3 \Dt^{-3} \phi  
\cmat
\left(\frac{\Dt}{\Do+\Dt\hsom\xi}\right)^2
.
\ee
The Cholesky decomposition
\be\label{choldef}
\Tbo\ofk=\Lbo\ofk\Lbo\Trm\ofk
\ee
reverts to the Cholesky decomposition of $\Tb\ofk$ when
$r=0$. Explicit expressions appear in Appendix~\ref{specapp}. Because of the
above relationships the transformed quantities
\be\label{ZH}
d\bmcZo\ofk =\cSoo^{-1/2}(k)\hsom\Lboinv\ofk \hsom d\bmcHo\ofk 
\ee
have a spectral matrix that is the 2$\times$2 identity,
\be
\langle
d\bmcZo\ofk \hsom d\bmcZoH\ofk 
\rangle=\Ib\dtbk\dkkp
.
\ee

\subsubsection{In actuality: finite length and discretely sampled}
\label{actuality}

We now define the vector of Fourier-transformed observations, derived
from the actual measurements in eq.~(\ref{fourierH}) and in
(\ref{fourierG}), through eq.~(\ref{gho2}),   
\be\label{newobs}
\Hbo\ofk=\Hovec
.
\ee
With~$W_K\ofk$ the Fourier transform of the applied window defined in
eq.~(\ref{Wdefine}), and by comparison with
eqs~(\ref{specblur})--(\ref{fourierG}), the covariance  
\be\label{covHH}
\langle\Hbo\ofk\hsom \Hbo\Hrm\!\ofkp\rangle=
\intnyq W^{}_K(\kb-\kb'')W_K^*\hsomm(\kb'-\kb'')\hsom\bmcSo\ofkpp\dbk''
\approx 
\bbmcSo\ofk\,\dkkp
.
\ee
In comparison to eq.~(\ref{specdensoij}) and eqs~(\ref{S0def})
or~(\ref{split}), the finite observation window introduces spectral
blurring, the loss of separability of the spectral and lithospheric
portions, and small correlations between wave vectors. These we
ignored when writing the last, approximate equality, introducing the
blurred quantity (for a specific window $w_K$, as opposed to
eqs~\ref{infinite}--\ref{finitea} where we first used the overbar notation)
% The explanation is that we DEFINED something and then GOT something
% which doesn't necessarily exactly decorrelate, though we are trying!  
\be
\label{Hcovar}\label{Sdefinebar}
\bbmcSo\ofk 
=
\intnyq
\left|W_K(\kb-\kb')\right|^2
\bmcSo\ofkp
\dbk'
.
\ee
We denote the Cholesky decomposition of~$\bbmcSo$ as 
\be\label{choleskybar}
\bbmcSo\ofk=
\bLbo\ofk\hsom\bLbo\Trm\ofk
,
\ee
such that the transformed variable
\be\label{Zdef}
\Zbo\ofk=\bLboinv\ofk\Hbo\ofk
\ee
has unit variance
\be\label{Idef}
\langle
\Zbo\ofk \hsom \Zbo\Hrm\ofk \rangle=\Ib
.
\ee

\begin{figure}\centering
\includegraphics[width=\abit\textwidth]{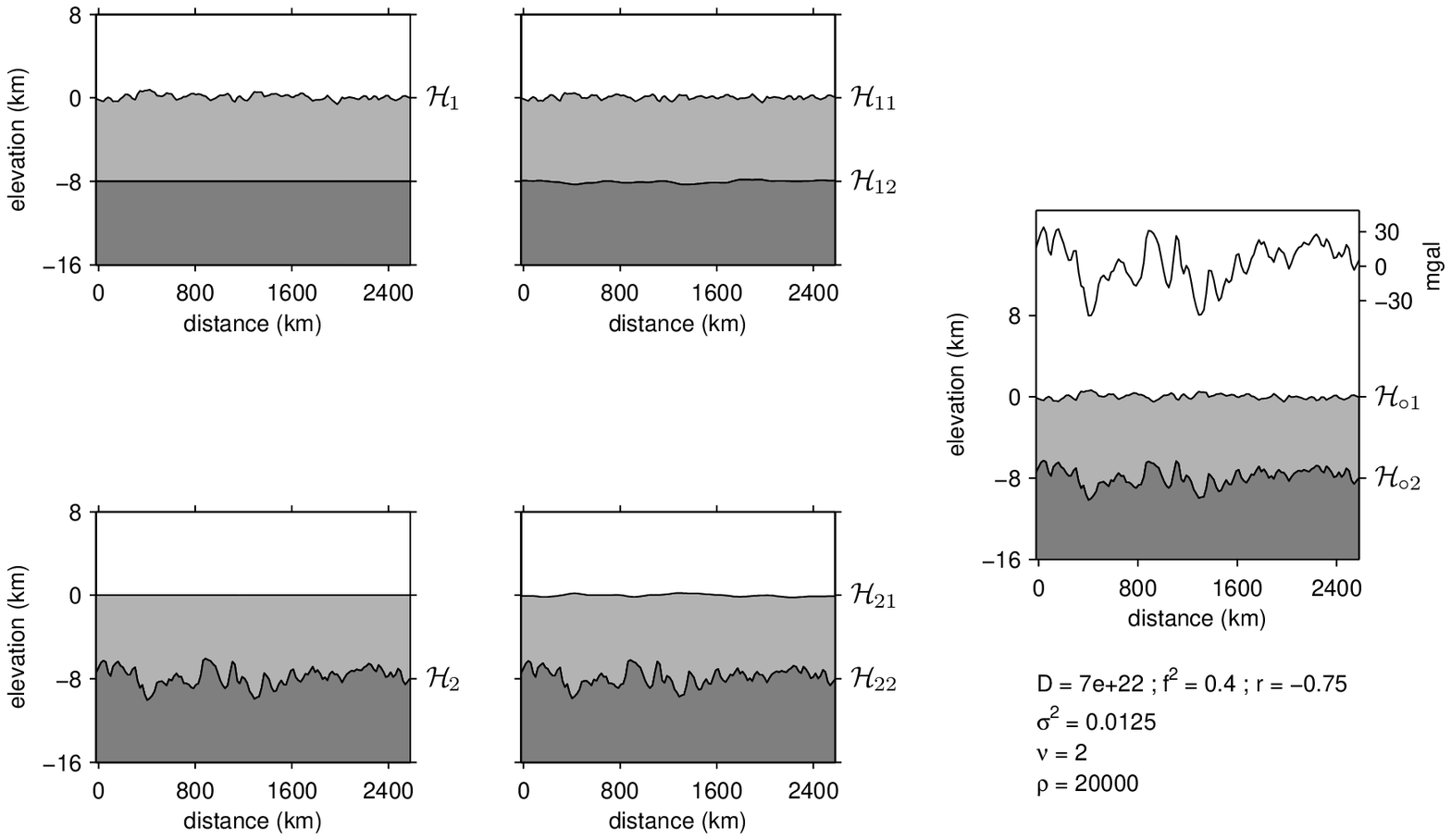}
\caption{\label{oh1}Synthetic data representing the standard model,
  identifying the initial, $\Hj$, equilibrium, $\Hij$, and final
  topographies, $\Hnj$, emplaced on a lithosphere with flexural
  rigidity~$D$. The initial loads were generated from the Mat\'ern
  spectral class with parameters $\st$, $\rho$ and $\nu$; they were
  negatively correlated, $r=-0.75$, and the spectral proportionality
  was $\ft$, as indicated in the legend. Also shown, by the black line,
  is the Bouguer gravity anomaly,~$\Gpn$. The density contrasts used were
  $\Do=2670$~kg$\hsom$m$^{-3}$ and $\Dt= 630$~kg$\hsom$m$^{-3}$,
  respectively. All symbols are listed and explained in
  Table~\ref{tablelitho}.}
\end{figure}

\subsubsection{In simulations: how to go from the continuous to the
  discrete formulation} \label{simulations}

Correctly generating a data set~$\Hbo$ that is a realization from a
theoretical spectral process~$d\bmcHo$ with the prescribed
spectral density~$\bmcSo$ requires ensuring that when we
observe a finite sample of it, and we form the (tapered)
periodogram of this, we get the correctly blurred spectral
density
\cite[]{Percival92,Chan+99,Dietrich+93,Dietrich+97,Thomson2001,Gneiting+2006}
in our case eq.~(\ref{Sdefinebar}).  
Stability considerations require that should we simulate data on one
discrete grid and then extract a portion on another discrete grid, we
replicate the correct covariance structure everywhere in space and
always produce the correct blurring upon analysis. Failure to
acknowledge the grid properly at the simulation stage can lead to
severely compromised results as will be readily experienced but has
not always been consciously acknowledged in the (geophysical)
literature \cite[]{Peitgen+88,Robin+93}. The method that we
outline here is variously known as \cite{Davies+87} or circulant
embedding \cite[]{Wood+94,Craigmile2003}. 

Let us assume that we have a spatial grid~$\xb$ as in
eq.~(\ref{xgrid}), and a half-plane Fourier grid~$\kb$ as in
eq.~(\ref{kgrid}). On the~$K$ entries of the latter we generate
(complex proper) Gaussian variables~$\Zbo\ofk$ and then transform
these as suggested by eqs~(\ref{Zdef})--(\ref{Idef}),   
% MORE ABOUT COMPLEX PROPER IS TO ANNOUNCED
\be\label{Hogen}
\Hbo\ofk=\bLbo\ofk\Zbo\ofk
,
\ee
whereby $\bLbo$ is the Cholesky decomposition expressed on the
grid~$\kb$, of eq.~(\ref{Hcovar}) calculated on a much finer
grid~$\kb'$. In other words, 
\be\label{simublur}
\bLbo\ofk=\mathrm{chol}\!\left[
\mathrm{conv}\!\left\{\left|F\ofkp\right|^2\!,\bmcSo\ofkp\right\}
\right]
=\mathrm{chol}\!\left[
\intnyq
\left|F(\kb-\kb')\right|^2
\bmcSo\ofkp
\dbk'\right]
=\mathrm{chol}\!\left[\bbmcSo(\kb)
\right]
,
\ee
whereby $|F\ofk|^2$ is the unmodified periodogram of the %normalized
spatial boxcar function % $f\ofx = (2K)^{-1}$ 
that defines the simulation grid.
%: $F\ofk$ is Dirichlet's kernel, $|F\ofk|^2$ Fej\'er's, see PW
The convolution in eq.~(\ref{simublur}) is to be implemented
numerically, with care taken to preserve the positive-definiteness of
the result. We now define the discrete inverse Fourier
transform of this particular set of variables for this fixed  set of
wave vectors~$\kb$ to be equal to the integral that we 
introduced in eq.~(\ref{fourierfirst}), 
\be
\label{thetrick1}
\bmcHo\ofx=\intnyq e^{i\kb\cdot\xb}\,d\bmcHo\ofk
\equiv\frac{1}{K}\sum_\kb e^{i\kb\cdot\xb}\hsom\Hbo\ofk
,
\ee 
which holds, in fact, for any $\xb\in \mathbb{R}^2$, and is 
consistent with  eq.~(\ref{fourierH}) which holds for
the area of interest picked out by the boxcar
window. We generate synthetic data sets~$\bmcHo\ofx$ via
eqs~(\ref{Hogen})--(\ref{thetrick1}): by this procedure the 
covariance between any two points $\xb$ and $\xb'$ in any portion of
space identified as our region of interest is now determined to be
\be\label{thetrick2}
\langle
\bmcHo\ofx\bmcHot\ofxp\rangle=\intnyq 
e^{i\kb\cdot(\xb-\xb')} 
\bmcSo(\kb)\dbk
=\bmcC_0(\xb-\xbp)
\approx
\frac{1}{K}\sum_\kb
e^{i\kb\cdot(\xb-\xbp)} 
\bbmcSo(\kb)
,
\ee
which follows from eqs~(\ref{thetrick1}), (\ref{cSodefdef})
and~(\ref{covHH}) with the small correlations between wave vectors
neglected, and using the notation introduced in eq.~(\ref{Ciso}).  Now
eq.~(\ref{thetrick2}) is equal to the universal expression in
eq.~(\ref{specdensx}), consistent with
eqs~(\ref{infinite})--(\ref{finiteb}), and since the dependence is only
on the separation $\xb-\xb'$, stationarity is guaranteed.  With
$\xb=\xb'$ eq.~(\ref{thetrick2}) states Parseval's theorem: at every
point in space the variance of~$\bmcHo$ is equal to all of its
spectral energy. Of course in the isotropic case considered here,
$\bmcC_0(\xb-\xbp)=\bmcC_0(\|\xb-\xbp\|)$, depending only on distance.  

Should we now take the finite windowed Fourier transform of
such synthetically generated spatial data~$\bmcHo\ofx$ on a
different spatial patch (e.g. a subportion from the master
set), while using any arbitrary window or taper~$w_{K'}\ofx$, we will be
seeing the correctly blurred version of the theoretical spectral
density~$\bmcSo$, as required to ensure stability. Indeed, when
forming a new set of modified Fourier coefficients~$\Hbo'\ofk$,
distinguished by a prime,  
\be\label{Hogen2}
\Hbo'\ofk=\sum_\xb w_{K'}\ofx \bmcHo\ofx e^{-i\kb\cdot\xb}
,
\ee
their covariance now must be, as follows directly from
eqs~(\ref{Hogen2}), (\ref{thetrick2}) and~(\ref{Wdefine}), the blurred
quantity
\ber
\langle
\Hbo'\ofk\hsom \Hbo\Hbfp\ofkp 
\rangle&=&
\sum_\xb w_{K'}\ofx e^{-i\kb\cdot\xb} 
\sum_\xbp w^*_{K'}\hsomm\ofxp e^{i\kb'\cdot\xb'} 
\langle\bmcHo\ofx\bmcHot\ofxp\rangle\\
&=& 
\intnyq
\sum_\xb w_{K'}\ofx e^{-i(\kb-\kb'')\cdot\xb} 
\sum_\xbp w^*_{K'}\hsomm\ofxp e^{i(\kb'-\kb'')\cdot\xb'}
\bmcSo(\kb'')\dbk''\\
&=& 
\intnyq
W_{K'}(\kb-\kb'')
W^*_{K'}\hsomm(\kb'-\kb'')\hsom
\bmcSo(\kb'')\dbk''
,
\eer
which is exactly as we have wanted it to be consistent with
eq.~(\ref{covHH}). We will continue to neglect the small correlations
between wave vectors, but fortunately this will have limited
impact~\cite[]{Varin2008,Varin+2011}.  

Fig.~\ref{oh1} shows a realization of a simulation produced with the
method just described. In contrast to Fig.~\ref{oh0} we now show the
result of the case where the initial-loading topographies are indeed
(negatively) correlated. Evidence for the loading correlation is not
apparent to the naked eye.

\subsection{The log-likelihood function,  $\mcL$}
\label{loglisec}

Conditioned upon higher-order moments of the space-domain data being
finite \cite[]{Brillinger74}, their Fourier components are
near-Gaussian distributed, and for stationary processes, there are no
correlations between the real and imaginary parts of the Fourier
transform, which are independent. Writing $\mcN$ for the Gaussian and
$\mcNC$ for the proper complex Gaussian distributions
\cite[]{Miller69,Neeser+93}, and dropping more wave vector
dependencies as arguments than before, the observation
vectors~$\Hbo\ofk$ in eq.~(\ref{newobs}) and the rescaled~$\Zbo\ofk$
of eq.~(\ref{Zdef}) are thus characterized at each wave vector~$\kb$
by the probability density functions
\be\label{hnc}
\pHbo=
\NCOSbo,\qquad
p_{\Zbo}=
\NCOI,\qquad
{\mathcal{R}e}\{\Zbo\}\sim
\NOh
,\also
{\mathcal{I}m}\{\Zbo\}\sim
\NOh
.
\ee
As we have noted at the end of Section~\ref{Cramer}, at the Nyquist and
zero wave numbers these quantities are real with unit variance. In so
writing the observation vector is treated as a random variable,
but we are interested in the likelihood of
observing the particular data set at hand given the
model, which for us means an evaluation at the data in
function of the deterministic parameters $\st$, $\rho$, $\nu$, $D$,
$\ft$, $r$. This quantity, $\Lbar\oft$, receives contributions from
each wave vector~$\kb$ that, once the number $K$ of considered wave
vectors is large enough, can be considered independent from one
another \cite[]{Dzhamparidze+83}. The log-likelihood is thus,
up to a constant, given by the standard form 
\be\label{firstlbar}
\Lbar\oft= % Changed to script_o quantities
\frac{1}{K}\left[\ln\prod_\kb\fracd{
\exp(-\qform{\bbmcSoinv})
}
{\det\bbmcSo}\right]
=-\norml
\left[
\ln(\det\bbmcSo)
+\qform{\bbmcSoinv\hsomm}\right]
=\norml \Lbark\oft
.
\ee
While we know that there is in fact correlation between the terms
$\Lbark\oft$, only at very small sample sizes~$K$ will this produce
inefficient estimators, as the accrued effects of the correlation
diminish in importance with increasing sample sizes. At moderate to large sample  
sizes there is  considerable gain in computational efficiency and no loss of
statistical efficiency due to the fast spectral decay of the blurring kernel
functions involved. Our objective function, the log-likelihood,
remains simply the average of the contributions at each wave vector in
the half plane. Of course eqs~(\ref{hnc})--(\ref{firstlbar}) contain
the blurred spectral forms~$\bbmcSo\ofk$ that we defined in 
eq.~(\ref{Sdefinebar}), in acknowledgment of the fact that the
variance experiences the influence from nearby wave vectors: the
approximation made asymptotically is that of eq.~(\ref{covHH}), but
eq.~(\ref{Sdefinebar}) is exact. 

While we cannot ignore this blurring for finite sample size and for
the particular data tapers used to obtain the windowed Fourier
transforms, for very large data sets and well-designed, fast-decaying,
window functions \cite[e.g.][]{Simons+2011a} the observation vectors~$\Hbo$
will converge `in law' \cite[]{Ferguson96} to random variables~$\Hbo'$ that
are distributed as complex proper Gaussian with an unblurred variance,
\be
\Hbo\inlaw\Hbo'\sim\mcNC\!\!\left(\bzero,\bmcSo\right)
\ee
in which case we would simply write 
\be\label{hncno}
\pHbo=
\mcNC\!\!\left(\bzero,\bmcSo\right)
. 
\ee
Working with this distribution is mathematically more convenient
since all of the subsequent calculations can be done analytically,
and, per eq.~(\ref{split}), separably in the lithospheric and spectral
parameters, so we will adhere to it until further notice. In this case
the log-likelihood is 
\be\label{firstl}\label{firstlt}
\mcL\oft=
\frac{1}{K}\left[\ln\prod_\kb\frac{\exp(-\qform{\bmcSoinv})
}
{\det\bmcSo}\right]
=
-\norml
\left[
2\ln\cSoo
+\ln(\det \Tbo)
+\hform{\hsom\Tboinv\hsomm}\right]
=\norml \mcLk\oft
.
\ee
While algorithms for simulation and data analysis will be based on
eq.~(\ref{firstlbar}), we will use eq.~(\ref{firstl}) to study the
properties of the solution, ultimately (in Section~\ref{numexper} and
Appendix~\ref{numex}) demonstrating why such an approach is justified.
On par with eq.~(\ref{firstlt}) we introduce an equivalent likelihood
in whose formulation the correlation coefficient~$r$ does not appear,
with the notation of eqs~(\ref{btheta})--(\ref{bthetawi}) and
eqs~(\ref{split})--(\ref{dTdefagain}), namely
\be\label{firstlw}
\mcLw\oftw=-\norml
\left[
2\ln\cSoo
+\ln(\det \Tb)
+\hform{\hsom\Tbinv\hsomm}\right]
.
\ee

\subsection{The maximum-likelihood estimator, $\hbt$}

The gradient of the log-likelihood, the score function, is the vector   
\be\label{score2}
\bgamma\oft=
\left[
\frac{\pl\mcL}{\pl D}\quad
\frac{\pl\mcL}{\pl\ft}\quad
\frac{\pl\mcL}{\pl r}\quad
\frac{\pl\mcL}{\pl\st}\quad
\frac{\pl\mcL}{\pl\nu}\quad
\frac{\pl\mcL}{\pl\rho}
\right]\Tit
,
\ee
with generic elements, never to be confused with the
coherence functions~(\ref{gto})--(\ref{gtf}), that we shall denote as  
\be\label{scoregen}
\gth=\frac{\pl\mcL}{\pl\theta}
=\norml\frac{\pl\mcLk}{\pl\theta}
=\norml\gtk 
.
\ee
Following standard theory
\cite[]{Pawitan2001,Davison2003} we define the maximum-likelihood
estimate as that which maximizes~$\lth$, thus $\hbt$~is the vector of
the maximum-likelihood estimate of the parameters, for which 
\be\label{score1}
\bgamma\ofth =\bzero
.
\ee
Contingent upon the requisite second order conditions being satisfied
\cite[]{Severini2001}, this is also assumed to be the global maximum
of~(\ref{firstl}) in the range of parameters that~$\btheta$ is allowed
to take. We now let~$\btruth$ be the vector containing the true,
unknown values, and have a certain~$\btheta'$ lie somewhere inside a
ball of radius $\|\hbt-\btruth\|$ around it. Then we may expand the
score with a multivariate Taylor series expansion, using the Lagrange
form of the remainder, to arrive at the exact expression
\be\label{ball}
\bgamma \ofth
=
\bgamma \oftr
+\Fb(\btheta')(\hbt-\btruth),
\for
\|\btheta'-\btruth\|<\|\hbt-\btruth\|
.
\ee
The random matrix~$\Fb$ is the Hessian of the log-likelihood function,
with elements defined by
\be\label{fobs}
\Hththp=
\frac{\pl\gamma_{\theta'}}{\pl\theta}=
\frac{\pl^2\hsomm\hsomm\mcL}{\pl\theta\,\pl\theta'}
,
\ee
and an expected value~$-\bmcF$, the Fisher `information matrix', 
\be\label{fisher}
\bmcF\oft=
-\langle\Fb\oft\rangle
,\quad\mbox{with elements}\quad
\Fththp=
-\Big\langle\frac{\pl^2\hsomm\hsomm \mcL}
{\pl \theta\, \pl \theta'}\Big\rangle
.
\ee
Hence the name `observed Fisher matrix' which is sometimes used
for the Hessian. If it is invertible we may rearrange eq.~(\ref{ball})
and write  
\be
\label{MLE}
\hbt=\btruth-\Fb^{-1}\!
(\btheta')\,\bgamma\oftr
.
\ee
For this exponential family of distributions the random
Hessian converges `in probability' to the constant Fisher matrix 
\be\label{inprob}
\Fb\oft\inprob-\bmcF\oftr
.
\ee
This is more than a statement about means: the fluctuations of~$\Fb$
about its expected value also become smaller and smaller. Thus, no
matter where we evaluate the Hessian, at~$\btheta'$ or at~$\btruth$,
both tend to the constant matrix~$\bmcF$. The distributional
properties of the maximum-likelihood estimator~$\hbt$ can be deduced
from eqs~(\ref{MLE})--(\ref{inprob}), which are also the basis for
Newton-Raphson iterative numerical
schemes~\cite[e.g.][]{Dahlen+2008}. We thus need to study the behavior
of~$\bgamma$, $\Fb$, and~$\bmcF$. The symbols of the statistical
apparatus that we have assembled so far are listed in
Table~\ref{tablespectral}.  

\subsection{The score function, $\bgamma$}

Per eqs~(\ref{score2})--(\ref{score1}) the derivatives of the
log-likelihood function~$\mcL$ vanish at the maximum-likelihood
estimate~$\hbt$. With our representation of the unknowns of our
problem by the parameter sets $\bthetaL$ and $\bthetaS$ we are in the
position to calculate the elements of the score function~$\bgamma$
explicitly. We remind the reader that these are not for use in the
optimization using real data sets where the blurred
likelihood~$\Lbar$ is to be maximized instead. In that case the
scores of~$\Lbar$ will need to be calculated numerically. However,
the scores of the unblurred likelihood~$\mcL$ that we present
here will prove to be useful in the calculation of the variance of the
maximum-blurred-likelihood estimator.  
Combining eqs~(\ref{firstl}) through~(\ref{scoregen}) we see that the
general form of the elements of the score function will be given by
\be
\label{score}
\gth
=
\norml
\gtk=
-\norml
\left[2\mth\ofk+\hform{\Abth}\right]
.
\ee
For the lithospheric and spectral parameters, respectively, we will have
\ber\label{mthAL}
\mthL\hsomm\ofk=\frac{1}{2}\frac{\pl\hsomm\ln(\det\Tbo)}{\pl\thetaL},&&
\AbthL=\frac{\pl\Tboinv}{\pl{\thetaL}},\\
\label{mthAS}
\mthS\hsomm\ofk=\cSoom\frac{\pl\cSoo}{\pl{\thetaS}},\hspace{2em}&&
\AbthS=-\mthS\hsomm\ofk\hsom\Tboinv
.
\eer
The explicit expressions can be found in
Appendices~\ref{scorelapp}--\ref{scoresapp}. For 
completeness we note here that 
$
\pl\Soom/\pl\thetaS=-\mthS\Soom
$.

To determine the sampling properties of the maximum-likelihood
estimation procedure we use eqs~(\ref{hncno})--(\ref{scoregen}) to 
make the identifications 
\be\label{logp}
\mcLk=\ln \pHbo\also
\gtk=\frac{1}{\pHbo}\frac{\pl \pHbo}{\pl\theta}
,
\ee
to obtain the standard result that the expectation of the score over
multiple hypothetical realizations of the observation vector vanishes,
as
\be\label{blabla}
\langle
\gtk
\rangle=
\int \gtk\, \pHbo\,d\Hbo=
\int \left(\frac{\pl \pHbo}{\pl \theta}\right)\,d\Hbo=
\frac{\pl}{\pl \theta}\left(\int  \pHbo\,d\Hbo\right)=
\frac{\pl}{\pl \theta}\big(1\big)
=0.
\ee
In the treatment that is to follow~\cite[]{Johnson+73}, we
will need to perform operations on multiple similar forms as in
eq.~(\ref{score}), namely  
\be\label{similar}
\gtk=-2\mth\ofk-\hform{\Abth}
.
\ee
To facilitate the development for the second term in
eq.~(\ref{similar}) we use eq.~(\ref{Hogen}), but again
without the complications of spectral blurring, see eq.~(\ref{ZH}),
and proceed by eigenvalue decomposition of the symmetric matrices  
$\Lbo\Trm\hsomm \Abth\Lbo$ to yield  
\ber\label{SHOH}
\hform{\Abth}&=&\Zbo\Hrm(\lform{\Abth})\Zbo
=
\Zbo\Hrm(\Pb_\theta\Hrm \blt \Pb_\theta) \Zbo
=
(\Pb_\theta \Zbo)\Hrm\blt(\Pb_\theta \Zbo)
=
\tilde{\Zb}_\theta\Hrm\blt\tilde{\Zb}_\theta\\
&=&\label{SHAH}\lambda\unth\rmo\ofk
\left|\tilde{\mcZ}\unth\rmo\ofk\right|^2+
\lambda\unth\rmt\ofk\left|\tilde{\mcZ}\unth\rmt\ofk\right|^2 
,
\eer
where $\lambda\unth\rmo\ofk$ and $\lambda\unth\rmt\ofk$ are the two
possibly degenerate eigenvalues of $\lform{\Abth}$ constructed by
combining eqs~(\ref{choldef}) and~(\ref{mthAL})--(\ref{mthAS}),
\be
\lambda_\theta^\pm=
\mathrm{eig}\left(
\lform{\Abth}
\right)
.
\ee
Since the matrix $\Pb_\theta$ is orthonormal, $\Zb_\theta$
and~$\tilde{\Zb}_\theta$ are identically distributed and thus we find
through eq.~(\ref{hnc}) that eq.~(\ref{SHAH}) is a weighted sum of
independent random variables, each exponentially 
%Erlang
distributed, $\chit_2/2$,
with unit mean and variance. In summary, we have the convenient form
for the contributions to the score~(\ref{score}) from each individual
wave vector,
\be 
\label{scoreinthetaj}
\gtk=
-2\mth\ofk-
\lambda\unth\rmo\ofk
\left|\tilde{\mcZ}\unth\rmo\ofk\right|^2-
\lambda\unth\rmt\ofk\left|\tilde{\mcZ}\unth\rmt\ofk\right|^2
.
\ee
Since $\mth$ is  nonrandom we thus have an expectation for the
contributions to the score that confirms eq.~(\ref{blabla}), namely
\be\label{meanscore}
\langle\gtk\rangle
=
-2\mth\ofk
-
\lambda\unth\rmo \ofk
-\lambda\unth\rmt\ofk=0
,
\ee
and a variance given by 
\be\label{varscore}
\langle\gtk\gtk\rangle
=\left[\lambda\unth\rmo\ofk\right]^2
+\left[\lambda\unth\rmt\ofk\right]^2
=\var\{\gtk\}.
\ee
We also retain the useful expression 
\be\label{retain}
\langle\hform{\Abth}\rangle=
\tr(\lform{\Abth})=
\lambda\unth\rmo\ofk+\lambda\unth\rmt\ofk=-2\mth\ofk.
\ee
Eq.~(\ref{varscore}) gave us the variance of the derivatives
of the log-likelihood function with respect to the parameters of
interest, which was written in terms of the eigenvalues of the
non-random matrix $\lform{\Abth}$. More specifically, for the
variances of the scores in the 
lithospheric parameters $\tl$ in $\bthetaL=\bthetaLfullr$, we will find
\be\label{varagain}
\var\{\gtlk\}
=\left[\lambda\unthL\rmo\ofk\right]^2+\left[\lambda\unthL\rmt\ofk\right]^2,
\ee
whereas for the variances of the scores in any of the three spectral
parameters $\ts$ in $\bthetaS=\bthetaSfull$, judging from
eq.~(\ref{mthAS}), we will need the sum of the squared 
eigenvalues of $-\mthS\,\lform{\Tboinv}$ and since~$\Lbo$ is the
Cholesky decomposition of~$\Tbo$, we have  $\Tboinv=\Lbo\Trminv\Lboinv$ and 
\be\label{varagain2}
\var\{\gtsk\}
=2\mthS^2\hsomm\ofk
.
\ee

\begin{table*}
\centering
\begin{tabular}{clc}\hline 
  & description &  eq.\\\hline
$\btruth$ & the true, unknown, parameter set of the problem,
  consisting of lithospheric and spectral parameters & (\ref{ball}) \\[0.5em]
$\hbt$ & the maximum-likelihood estimate of the parameter set & (\ref{score1})\\
$\btheta,\btheta'$ & generic occurrences of the parameter set & (\ref{btheta})\\
$\bthetaL$ & the lithospheric parameter set of the estimation procedure, containing $D$, $\ft$ and $r$ listed in Table~\ref{tablelitho} & (\ref{thetaL})\\
$\bthetaS$ & the spectral parameter set of the estimation procedure,
  containing $\st$, $\rho$ and $\nu$ listed below & (\ref{thetaS})\\
$\bthetaw$ & the parameter set not including the correlation coefficient~$r$  & (\ref{bthetawi})\\[0.75em]
$d\bmcHo$ & the ``theoretical'' observation vector, containing final topographies $\Hno$ and $\Hnt$ at both interfaces &  (\ref{vecdH})\\
$d\bmcH$ & the theoretical  vector containing initial topographies $\Hoo$ and $\Htt$ at both interfaces &  (\ref{vecdH})\\
$\bmcS$ & the spectral matrix containing the (cross-)spectral densities of the theoretical initial topographies  & (\ref{specdensij}) \\
$\bmcSo$ & the spectral matrix containing the (cross-)spectral densities of the theoretical final topographies  & (\ref{specdensoij}) \\
$\bmcMD$ & the matrix that maps the initial-loading spectral matrix $\bmcS$ to the final-observed spectral matrix $\bmcSo$ & (\ref{Mdef}) \\[0.75em]
$\cSoo$ & the power spectral density of the top-loading process, here assumed to be isotropic & (\ref{materndef}) \\
$\st$ & the first quantity in the parameterized Mat\'ern form of the spectral density~$\cSoo$, to be estimated  & (\ref{materndef})\\
$\rho$ & the second quantity in the parameterized Mat\'ern form of the spectral density~$\cSoo$, to be estimated &  (\ref{materndef}) \\
$\nu$ & the third quantity in the parameterized Mat\'ern form of the spectral density~$\cSoo$, to be estimated &   (\ref{materndef}) \\[0.75em]
$\Tbo$ & the ``spectral'' matrix after factoring the power spectrum of the top-loading process, $\cSoo$, out of $\bmcSo$ & (\ref{split}) \\ 
$\Tb$ & the part of $\Tbo$  that is independent of the correlation coefficient $r$ between the loads  &  (\ref{Tdefagain}) \\ 
$\dTb$ & the part of $\Tbo$  that depends on the correlation coefficient $r$ between the loads  & (\ref{dTdefagain}) \\ 
$\Lbo$ & a lower-triangular matrix forming the Cholesky decomposition of $\Tbo$ & (\ref{cholesky}) \\[0.75em]  
$\Hbo$ & the ``observed'' observation vector, containing final topography  $\nHno$ and $\nHnt$ at both interfaces &  (\ref{newobs})\\
$\bbmcSo$ & the ``blurred'' spectral matrix, containing the (cross-)spectral densities of the actual final topographies  & (\ref{Sdefinebar}) \\
$\bLbo$ & a lower-triangular matrix forming the Cholesky decomposition of $\bbmcSo$ & (\ref{choleskybar}) \\[0.75em]  
%$\mcH,\mcG$ & generic univariate terrestrial quantity such as topography or gravity & (\ref{fourierfirst}) \\
$\bar\mcL\oft$ &  the likelihood of observing Bouguer gravity and topography under the two-layer flexural model & (\ref{firstlbar})\\
$K$ & total number of all wave vectors considered, covering the upper half-plane of spectral space  & (\ref{kgrid})\\
$k,k'$ & generic wavenumbers from the wave vectors $\kb$, $\kb'$  &
  (\ref{materndef}) \\[0.75em]
$\mcL\oft$ &  the likelihood of observing Bouguer gravity and topography neglecting spectral blurring & (\ref{firstl})\\
%$\mcLk\oft$ &  the unblurred likelihood of observing gravity and topography under the model at a single wave vector~$\kb$ & (\ref{firstl})\\
$\mcLw\oftw$ &  the likelihood of observing Bouguer gravity and topography neglecting spectral blurring and load correlation & (\ref{firstlw})\\
$\gamma_\theta$ & an element of the gradient of the likelihood~$\mcL$, or the
  score function, $\bgamma$ & (\ref{scoregen})\\
%$\gamma_\theta\ofk$ & an element of the gradient of the likelihood~$\mcLk$ at a single wave vector~$\kb$ & (\ref{scoregen})\\
$\Hththp$ &  an element of the Hessian of the likelihood~$\mcL$, or the observed Fisher matrix, $\Fb$ & (\ref{fobs})\\
% CHECK WITH SOFIA AS TO NEGATIVE
$\Fththp$ &  an element of the negative expectation of the Hessian, or the Fisher information matrix, $\bmcF$  & (\ref{fisher})\\
$\Jththp$ &  an element of the inverse of the Fisher information matrix, $\bmcJ$  & (\ref{Jdef})\\
$X_0$ & quadratic residual surface obtained after maximizing the likelihood  & (\ref{Xok})\\\hline 
$\mcS$ & generic isotropic Mat\'ern spectral density for univariate fields &
  (\ref{maternredux}) \\
$\mcLS$ &  the likelihood of observing univariate data under the isotropic Mat\'ern model & (\ref{secondl})\\
$\gamma\unS$ &  the score of the likelihood $\mcLS$ & (\ref{scoreS})\\
$\mcF\unS$ &  the Fisher matrix of the likelihood $\mcLS$ & (\ref{fisherS})\\
$X$ &  maximum-log-likelihood ratio test statistic to evaluate the need for initial-loading correlation & (\ref{X})\\
\hline\end{tabular}
\caption{\label{tablespectral}Some of the symbols used for the statistical theory
presented in this paper, their short description, and equation numbers
for context.}
\end{table*}

\noindent As to the covariance of the scores in the different parameters  we use
eqs~(\ref{logp})--(\ref{blabla}) to write 
\be
0=
\frac{\pl }{\pl \theta}\left[\int\gtpk \,\pHbo d\Hbo\right]
=
\int \frac{\pl }{\pl \theta}[\gtpk \,\pHbo]\,d\Hbo
=\int\left[
\frac{\pl}{\pl \theta}\gtpk
\right] \pHbo d\Hbo+
\int \left[\gtk\gtpk\right]\pHbo d\Hbo
,
\label{secondder}
\ee
and thereby manage to equate the variance of the score to the
expectation of the negative of its derivative, 
\be\label{handy}
\langle\gtk\gtpk\rangle
=-\Big\langle
\frac{\pl}{\pl \theta}\gtpk
\Big\rangle=
-\Big\langle
\frac{\pl^2\hsomm\hsomm\mcLk}{\pl\theta\pl\theta'}
\Big\rangle=
\cov\{\gtk,\gtpk\}
,
\ee
which should of course specialize to verify eq.~(\ref{varscore}),
giving us two calculation methods for the variance terms. We do not
consider any covariance between the scores at non-equal wave vectors.  

From eqs~(\ref{score}) and~(\ref{scoreinthetaj}) we have learned that
the full score~$\gth$ is a sum of random variables $\gtk$ or indeed
the~$|\tilde{\mcZ}\unth^\pm\ofk |^2$, which belong to the exponential
family. Between those we consider no correlations at different wave
vectors, and eqs~(\ref{meanscore}) and~(\ref{handy}) have given us
their mean and covariance, respectively. Lindeberg-Feller central
limit theorems apply~\cite[]{Feller68}, and so the distribution
of the score~$\gth$ will be Gaussian with mean zero and covariance
\be\label{CLT}\label{sumint}
\cov\{\gth,\gthp\}
=\normlt\cov\{\gtk,\gtpk\}
.
\ee
Using eqs~(\ref{handy}), (\ref{firstl})
and~(\ref{fobs})--(\ref{fisher}) we can rewrite the above expression
in terms of the diagonal elements of the Fisher matrix, 
\be\label{kvar}
K\cov\{\gth,\gthp\}
=
-\Big\langle
\frac{\pl^2\hsomm\hsomm\mcL}{\pl\theta\pl\theta'}
\Big\rangle
=-\langle \Hththp\rangle
=\Fththp
.
\ee

We can summarize all of the above by stating that, for $K$
sufficiently large, ignoring wave vector correlations, and through the
Lindeberg-Feller central limit theorem, the vector with the
scores in the individual parameters converges in law to what is
distributed as 
\be\label{inlaw} 
\sqrt{K}\bgamma\oft\sim\mcN(\bzero,\bmcF\oft)
.
\ee

\subsection{The Fisher information matrix, $\bmcF$} 
\label{Fisher}

From the definition in eq.~(\ref{fisher}) we have that the elements of
the Fisher matrix~$\bmcF$ are given by the negative expectation of the
elements of the Hessian matrix~$\Fb$, which themselves are the second
derivatives of the log-likelihood function~$\mcL$ with respect to the
parameters of interest~$\btheta$.  Per eq.~(\ref{kvar}) the Fisher
matrix scales to the covariance of the score~$\bgamma$, and by
combining eqs~(\ref{varagain})--(\ref{varagain2}) with
eq.~(\ref{score}) or, ultimately, eqs~(\ref{varscore})
and~(\ref{sumint}), we thus find a convenient expression for the
diagonal elements of the Fisher matrix, namely
\be\label{diago1}
\Fthth=\norml\var\{\gtk\}=\norml\big\{\left[\lambda\unth\rmo\ofk\right]^2
+\left[\lambda\unth\rmt\ofk\right]^2\big\},
\ee
which, for the spectral parameters specializes to the more easily
calculated expression 
\be\label{diago2}
\Fthsths=\twnorml\mthS^2\hsomm\ofk.
\ee
%The Lindeberg conditions
%~(\ref{lindeberg}) is
%are satisfied throughout the
%range of~$\kb$. 

For the cross terms, rather than combining eqs~(\ref{scoreinthetaj})
and~(\ref{sumint}), we proceed via eq.~(\ref{kvar}) and thus require
expressions for the elements of the Hessian. From eqs~(\ref{fobs})
and~(\ref{score}) we derive that the general expression for the
elements of the symmetric Hessian matrix are 
\be\label{fththp}  % VERIFIED AGAIN AND AGAIN 
\Hththp=
\frac{\pl\gamma_{\theta'}}{\pl\theta}
=
-\norml
\left[
2\frac{\pl m_{\theta'}\hsumm\ofk}{\pl\theta}
-\left(\cSoom\frac{\pl\cSoo}{\pl\theta}\right)
\hform{\Abthp}
+\hform{\left(\frac{\pl\Abthp}{\pl\theta}\right)}
\right]
.
\ee
Unless we use it in the numerical optimization of the
log-likelihood we only need the negative expectation of
eq.~(\ref{fththp}), the Fisher matrix
\be\label{mcfththp}  % VERIFIED AGAIN AND AGAIN
\Fththp=
-\langle\Hththp\rangle
=
\norml
\left[
2\frac{\pl m_{\theta'}\hsumm\ofk}{\pl\theta}
+2\left(\cSoom\frac{\pl\cSoo}{\pl\theta}\right)
\mthp\ofk
+\tr\left\{\lform{\left(\frac{\pl\Abthp}{\pl\theta}\right)}\right\}
\right]
,
\ee
where we have used eq.~(\ref{retain}). Of course, when
$\theta=\theta'$, the general eq.~(\ref{mcfththp}) specializes to the
special case~(\ref{diago1}) discussed before. 
Ultimately this equivalence is a consequence of
eq.~(\ref{handy}) which held that in expectation,
the product of first derivatives of the log-likelihood is equal to its
second derivative. 

The explicit forms are listed in Appendix~\ref{hessapp}, but looking ahead, we
will point to two special cases that result in simplified expressions.
It should be clear from the separation of lithospheric and spectral
parameters achieved in eq.~(\ref{split}) and from
eqs~(\ref{mthAL})--(\ref{mthAS}) that the mixed derivatives of one
lithospheric and one spectral parameter,
$\pl\unthL\mthS=\pl_{\thetaS}\mthL=0$ and $\pl\unthL\cSoo=0$,
both vanish, and that we thereby have 
\be\label{FLS} % CORRECT AUG 9 2011
F_{\thetaL\hsumm\thetaS}=\norml
\left(\hsom
\hform{\AbthL\hsomm}
\right)\mthS\hsomm\ofk
,\qquad
\mcF_{\thetaL\hsumm\thetaS}=
\twnorml \mthL\hsomm\ofk\hsom\mthS\hsomm\ofk
.
\ee
Finally, we also easily deduce that
\be\label{FSS}
F_{\thetaSu\hsumm\thetapS}=-\norml\left[
2\frac{\pl m_{\thetapS}\hsomm\ofk}{\pl\thetaS}
+\left\{m_{\thetaSu}\hsomm\ofk m_{\thetapS}\hsomm\ofk
-\frac{\pl m_{\thetapS}\hsomm\ofk}{\pl\thetaS}\right\}
\left(\hsom
\hform{\hsom\Tboinv\hsomm}
\right)
\right]
,\qquad
\mcF_{\thetaSu\hsumm\thetapS}=
\twnorml 
m_{\thetaSu}\hsomm\ofk\hsom m_{\thetapS}\hsomm\ofk 
,
\ee
where we have used the previously noted special case of
eq.~(\ref{retain}) by which 
$\langle\hsom\hform{\hsom\Tbinv\hsomm}\hsom\rangle=
\tr\left(\lform{\Tboinv}\right)=2$. The previously encountered
eq.~(\ref{diago2}) is again a special case of eq.~(\ref{FSS}) when $\thetaS=\thetaS'$. 
Both expressions (\ref{FLS}) and~(\ref{FSS}) are of an appealing
symmetry. Between them they cover the majority of the elements of
the Fisher matrix, which will thus be relatively easy to compute. 

\subsection{Properties of the maximum-likelihood estimate, $\hbt$}
\label{properties}

We are now ready to derive the properties of the maximum-likelihood
estimate given in eq.~(\ref{MLE}), which we repeat here, as
\be
\hbt=\btruth-\Fb^{-1}\! (\btheta')\,\bgamma\oftr.
\ee
From eq.~(\ref{inlaw}) we know that the score~$\bgamma$ converges to a
multivariate Gaussian, and from eq.~(\ref{inprob}) we know that the
Hessian~$\Fb$ converges in probability to the Fisher 
matrix~$\bmcF$. A Taylor expansion allows us to replace~$\btheta'$
by~$\btruth$ as in standard statistical practice \cite[]{Cox+74}. % P 109
Thus, by Slutsky's lemma~\cite[]{Severini2001,Davison2003} the distribution
of~$\hbt$ is also a multivariate Gaussian. Its expectation will be  
\be
\label{unbiased}
\langle\hbt\rangle=
\btruth
,
\ee
showing how our maximum-likelihood estimator is unbiased. 
Its covariance is
\be\label{MLE2}
\cov\{\hbt\}=
\bmcF^{-1}\oftr\,
\cov\{\bgamma\oftr
\}\,
\bmcF\Tcalinv\oftr
.
\ee
From eq.~(\ref{kvar}) we retain that
$K\cov\{\bgamma\oftr\}=\bmcF\oftr$ and with~$\bmcF=\bmcF\Tcal$ a
symmetric matrix, we conclude that the covariance of the
maximum-likelihood estimator is given by
\be\label{Jdef}
K\cov\{\hbt\}=\bmcF^{-1}\oftr,\qquad
\mbox{or indeed}\qquad
K\cov\{\hth,\hth'\}=\Jththp\oftr
,\where
\bmcJ\oftr=\bmcF^{-1}\oftr.
\ee

In summary, we have shown that
\be\label{summary2}
\sqrt{K}(\hbt-\btruth) \sim \mcN(\bzero,\bmcF^{-1}\oftr)
=
\mcN(\bzero,\bmcJ\oftr)
,
\ee
which allows us to construct confidence intervals on the parameter
vector~$\btheta$.  Denoting the generic diagonal element of the
inverse of the Fisher matrix evaluated at the truth~$\btruth$
as~$\Jthth\oftr$, this equation shows us that each element of the
parameter vector is distributed as
\be\label{confidence}
\frac{\sqrt{K}}{\Jthth\shalf\oftr}\left(\hth-\truth\right)\sim
\mcN(0,1),
\ee
As customary, we shall replace the needed values~$\btruth$ with the
estimates~$\hbt$ and quote the 100$\times\alpha$~\% confidence interval
on~$\truth$ as given by 
\be\label{conflocation}
\hth-z_{\alpha/2}\frac{\Jthth\shalf\ofth}{\sqrt{K}}
\le
\truth
\le
\hth+z_{\alpha/2}\frac{\Jthth\shalf\ofth}{\sqrt{K}}
,
\ee
where $z_{\alpha}$ is the value at which the standard normal reaches a
cumulative probability of $1-\alpha$, i.e. $z_{\alpha/2}\approx 1.96$
for a 95\% confidence interval.

These conclusions, which are exact for the case under consideration,
will hold asymptotically when in practice we use the blurred
likelihood~(\ref{firstlbar}) instead of eq.~(\ref{firstl}). In the
blurred case and for all numerical optimization procedures, we expect
to have to amend eqs~(\ref{unbiased}) and~(\ref{MLE2}) by correction
factors on the order of $K^{-1}$ and $K^{-2}$,
respectively. Eq.~(\ref{conflocation}) would receive extra correction
terms starting with the order $K^{-1}$, which would be immaterial
given the size of the confidence interval. 

In some sense, eq.~(\ref{conflocation}) concludes the analysis of our
maximum-likelihood solution to the problem of flexural-rigidity
estimation. It makes the important statement that each of the
estimates of flexural rigidity~$D$, initial-loading ratio~$\ft$, and
load correlation coefficient~$r$, will be normally distributed
variables centered on the true values and with a standard deviation
which will scale with the inverse square-root of the physical data
size~$K$. Obtaining the variance on the estimates of effective
elastic thickness~$T_e$ from the estimates of~$D$ will be made through
eq.~(\ref{tedef}) via the ``delta method'' \cite[]{Davison2003}. This
implies that the estimate of the effective elastic thickness is
approximately distributed as
\be
\widehat{T_e}\sim
\mcN\left(s^{1/3}D_0^{1/3},
\frac{1}{9}s^{2/3}D_0^{-4/3}\var\{\hat{D}\}\right)
\where
s=12(1-\nu^2)/E
.
\ee

\subsection{Analysis of residuals}
\label{analres}

\begin{figure}\centering
\rotatebox{-90}{\includegraphics[height=0.91\textwidth]{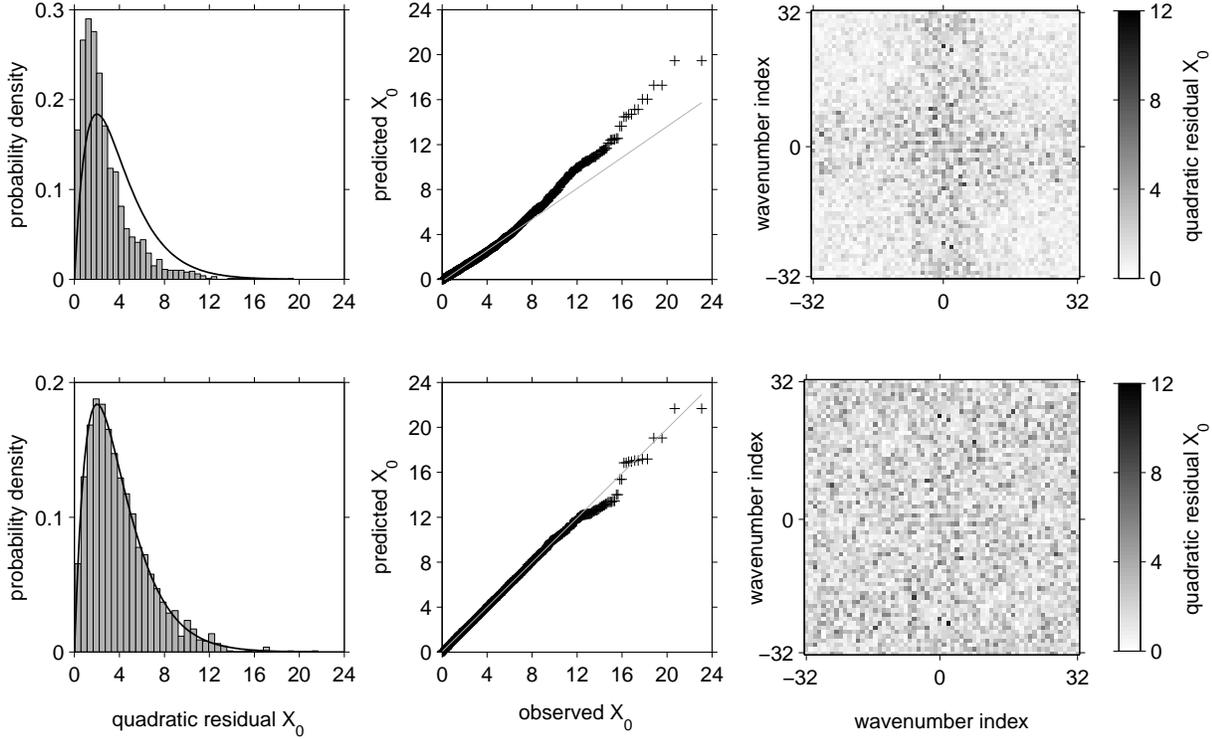}}
\caption{\label{oh3}The behavior of the quadratic residuals~$X_0\ofk$
  defined in eq.~(\ref{Xok}) in a recovery simulation for correlated
  loading. (\textit{Left column}) Observed (histogram) and
  theoretical~$\chi^2_4/2$ distribution (black curve) of the
  residuals~$X_0\ofk$ across all wave vectors~$\kb$. (\textit{Middle
    column}) Quantile-quantile plot of the observed~$X_0\ofk$ compared
  to their theoretical~$\chi^2_4/2$ distribution across all wave
  vectors~$\kb$. (\textit{Right column}) Plot of the observed
  residuals~$X_0\ofk$ in the wave vector plane.  The examples are for
  a case where~$K=2\times32\times32$, $\Do=2670$~kg$\hsom$m$^{-3}$,
  $\Dt= 630$~kg$\hsom$m$^{-3}$, $z_2=35$~km, and the sampling
  intervals were 20~km in each direction. The true model is for the
  correlated case where the lithospheric parameters are $D=1\times
  10^{24}$~N$\hsom$m, $\ft=0.8$ and $r=0.75$, and the spectral
  parameters $\st=2.5\times 10^{-3}$, $\nu=2$, $\rho=4\times 10^{4}$.
  (\textit{Top row}) A ``bad'' example where the residuals do not
  follow the predicted distribution and continue to show too much
  structure in the wave vector domain.  For this example, the poor
  estimate is given by $\hat{D}=1.5\times 10^{24}$~N$\hsom$m,
  $\hat{\ft}=0.915$ and $\hat{r}=0.656$, $\hat{\st}= 1.9\times
  10^{-3}$, $\hat{\nu}=1.5$, $\hat{\rho}=4.25\times 10^{4}$, and the
  blurred log-likelihood $\Lbar=-18.591$. (\textit{Bottom row}) A
  ``good'' example which indicates that the estimate will be accepted
  as a fair representation of the truth, which in this case is
  $\hat{D}=1.326\times 10^{24}$~N$\hsom$m, $\hat{\ft}=0.790$ and
  $\hat{r}=0.741$, $\hat{\st}= 2.415\times 10^{-3}$, $\hat{\nu}=2.00$,
  $\hat{\rho}=3.974\times 10^{4}$. The blurred log-likelihood
  $\Lbar=-18.2883$. No structure is detected in the residuals: the
  model fits.}
\end{figure}

Once the estimate $\hbt=[\hbt\unL\,\,\,\hbt\unS]\Tit$ has been found,
we may combine it with our observations, and  
through %eq.~(\ref{ZH}),
eq.~(\ref{Zdef}), form the variable
\be\label{nope}
\Zbho\ofk=\bLboinv(\kb)\big|_{\hbt}\hsom\hsom\Hbo\ofk
,
\ee
which should be distributed as the standard complex proper Gaussian
$\NCOI$. Equivalently, and as a special case of
eqs~(\ref{firstlbar}) and~(\ref{SHAH}),
\be\label{Xok}
X_0\ofk=\Zbho\Hrm\ofk\Zbho\ofk
=\qform{\bbmcSoinv\big|_{\hbt}}
\sim\chit_4/2
,
\ee
and these variables should be approximately independent. 
We can rank order them according to their size,
\be
\label{bqqplot}
X_0^{(1)}=\min\{X_0\ofk\} \le 
X_0^{(2)} \le\dots\le
X_0^{(K)}=\max\{X_0\ofk\},
\ee
and inspect the quantile-quantile plot \cite[]{Davison2003} whereby
the $X_0^{(j)}$, for all $j=1,\dots,K$, are plotted versus the inverse
cumulative density function of the~$\chi^2_4/2$ distribution,
evaluated at the argument $j/(K+1)$. If, apart from at very low and
very high values of~$j$, this graph follows a one-to-one line, there
will be no reason to assume that our model is bad for the data. This
can then further be formalized by a chi-squared test
\cite[]{Davison2003}, but a plot of the residuals as a function of
wave vector will be more informative to determine how the model is
misfitting the data. In particular it may diagnose anisotropy of some
form, or identify particular regions of spectral space that poorly
conform to the model and for which the latter may need to be revised.
Fig.~\ref{oh3} illustrates this procedure on a recovery simulation
under correlated loading.

If the method holds up to scrutiny of this type, then because ours is
a maximum-likelihood estimator, it will be asymptotically efficient,
with a mean-squared error that will be as small or smaller than that
of all other possible estimators, converging to the optimal estimate
as the sample size grows to infinity.

\subsection{Admittance and coherence return, briefly}
\label{adcohreturn}

The theoretical admittance~$\Qo$ and coherence~$\gto$ are nothing but
one-to-one functions of our parameters of
interest. Consequently~\cite[]{Davison2003}, maximum-likelihood
estimates for either~$\Qo$ or~$\gto$ are obtained simply by evaluating the
functions~(\ref{Qo}) or~(\ref{gto}) at the maximum-likelihood
estimate of the parameters. The equivalence is easy to appreciate by
expanding the score in the desired function, e.g. $\gto$, as a total
derivative involving the parameters $D$, $\ft$ and $r$, 
\be\label{totaldif}
\frac{\pl\mcL}{\pl\gto}=
\frac{\pl\mcL}{\pl D}
\frac{\pl D}{\pl\gto}+
\frac{\pl\mcL}{\pl\ft}
\frac{\pl\ft}{\pl\gto}+
\frac{\pl\mcL}{\pl r}
\frac{\pl r}{\pl\gto}
.
\ee
The score in $\gto$ vanishes when  $\pl\mcL/\pl D=\pl\mcL/\pl
\ft=\pl\mcL/\pl r=0$ as long as each of $\pl\gto/\pl D$, $\pl\gto/\pl
\ft$ and $\pl\gto/\pl r$ are non-zero. Thus the maximum-likelihood
estimates $\widehat{\Qo}$ and  $\widehat{\gto}$ are obtained at the
maximum-likelihood values $\hat{D}$, $\widehat{\ft}$ and $\hat{r}$,
and are computed without difficulty, as we will illustrate
shortly. See Appendix~\ref{cramerrao} for a few additional
considerations. 

\section{T~E~S~T~I~N~G{\hsps}T~H~E{\hsps}M~O~D~E~L}
\label{modeltest}

In the previous section we discussed the question whether the
``model'' to which we have subscribed is at all ``valid'' in very
general terms. Here, we will address two possible concerns more
specifically.  The main ingredients of our model are the flexural
equations~(\ref{biharm0}), correlation~(\ref{corrdef}) and
proportionality~(\ref{ftdef}) of the initial topographies, and the
isotropic spectral form~(\ref{materndef}) that we assumed for the
loading terms. Other than that, we have introduced a certain fixed
two-layer density structure $\Do$, $\Dt$ and $z_2$, and an approximate
way of computing gravity anomalies by way of eq.~(\ref{fanom}). When
working within this framework, we showed in Section~\ref{analres} how
to assess the quality of the data fit, and in
Section~\ref{adcohreturn} how to hindcast the traditional observables
of admittance and coherence. However, what we have not addressed is
the relative merits of alternative models. How appropriate is the
Mat\'ern class, especially in its isotropic form?  How different would
an analysis that does not consider correlated loading be from one that
does? What would be the effect of modifying or adding additional terms
to the flexural equations, as could be appropriate to consider more
complex tectonic scenarios, elastic non-linearities, elastic
anisotropy, or alternative rheologies~\cite[as, for
example,][]{Stephenson+80,Stephenson+85a,Ribe82,Swain+2003a,McKenzie2010}?
We cannot, of course, address all of these questions with any hope for
completeness, but in this section we introduce two specific
considerations that will speak to these issues.

The first, detailed in Appendix~\ref{appretrieval}, involves a
stand-alone methodology to recover the spectral parameters in the
Mat\'ern form given univariate multi-dimensional data. This will help
us build well-suited data synthetics; it will also enable the study of
terrestrial and planetary surfaces \textit{per se}, e.g. to measure the
roughness of the ocean floor or the lunar
surface~\cite[e.g.][]{Goff+2010,Rosenburg+2011}. Even more broadly, it
is an approach to characterize texture~\cite[]{Haralick79,Cohen+91} in
the context of geology and geophysics. Although our chosen
parameterization~(\ref{materndef}) permits a wide variety of spectral
shapes, we are of course limiting ourselves by only considering
isotropic loading models. In future work, anisotropic spectral shapes
for the loading terms will be considered.

The second, in Appendix~\ref{appratio}, is a worked example of how,
specifically, the inclusion or omission of the initial-loading correlation
coefficient, $r$, may influence the confidence that we should have in
our maximum-likelihood estimates obtained with or without it. We might
construct a likelihood~$\mcL\oft$, as in eq.~(\ref{firstl}) with all
terms~(\ref{split})--(\ref{dTdefagain}) present, or instead we might
force the initial-loading correlation to $r=0$. This would result in a
simpler form that we have called~$\mcLw\oftw$ in eq.~(\ref{firstlw}),
whereby the parameter~$r$ is lacking altogether from the vector 
\be\label{bthetaw}
\bthetaw=\bthetafull , 
\ee 
to be compared with the expression
for $\btheta$ in eq.~(\ref{btheta}). Since $\bthetaw\subset\btheta$,
both models are `nested': the less complicated model can be obtained by
imposing constraints on the more complicated model, so that the
simpler model is a special case of the more complicated one. In that
case the likelihood-ratio test~\cite[]{Cox+74,Severini2001} that we
describe in Appendix~\ref{appratio}  is applicable. It is inappropriate to
compare models using likelihood ratios if they are not nested, even if
special exceptions exist to that rule \cite[see,
e.g.,][]{Vuong89,Fan+2001}.

What we have not done is incorporate the effect of downward
continuation in eq.~(\ref{gho2}) into the analysis. The `data' that we
will generate and analyze in our synthetic experiments will have been
`perfectly' downward continued to the single `appropriate' interface
at depth, from `noise-free' gravity observations, which remains a very
idealized situation. Some problems anticipated with numerical
stability might be remediated through dedicated robust deconvolution
methods, but more generally, giving up this level of idealization for
real-world data analysis will cause complications that require special
treatments. Absent these, our theoretical error estimates will be
minimum bounds. Keeping in mind that the complications of this kind
are shared by other gravity-based methods, we feel justified in not
exhaustively discussing all of our options here. Nevertheless, we can
look ahead at addressing the downward continuation of the gravity
field within the framework of our maximum-likelihood method by
considering what would happen if we took the surface topography and
the gravity anomaly as the primary observables, rather than the
surface and (deconvolved) subsurface topography as we now have, in
eq.~(\ref{vecdH}). We would, essentially, continue to carry the
factors~$\chik$ from eq.~(\ref{gho2}) throughout the development. In
the application of the blurred data analysis~(\ref{simublur}) those
factors would appear inside the convolutional integrals, to appear in
Appendix~\ref{numex}, of the kind~(\ref{detbarS0}), and their
appearance there would no doubt regularize the gravity deconvolution
by stabilizing the inverse~(\ref{invbarS0}) and its
derivatives~(\ref{dthinvbarS0}) as actually used by the optimization
algorithm. However, the variance expressions for the
maximum-likelihood estimates, which we derive based on the unblurred
likelihoods, would presumably be farther from their blurred 
equivalents once the deconvolution is also part of the estimation in
this way, and it would require much detailed work to arrive at a
complete understanding of such a procedure. At the end of the day, we
would still not have remediated the geophysical problems of
measurement and data-reduction noise in obtaining the Bouguer gravity
anomalies, nor handled possible departures from the two-layer model
that may exist in the form of internal density anomalies. The list of
caveats is long but again shared among other gravity-based methods,
over which the maximum-likelihood method has a clear advantage, as we
have seen, theoretically, above, and are about to show, via
simulation, in what follows.

\section{N~U~M~E~R~I~C~A~L{\hsps}E~X~P~E~R~I~M~E~N~T~S}
\label{numexper}

Numerical experiments are straightforward. We generate synthetic
data using the procedure established in
Sections~\ref{theory}--\ref{simulations}, and then employ an
iteration scheme along the lines of eqs~(\ref{MLE})--(\ref{inprob}):
starting from an initial guess we proceed through the 
iterations $k=0,\dots$~as
\be
\label{MLEalgorithm}
\hbt^{k+1}=\hbt^{k}-\Fb^{-1}\!(\hbt^{k})\,\bgamma(\hbt^{k})
,
\ee
until convergence. In practice any other numerical scheme, e.g. by
conjugate gradients, can be used, the only objective being to
maximize (or minimize the negative) log-likelihood~(\ref{firstlbar})
by whichever iteration path that is expedient, and for which canned
routines are readily available. 

The important points to note are, first, that we do need to implement
the convolutional blurring step~(\ref{simublur}) in the generation of
the data, so as to reference them to a particular generation grid
while keeping the flexibility to subsample, section, and taper them
for analysis as in the real-world case. Second, we do need to maximize
the blurred log-likelihood~(\ref{firstlbar}) and not its unblurred
relatives~(\ref{firstl}) or~(\ref{firstlw}). The data-generation grid
and the data-inversion grid may be different. If these two
stipulations are not met, an ``inverse crime''
\cite[]{Kaipio+2005,Kaipio+2007,Hansen2010} will be committed, leading
to either unwarranted optimism, or worse, spectacular failure --- both
cases unfortunately paramount in the literature and easily reproduced
experimentally.

From the luxury of being able to do synthetic experiments we can
verify, as we have, the important relations derived in this paper,
e.g., the expectation of the Hessian matrices of eq.~(\ref{fisher}),
the distribution of the scores in eq.~(\ref{kvar}), of the residuals
in eq.~(\ref{Xok}), of the likelihood ratios in eq.~(\ref{xsimchi}) of
the forthcoming Appendix~\ref{numex}, and of course virtually all of
the analytical expressions listed in the Appendices. We can
furthermore directly inspect the morphology of the likelihood 
surface~(\ref{firstlbar}) for individual experiments and witness the
scaled reduction of the confidence intervals with data size predicted
by eq.~(\ref{conflocation}). Via eq.~(\ref{totaldif}) we can compare
coherence (and admittance) curves with those derived from perfect
knowledge, and contrast them with what we might hope to recover from
the traditional estimates of the admittance and coherence. We do
stress again that even if we did have perfect estimates of admittance
and coherence, the problem of estimating the parameters of interest
from those would be fraught with all of the problems, encountered in
the literature, that led us to undertake our study in the first place.

Most importantly, we can check how well our theoretical distributions
match the outcome of our experiments. After all, in the real world we
will only have access to one data set per geographic area of interest,
and will need to decide on the basis of one maximum-likelihood
estimate which confidence intervals to place on the solution, and
which trade-offs and correlations between the estimated parameters to
expect. We were able to derive the theoretical distributions only by
neglecting the finite-sample size effects, basing our expressions on
the `unblurred' likelihood of eq.~(\ref{firstlt}) when using
eq.~(\ref{firstlbar}) would have been appropriate but analytically
intractable. In short, we can see how well we will do under realistic
scenarios, and check how much we are likely to gain by employing our
approach in future studies of terrestrial and planetary inversions for
the effective elastic thickness, initial-loading fraction and
load-correlation coefficient.  

\begin{figure}\centering
\rotatebox{-90}{\includegraphics[height=0.875\textwidth]{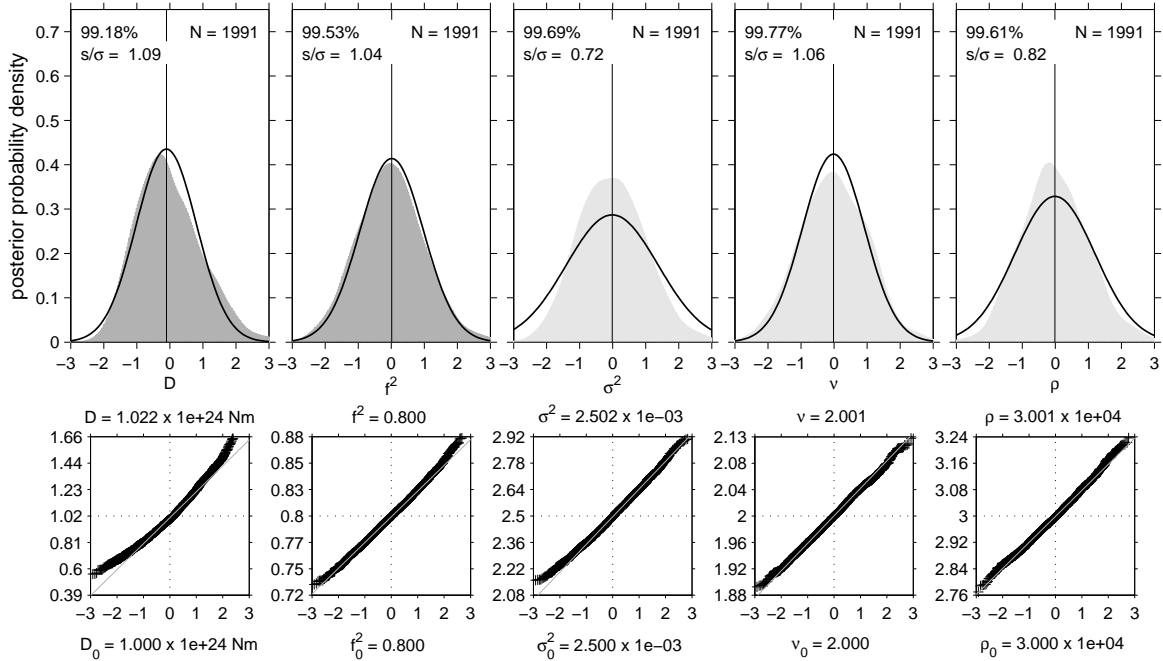}}
\caption{\label{syms1}Recovery statistics of simulations under
  uncorrelated loading on a 64$\times$64 grid with 20~km spacing in
  each direction. Density interfaces are at $z_1=0$~km, $z_2=35$~km,
  density contrasts $\Do=2670$~kg$\hsom$m$^{-3}$ and $\Dt=
  630$~kg$\hsom$m$^{-3}$. The top row shows the smoothly estimated
  standardized probability density function of the values recovered in
  this experiment of sample size~$N$, on which the theoretical distribution
  is superimposed (black line). The abscissas were truncated to
  within $\pm$3 of the empirical standard deviation; the percentage of
  the values captured by this truncation is listed in the top left of
  each graph. The ratio of the empirical to theoretical standard
  deviation is shown listed as $s/\sigma$. The bottom row shows the
  quantile-quantile plots of the empirical (ordinate) versus the
  theoretical (abscissa) distributions. The averages of the recovered
  values $D, \ft, \st, \nu$ and $\rho$ are listed at the top of the
  second row of graphs. The true parameter values $D_0, \fto, \sto,
  \nu_0$ and $\rho_0$ are listed at the bottom. Assuming Young's and
  Poisson moduli of $E=1.4\times 10^{11}$~Pa and $\nu=0.25$, the
  results imply a possible recovery of the parameters as
  $T_e=43.2\pm2.9$~km, $\ft=0.8\pm0.025$, $\st=(2.5\pm0.2)\times
  10^{-3}$, $\nu=2\pm0.039$, $\rho=(3\pm0.0967)\times 10^{4}$, quoting
  the true values plus or minus the theoretical standard deviation of
  their estimates, which are normally distributed and asymptotically
  unbiased.  }
\end{figure}

\begin{figure}\centering
\rotatebox{-90}{\includegraphics[height=0.885\textwidth]{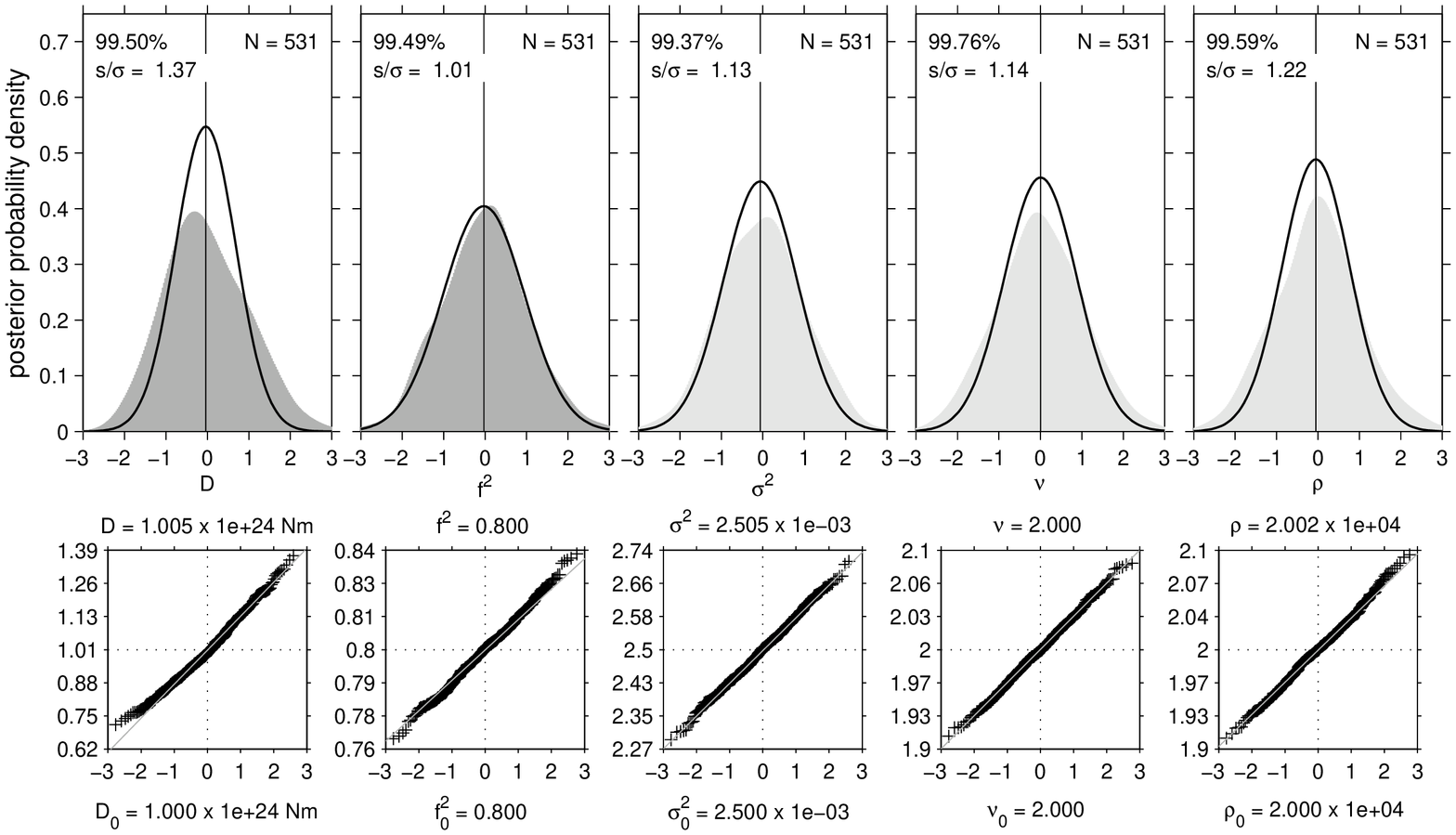}}
\caption{\label{syms2}Recovery statistics of simulations under
  uncorrelated loading with the same
  lithospheric
  parameters and shown in the same
  layout as in Fig.~\ref{syms1} but now carried out on a 128$\times$128
  grid. This roughly halves the standard deviation of the estimates, implying
  a theoretical recovery of $T_e=43.2\pm1.4$~km, $\ft=0.8\pm0.013$,
  $\st=(2.5\pm0.1)\times 10^{-3}$, $\nu=2\pm0.029$,
  $\rho=(2\pm0.0273)\times 10^{4}$. As in Fig.~\ref{syms1}, the
  experiments fit the theory encouragingly well.} 
\end{figure}

\begin{figure}\centering
\rotatebox{-90}{\includegraphics[height=0.885\textwidth]{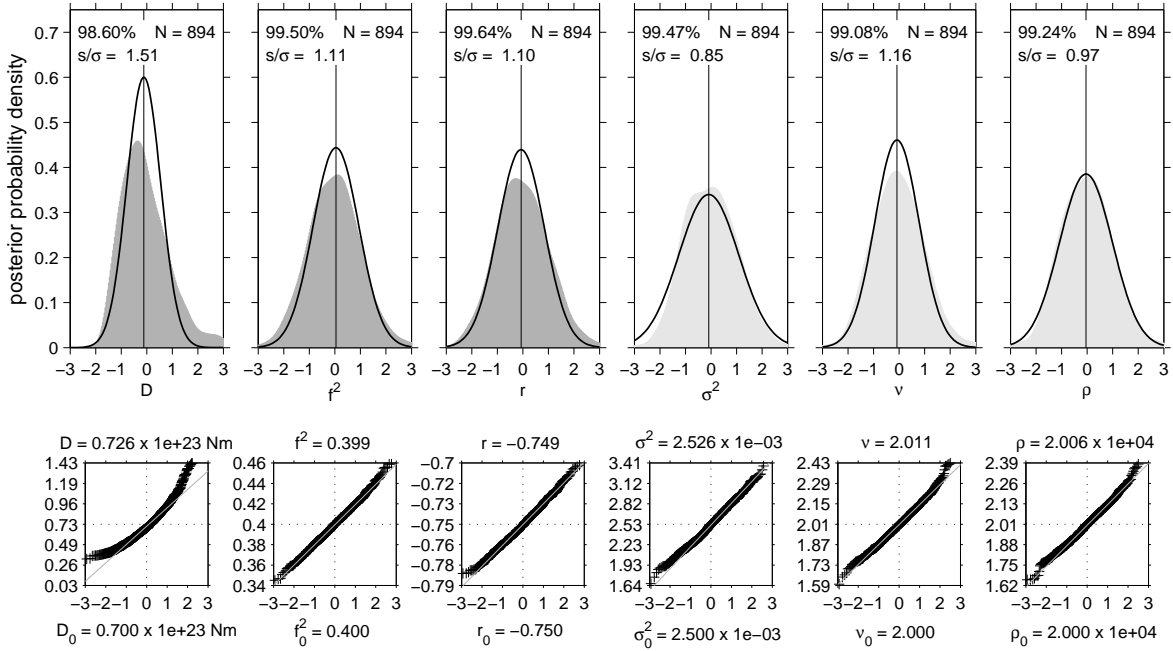}}
\caption{\label{syms3}Recovery statistics of simulations under
  correlated loading on a 32$\times$32 grid. Density interfaces are at
  $z_1=0$~km, $z_2=35$~km, density contrasts
  $\Do=2670$~kg$\hsom$m$^{-3}$ and $\Dt= 630$~kg$\hsom$m$^{-3}$. The
  top row shows the smoothly estimated standardized probability
  density function of the values recovered in this experiment of sample
  size~$N$, on which the theoretical distribution is superimposed
  (black line). The abscissas were truncated to within $\pm$3 of the
  empirical standard deviation; the percentage of the values captured
  by this truncation is listed in the top left of each graph. The
  ratio of the empirical to theoretical standard deviation is shown
   as~$s/\sigma$. The bottom row shows the quantile-quantile
  plots of the empirical (ordinate) versus the theoretical (abscissa)
  distributions. The averages of the recovered values~$D, \ft, r$, $\st,
  \nu$ and~$\rho$ are listed at the top of the second row of graphs.
  The true parameter values $D_0, \fto, r_0$, $\sto, \nu_0$
  and~$\rho_0$ are listed at the bottom. Assuming Young's and Poisson
  moduli of $E=1.4\times 10^{11}$~Pa and~$\nu=0.25$, the results imply
  a possible recovery of the parameters as $T_e=17.8\pm1.4$~km,
  $\ft=0.4\pm0.017$, $r=-0.75\pm0.014$, $\st=(2.5\pm0.3)\times
  10^{-3}$, $\nu=2\pm0.121$, $\rho=(2\pm0.1327)\times 10^{4}$, quoting
  the true values plus or minus the theoretical standard deviation of
  their estimates, which are very nearly normally distributed and
  asymptotically unbiased.}
\end{figure}

\begin{figure}\centering
\rotatebox{-90}{\includegraphics[height=0.875\textwidth]{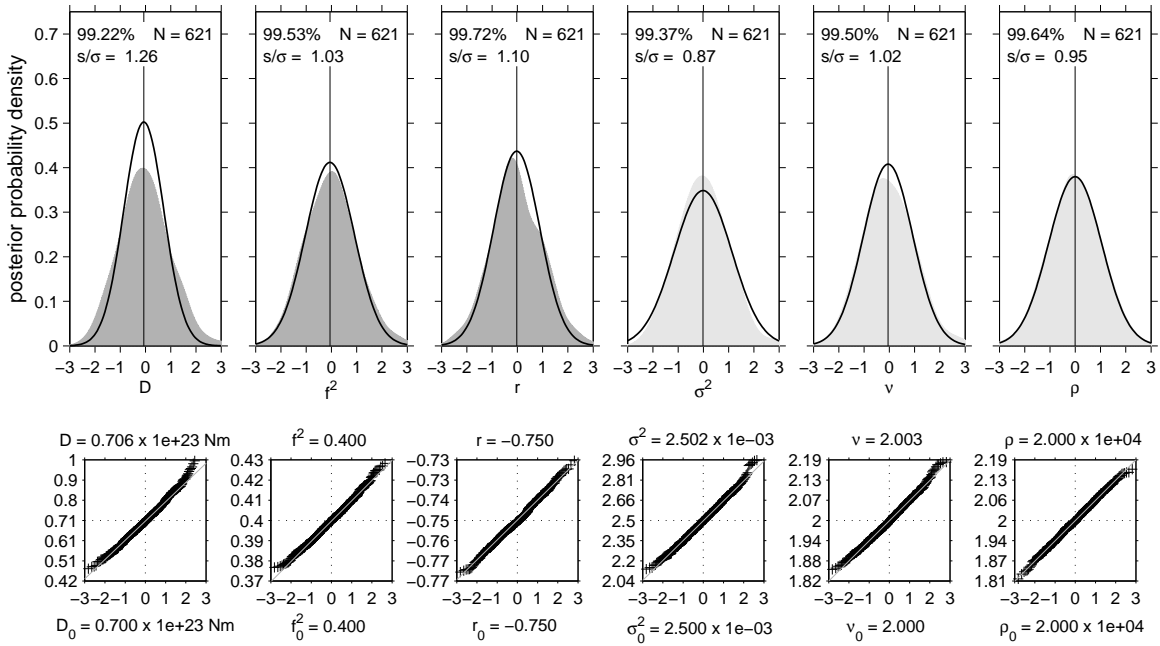}}
\caption{\label{syms4}Recovery statistics of simulations under
  correlated loading with the same parameters and shown in the same
  layout as in Fig.~\ref{syms3} but now carried out on a 64$\times$64
  grid. This roughly halves the standard deviation of the estimates,
  implying a theoretical recovery of $T_e=17.8\pm0.7$~km,
  $\ft=0.4\pm0.008$, $r=-0.75\pm0.007$, $\st=(2.5\pm0.2)\times
  10^{-3}$, $\nu=2\pm0.061$, $\rho=(2\pm0.0672)\times 10^{4}$. As in
  Fig.~\ref{syms3}, the experiments fit the theory very well.}
\end{figure}

Figs~\ref{oh0} and~\ref{oh2}--\ref{oh3} were themselves outputs of
genuine simulations to which the reader can refer again for visual
guidance.  Here we limit ourselves to studying the statistics of the
results on synthetic tests with simulated data. In
Figs~\ref{syms1}--\ref{syms4} we report on two suites of simulations:
one under the uncorrelated-loading scenario for two different data
sizes in Figs~\ref{syms1}--\ref{syms2}, and one under correlated
loading for two different data sizes in Figs~\ref{syms3}--\ref{syms4}.
Histograms of the outcomes of our experiments are presented in the
form of diffusion-based non-parametric `kernel-density estimates'
\cite[]{Botev+2010}, which explains their smooth appearance. The
distributions of the estimators are furthermore presented in the form
of the quantile-quantile plots as introduced in eq.~(\ref{bqqplot}),
which allows us to identify outlying regions of non-Gaussianity.
Figures of the type of Fig.~\ref{oh3} should help identify problems
with individual cases.

\begin{figure}\centering
\includegraphics[width=0.71\textwidth]{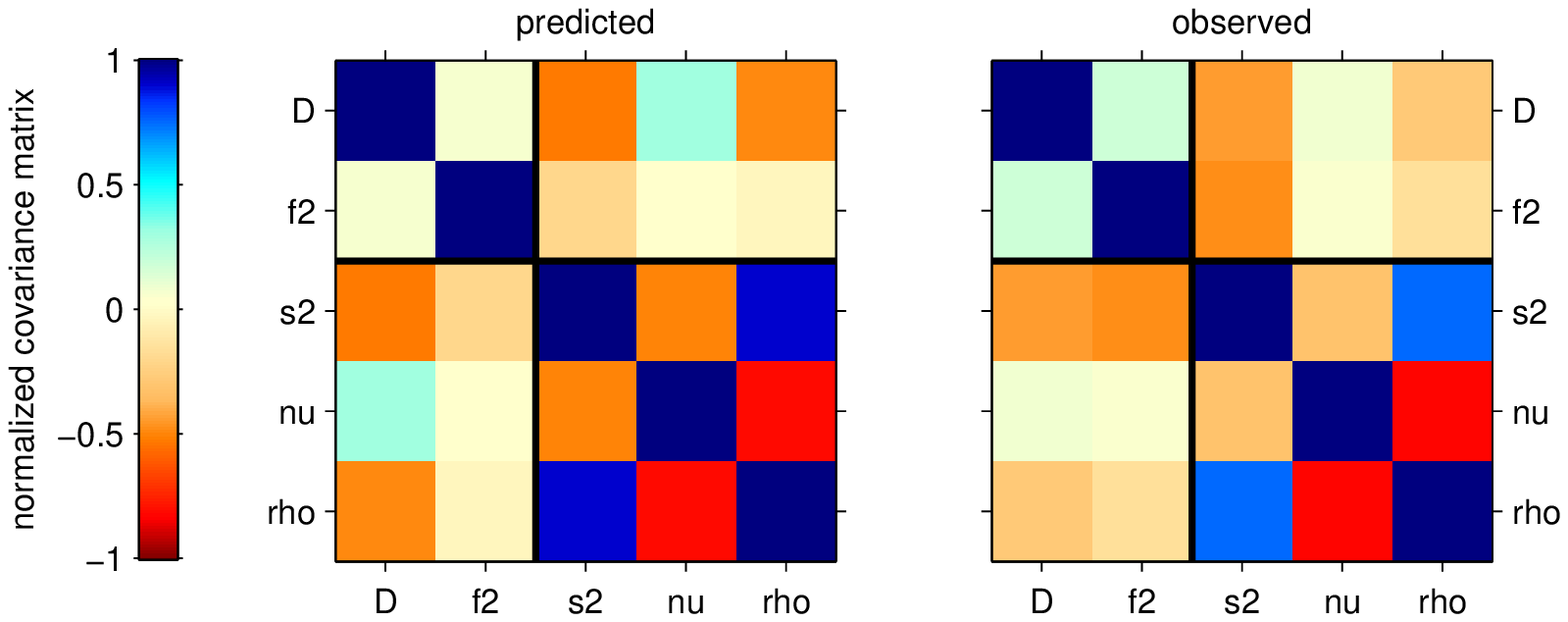}
\includegraphics[width=0.71\textwidth]{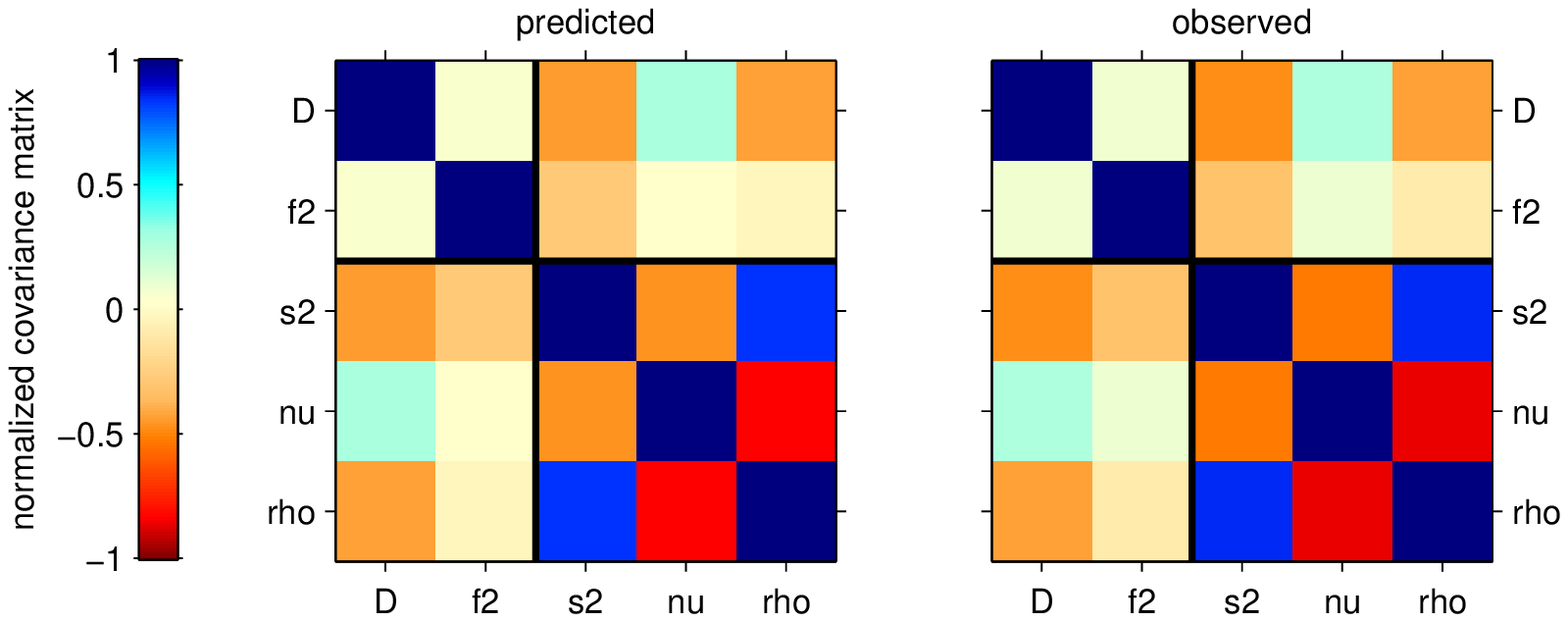}
\caption{\label{syms5}Correlation (normalized covariance) matrices for
  the uncorrelated-loading experiments previously reported in
  Figs~\ref{syms1} (\textit{top row}) and~\ref{syms2} (\textit{bottom
    row}). (\textit{Left column}) Theoretical correlation matrices.
  (\textit{Right column}) Empirical correlation matrices. There is
  some trade-off between the lithospheric ($D$, $\ft$) and the
  spectral parameters ($\st$, $\nu$, $\rho$), but virtually none among
  the lithospheric parameters, while the spectral parameters cannot be
  independently resolved.}
\end{figure}

For the uncorrelated-loading experiments shown in
Figs~\ref{syms1}--\ref{syms2} there are few meaningful departures
between theory and experiment. The predicted distributions match the
observed distributions very well, and the parameters of interest can
be recovered with great precision. Indeed, Fig.~\ref{syms1} shows us
that an elastic thickness $T_e=43.2$~km on a 1260$\times$1260~km$^2$
grid can be recovered with a standard deviation of $2.9$~km, with
similarly low relative standard deviations for the other parameters.
Fig.~\ref{syms2}, whose data grid is twice the size in each dimension,
yields standard deviations on the estimated parameters that are half
as big, in accordance with eq.~(\ref{conflocation}). What is
remarkable is that both theory and experiment, shown in
Fig.~\ref{syms5}, predict that the flexural rigidity~$D$ and the
initial-loading ratio~$\ft$ can be recovered without appreciable
correlation between them, and with little trade-off between them and
the spectral parameters $\st$, $\nu$ and $\rho$, even though the
trade-off between the spectral parameters themselves is significant.
This propitious ``separable'' behavior is not at all what the
entanglement of the parameters through the admittance and coherence
curves shown in Fig.~\ref{FF1} would have led us to believe, and it
runs indeed contrary to the experience with actual data as reported in
the literature. The likelihood contains enough information on each of
the parameters of interest to make this happen; the very act of
reducing this information to admittance and coherence curves virtually
erases this advantage by the collapse of their sensitivities.

\begin{figure}\centering
\includegraphics[width=0.71\textwidth]{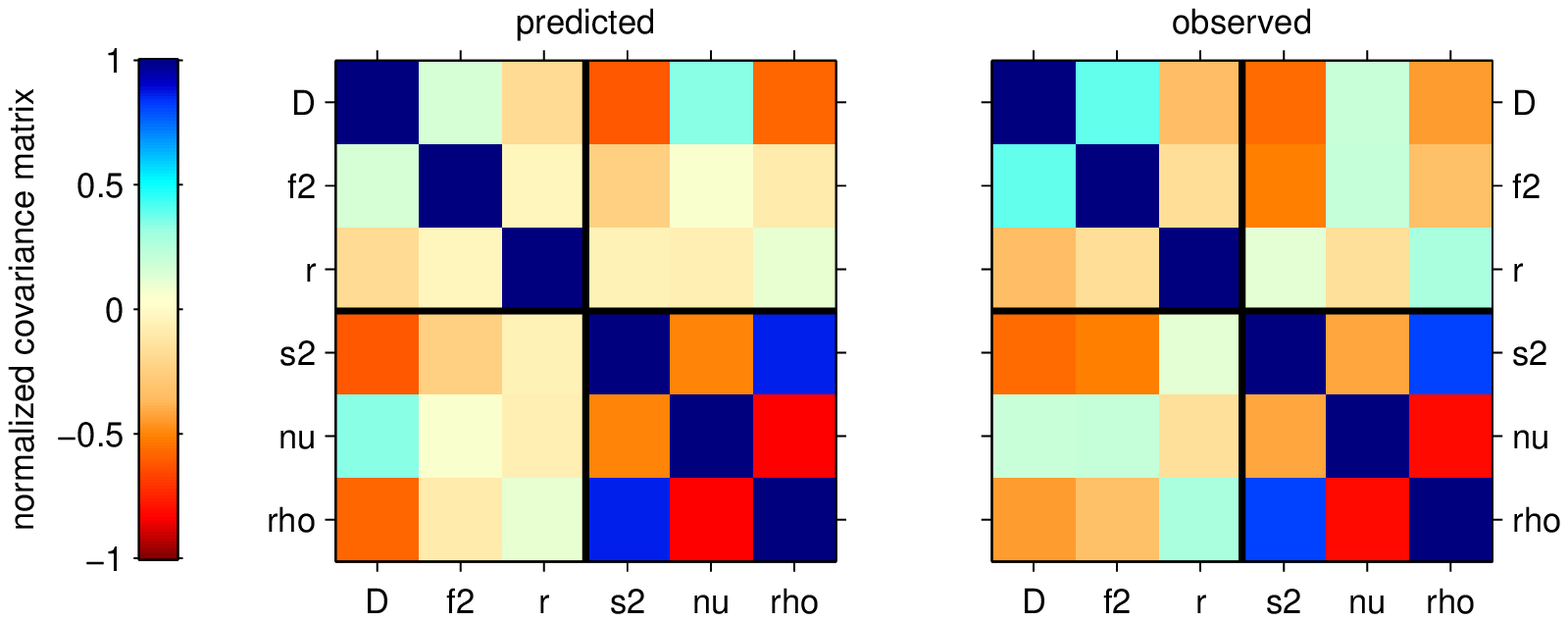}
\includegraphics[width=0.71\textwidth]{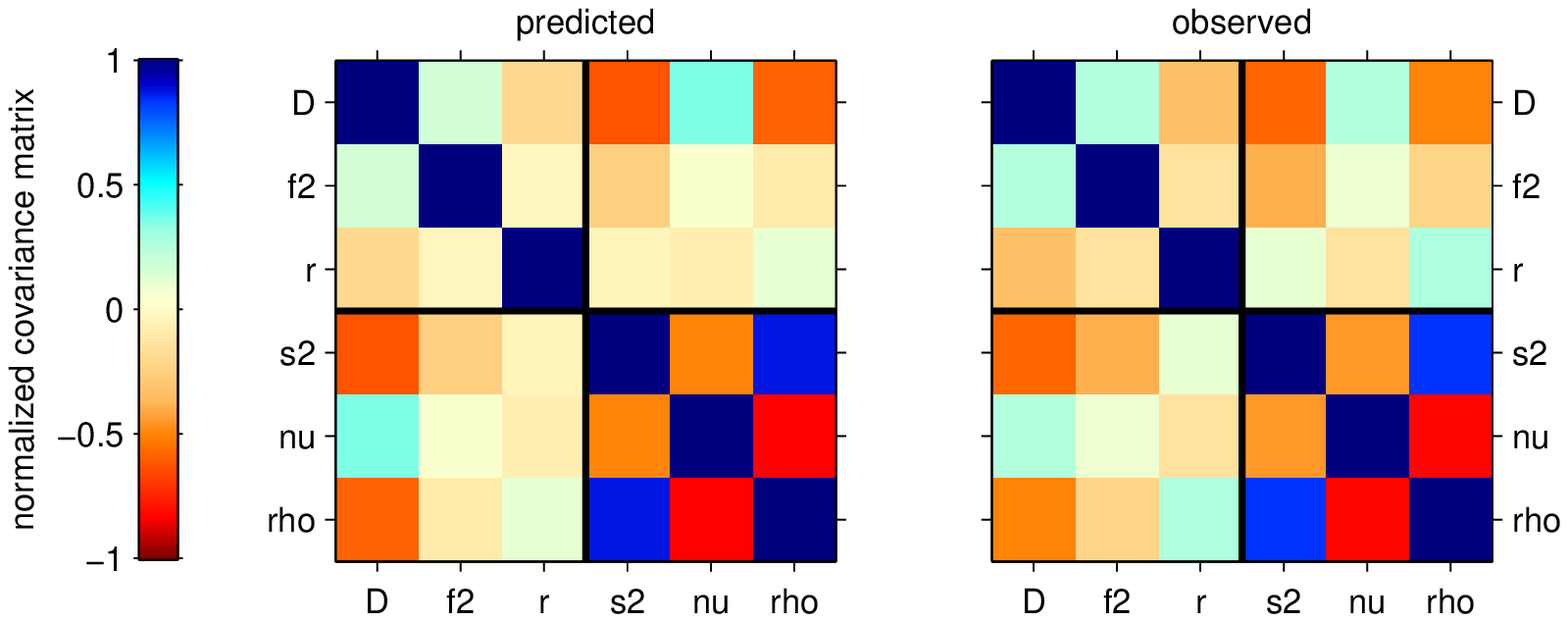}
\caption{\label{syms6}Correlation matrices for the correlated-loading
  experiments reported in Figs~\ref{syms3} (\textit{top row}) and \ref{syms4} 
  (\textit{bottom row}), with the layout as in Fig.~\ref{syms5}. The
  match between theory and experiment is on par with that found in the
  uncorrelated case. The covariances between parameters are similar in
  both cases, with a significant trade-off between the
  lithospheric parameter~$D$ and two of the spectral parameters, $\st$
  and $\rho$, which, themselves, cannot be independently resolved.}  
\end{figure}

For the correlated-loading experiments shown in
Figs~\ref{syms3}--\ref{syms4} the agreement between theory and
experiment is equally satisfactory. The introduction of the
load-correlation coefficient~$r$ contributes to making the
maximum-likelihood optimization `harder'. In our example we are
nevertheless able to estimate an elastic thickness $T_e=17.8$~km on a
1260$\times$1260~km$^2$ grid with a standard deviation of only
$0.7$~km, as shown in Fig.~\ref{syms4}. In contrast, Fig.~\ref{syms3},
whose data grid is half the size in each dimension, yields standard
deviations on the estimated parameters that are about twice as big, in
accordance with eq.~(\ref{conflocation}). Fig.~\ref{syms6} shows the
normalized covariance of the estimators. 

In all of our experiments as reported here we implemented the
finite-sample size blurring in the data analysis, but made predictions
based on the unblurred likelihoods, as discussed before. The figures
discussed in this section serve as the ultimate justification for the
validity of this approach, with further heuristic details deferred to
Appendix~\ref{numex}. When omitting the blurring altogether the
agreement between theory and practice becomes virtually perfect. As we
have argued, though, in those cases we commit the inverse crime of
analyzing the data on the same grid on which they have been generated,
which is unrealistic and needs to be avoided. We also note that in
designing practical inversion algorithms, care should be taken in
formulating an appropriate stopping criterion. The exactness of the
computations should match the scaling of the variances with the data
size, which we showed goes as $1/K$ in eq.~(\ref{kvar}). This is
difficult to tune, and some synthetic experiments might inadvertently
trim or `winsorize' the observed distributions by setting too
stringent a convergence criterion.

\begin{figure}\centering
\includegraphics[width=0.65\textwidth]{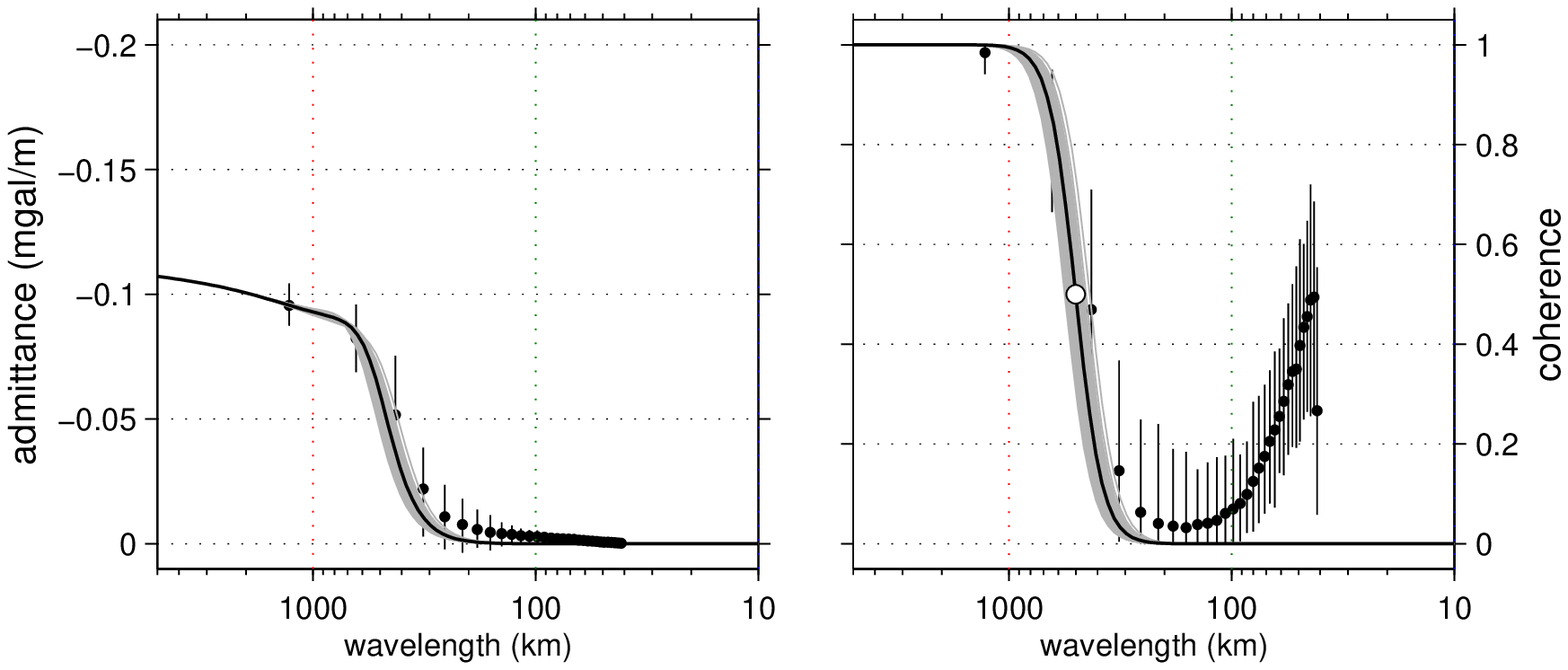}
\includegraphics[width=0.65\textwidth]{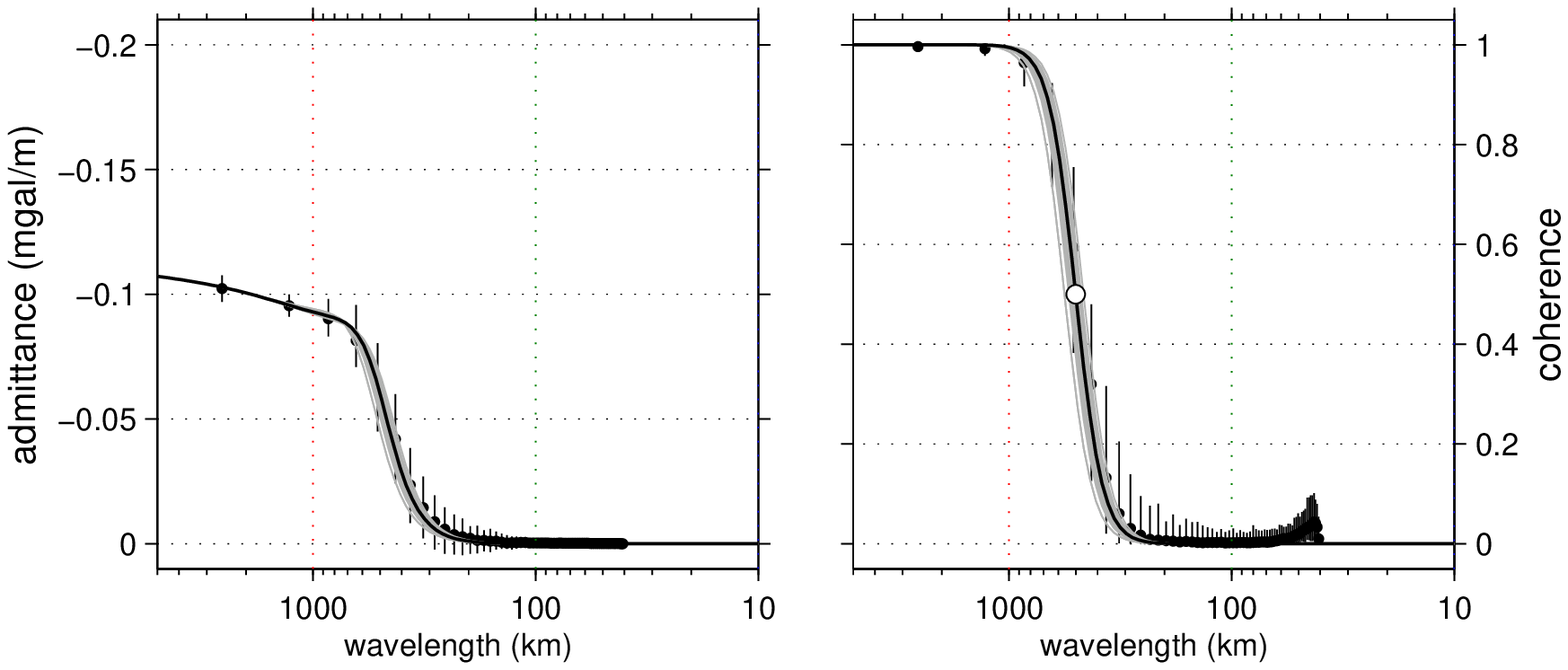}
\caption{\label{syms7}Admittance (\textit{left column}) and coherence
  curves (\textit{right column}) for the uncorrelated-loading
  experiments reported in Figs~\ref{syms1} (\textit{top row})
  and~\ref{syms2} (\textit{bottom row}). Black curves are the
  theoretical predictions. Superimposed grey curves, nearly perfectly
  matching the predictions, are one hundred examples of
  maximum-likelihood estimates selected at random from the
  experiments. Filled white circles show the ``half-coherence'' points
  calculated via eq.~(\ref{k4}). Underneath we show medians (black
  circles) and 2.5th and 97.5th percentile ranges (black bars) of two
  hundred ``traditional'' unwindowed Fourier-based estimates of
  admittance and coherence, highlighting the significant bias and/or
  high variance of such procedures.}
\end{figure}

\begin{figure}\centering
\includegraphics[width=0.65\textwidth]{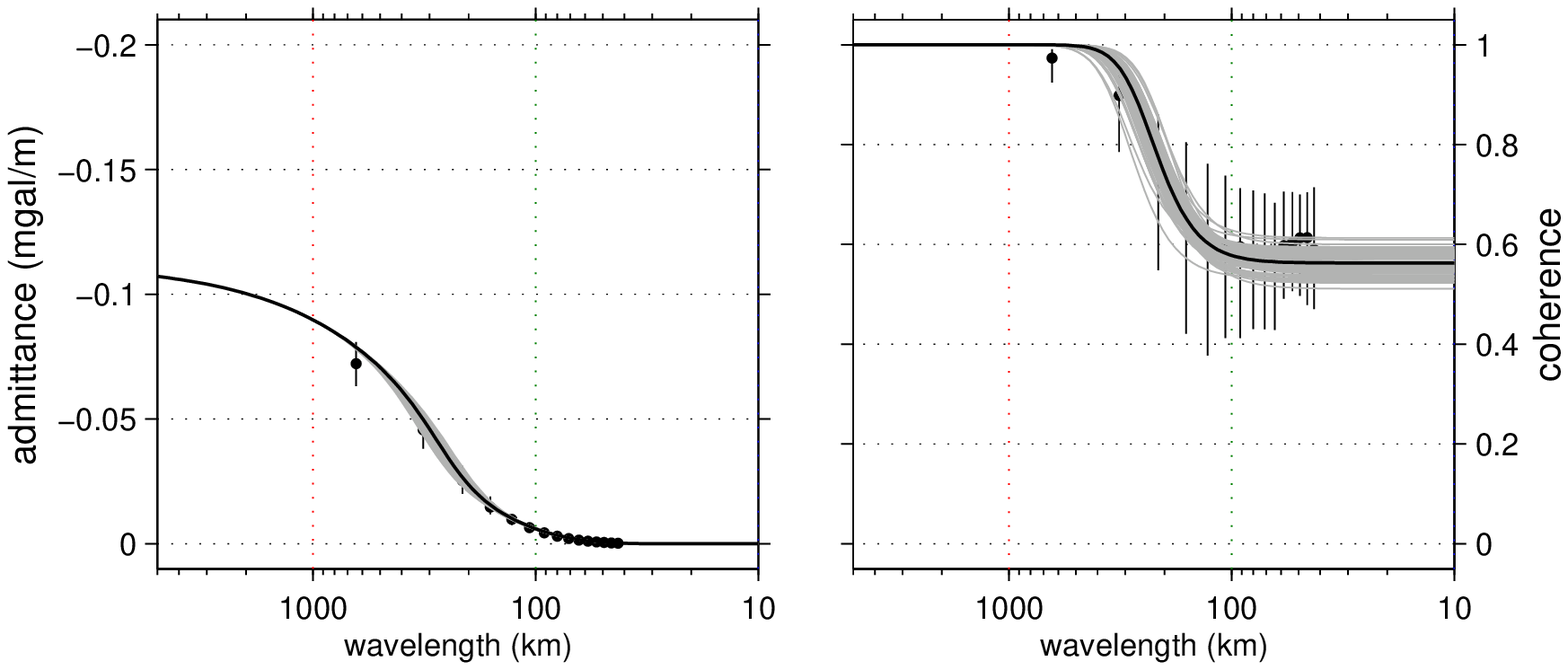}
\includegraphics[width=0.65\textwidth]{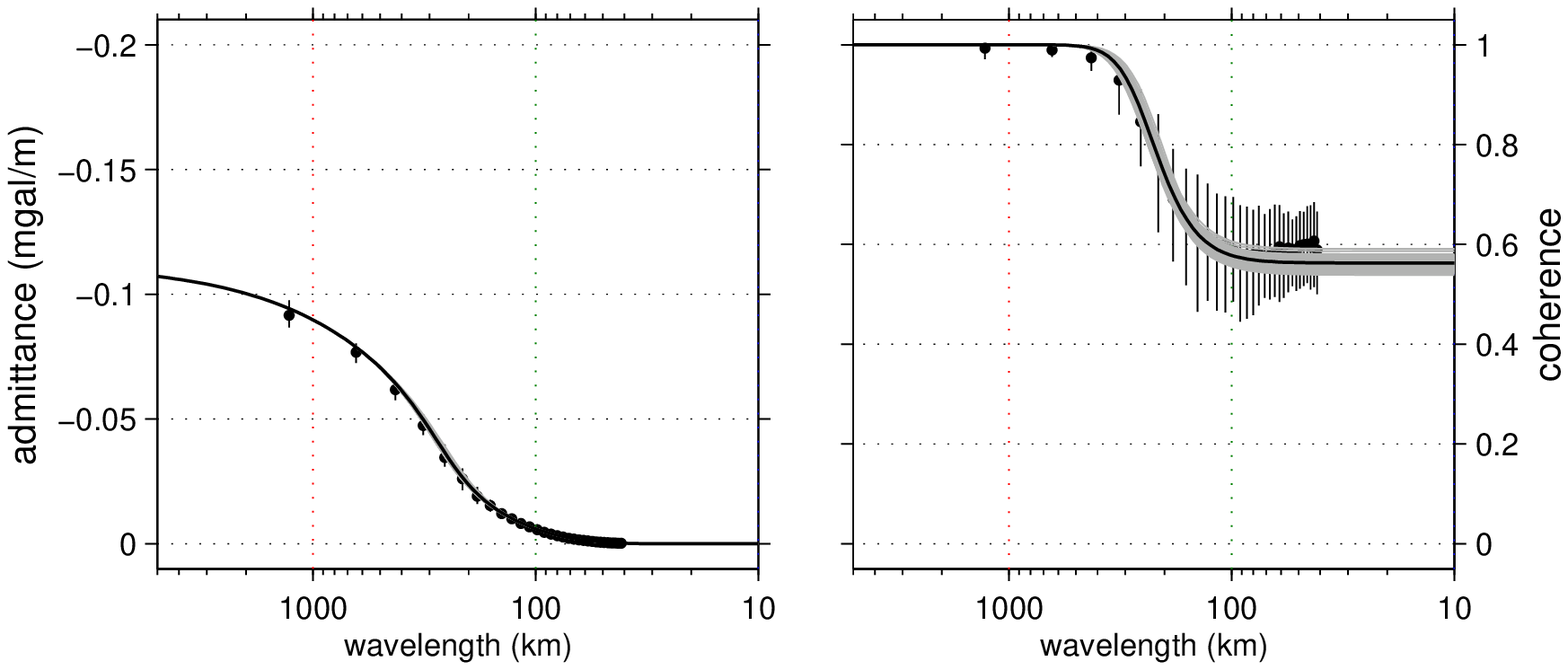}
\caption{\label{syms8}Admittance (\textit{left column}) and coherence
  curves (\textit{right column}) for the correlated-loading
  experiments reported in Figs~\ref{syms3} (\textit{top row})
  and~\ref{syms4} (\textit{bottom row}). The layout is as in
  Fig.~\ref{syms7}. The ``traditional'' admittance and coherence
  estimates (black circles, medians, and 2.5th to 97.5th percentiles)
  once again show significant bias and/or variance, although the
  admittance can be estimated much more accurately than the coherence
  using conventional Fourier methods.}
\end{figure}

Figs~\ref{syms7} and~\ref{syms8}, to conclude, show the distribution
of estimates of the admittance and coherence for the entire set of
experiments about which we have reported here. The maximum-likelihood
estimates agree very well with the theoretical curves, although the
effect of varying data size on the spread is understandably
noticeable. Our initial misgivings about the traditional admittance
and coherence estimates (obtained by Fourier transformation and
averaging over radial wavenumber annuli) are well summed up by their
behavior, which shows significant bias and large variance. While the
bias can be taken into account in comparing measurements with
theoretical curves, as it has been by various authors
\cite[]{Simons+2000,Perez+2004,Perez+2007,Perez+2009a,Kalnins+2009,Kirby+2011},
the high variance remains an issue.  Multitaper methods
\cite[]{Simons+2003a,Simons+2011a} reduce this variance but expand the
bias. The estimation of admittance and coherence is subservient to the
estimation of the lithospheric and spectral parameters that are of
geophysical value, and all methods that use admittance and coherence
estimates, no matter how good, as a point of departure for the
inversion for the geophysical parameters, will be deprived of the many
benefits that a direct maximum-likelihood inversion brings and that we
have attempted to illustrate in these pages.


\clearpage


\clearpage

%%%%%%%%%%% KEEP THE BELOW FIGURES TOGETHER ON ONE PAGE %%%%%%%%%%%%%%%%%%%%%

%%%%%%%%%%% KEEP THE ABOVE FIGURES TOGETHER ON ONE PAGE %%%%%%%%%%%%%%%%%%%%%

\clearpage

\clearpage

\section{C~O~N~C~L~U~S~I~O~N~S}

In this paper we have not answered the geophysical question ``What is
the flexural strength of the lithosphere?'' but rather the underlying
statistical question ``How can an efficient estimator for the flexural
strength of the lithosphere be constructed from geophysical
observations?''. Our answer was constructive: we derived the
properties of such an estimator and then showed how it can be found,
by a computational implementation of theoretical results that also
yielded analytical forms for the variance of such an estimate. We have
stayed as close as possible to the problem formulation as laid out in
the classical paper by~\cite{Forsyth85} but extended it by fully
considering correlated initial loads, as suggested
by~\cite{McKenzie2003}. The significant complexity of this problem,
even in a two-layer case, barred us from considering initial loads
with anisotropic power spectral densities, wave vector-dependent
initial-loading fractions and load-correlation coefficients,
anisotropic flexural rigidities, or any other elaborations on the
classical theory. However, we have suggested methods by which the
presence of such additional complexity can be tested through residual
inspection.

The principal steps in our algorithm are as follows. After collecting
the Fourier-transformed observations~(\ref{newobs}) into a
vector~$\Hbo\ofk$ we form the blurred Whittle likelihood of
eq.~(\ref{firstlbar}) as the average over the $K$~wavenumbers in the
half plane, the Gaussian quadratic form
\be\label{firstlbartwo}
\Lbar=
\frac{1}{K}\left[\ln\prod_\kb\fracd{
\exp(-\qform{\bbmcSoinv})
}
{\det\bbmcSo}\right]
,
\ee
whereby~$\bbmcSo$ is the blurred version, per eq.~(\ref{Sdefinebar}),
of the spectral matrix formulated in
eqs~(\ref{split})--(\ref{dTdefagain}). The likelihood depends on the
lithospheric parameters of interest, namely the flexural rigidity~$D$, the
initial-loading ratio~$\ft$, and the load-correlation coefficient~$r$,
and on the spectral parameters $\st$, $\nu$, $\rho$ of the Mat\'ern
form~(\ref{materndef}) that captures the isotropic shape of the power
spectral density of the initial loading. Maximization of
eq.~(\ref{firstlbartwo}) then yields estimates of these six
parameters. To appraise their covariance, we turn to the unblurred
Whittle likelihood of eq.~(\ref{firstl}),
\be\label{firstltwo}
\mcL=
\frac{1}{K}\left[\ln\prod_\kb\frac{\exp(-\qform{\bmcSoinv})
}
{\det\bmcSo}\right]
,
\ee
its first derivatives (the score),
\be
\label{scoretwo}
\frac{\pl\mcL}{\pl\theta}=
-\norml
\left[2\mth\ofk+\hform{\Abth}\right]
=\gth,
\ee
its second derivatives (the Hessian),
\be\label{fththptwo}
\frac{\pl^2\hsomm\hsomm\mcL}{\pl\theta\pl\theta'}
=
-\norml
\left[
2\frac{\pl m_{\theta'}\hsumm\ofk}{\pl\theta}
-\left(\cSoom\frac{\pl\cSoo}{\pl\theta}\right)
\hform{\Abthp}
+\hform{\left(\frac{\pl\Abthp}{\pl\theta}\right)}
\right]
=\Hththp
,
\ee
and their expectation (the Fisher matrix),
\be\label{mcfththptwo}
\left\langle\hsomm
\frac{\pl^2\hsomm\hsomm\mcL}{\pl\theta\pl\theta'}\hsomm
\right\rangle
=
-\norml
\left[
2\frac{\pl m_{\theta'}\hsumm\ofk}{\pl\theta}
+2\left(\cSoom\frac{\pl\cSoo}{\pl\theta}\right)
\mthp\ofk
+\tr\left\{\lform{\left(\frac{\pl\Abthp}{\pl\theta}\right)}\right\}
\right]=
\Fththp
,
\ee
whose inverse relates to the variance of the parameter estimates as
\be\label{summary2two}
\sqrt{K}(\hbt-\btruth) \sim \mcN(\bzero,\bmcF^{-1}\oftr)
=
\mcN(\bzero,\bmcJ\oftr)
.
\ee
With this knowledge we construct 100$\times\alpha$~\% confidence intervals
\be\label{conflocationtwo}
\hth-z_{\alpha/2}\frac{\Jthth\shalf\ofth}{\sqrt{K}}
\le
\truth
\le
\hth+z_{\alpha/2}\frac{\Jthth\shalf\ofth}{\sqrt{K}}
.
\ee

The problem of producing likely values of lithospheric strength,
initial-loading fraction and load correlation for a geographic region
of interest required positing an appropriate model for the
relationship between gravity and topography. The gravity field had to
be downward continued (to produce subsurface topography), and the
statistical nature of the parameter recovery problem had to be
acknowledged. There are many methods to produce estimators, and
depending on what can be reasonably assumed, different estimators will
result, all with different bias and variance characteristics. In
general one wishes to obtain unbiased and asymptotically efficient
estimators, i.e.~estimators whose variance is competitive with any
other method for increasing sample sizes. Our goal in this work has
been to whittle down the assumptions, while keeping the model both
simple and realistic.

If the parametric models that we have proposed are realistic then we
are assured of good estimation properties. Maximum-likelihood
estimators are both asymptotically unbiased and efficient \cite[often
with minimum variance, see, e.g.,][]{Portnoy77}. Should we use another
method, with more parameters, or even non-parametric nuisance terms,
unless those extra components in the model are necessary, we will
literally waste data points on estimating needless degrees of freedom,
and accrue an increased variance. Modeling the initial spectrum
non-parametrically is such an example, of wasting half of the data
points on the estimation. Producing the coherence or admittance
estimate as a starting point for a subsequent estimation of the
lithospheric parameters of interest is also highly suboptimal, and for
the same reason. If the parametric models that we have assumed are not
realistic then we will be able to diagnose this problem from the
residuals, and this will be a check on the methods we apply. Hence, if
the parametric models stand up to tests of this kind, then because of
the properties of maximum-likelihood estimators, asymptotically, no
other estimator will be able to compete in terms of variance. In that
case the confidence intervals that we have produced in this paper are
the best that could be produced.

\section{A~C~K~N~O~W~L~E~D~G~M~E~N~T~S}

This work was supported by the U.~S. National Science Foundation under
grants EAR-0710860, EAR-1014606 and EAR-1150145, and by the National
Aeronautics and Space Administration under grant NNX11AQ45G to
F.J.S., by U.~K. EPSRC Leadership Fellowship EP/I005250/1 to
S.C.O. She thanks Princeton University and he thanks
University College London for their hospitality over the course of
many mutual visits. In particular also, F.J.S. thanks Theresa Autino
and Debbie Fahey for facilitating his visit to London via Princeton
University account 195-2243 in 2011, and S.C.O. thanks the Imperial
College Trust for funding her sabbatical visit to Princeton in 2006,
where and when this work was commenced. We acknowledge useful
discussions with Don Forsyth, Lara Kalnins, Jon Kirby, Mark Wieczorek
and Tony Watts, but especially with Dan McKenzie. Two anonymous
reviewers and the Associate Editor, Saskia Goes, are thanked for their
helpful suggestions, which improved the paper. All computer code
needed to reproduce the results and the figures in this paper is made
freely available on \url{www.frederik.net}.

\begin{multicols}{2}
\bibliography{bib}
\bibliographystyle{gji}
\end{multicols}

\section{A~P~P~E~N~D~I~C~E~S}

\subsection{The spectral matrices~$\Tb$, $\dTb$ and~$\Tbo$}
\label{specapp}

We restate eqs~(\ref{S0def})--(\ref{dTdef}) or
eqs~(\ref{split})--(\ref{dTdefagain}), without any reference to
the dependence on wave vector or wavenumber, as   
\be\label{S0defapp}
\Tbo=\Tb+\dTb
,
\ee
\be\label{Tdefapp}
\Tb=
\omat{cc}
\xit+\ft\Do^2\Dt^{-2}
&
-\Dou\Dt^{-1}\xi-\ft\Do^3\Dt^{-3}\phi
\\
-\Dou\Dt^{-1}\xi-\ft\Do^3\Dt^{-3}\phi&
\Do^2\Dt^{-2}+\ft\Do^4\Dt^{-4}\phit
\cmat
\left(\frac{\Dt}{\Do+\Dt\hsom\xi}\right)^2
,
\ee
\be\label{dTdefapp}
\dTb=
r f
\omat{cc}
-2\Do\Dt^{-1}\xi & \Do^2\Dt^{-2}[\phixi+1]\\  
\Do^2 \Dt^{-2} [\phixi+1] & -2\Do^3 \Dt^{-3} \phi  
\cmat
\left(\frac{\Dt}{\Do+\Dt\hsom\xi}\right)^2
.
\ee
The Cholesky decomposition~(\ref{choldef}) of $\Tbo$ evaluates to 
\be\label{cholesky}
\Lbo=\frac
{(\Do+\Dt\hsom\xi)^{-1}}
{\sqrt{\Dt^2\hsom\xit+\ft\Do^2-2\hsom rf\Do\Dt\hsom\xi}}
\omat{cc}
\Dt^2\hsom\xit+\ft\Do^2 -2\hsom rf\Do\Dt\hsom\xi & 0 \\
-\Dou\Dt^{-1}[\Dt^2\hsom\xi+\ft\Do^2\phi]+rf\Do^2[\phixi+1] &
f\Do^2[\phixi-1]\hsom[1-r^2]^{1/2}
\cmat
.
\ee
For general reference we note the Cayley-Hamilton theorem
\cite[]{Dahlen+2002} for an invertible 2$\times$2
matrix~$\Ab$,  
\be\label{Cayley}
\Abinv=\frac{(\tr{\Ab})\Ib-\Ab}{\det\Ab}.
\ee
The determinants and inverses of $\Tbo$, $\Tb$ and $\dTb$ are given by
\ber\label{detTb} % Redone
\det\Tb&=&\ft\Do^{4}(\Do+\Dt\hsom\xi)^{-4}(\phixi-1)^2,\\
\label{invT}
\Tbinv&=&
\frac{\Do^{-2}(\Do+\Dt\hsom\xi)^2}{\ft(\phixi-1)^2}
\omat{cc}
1+\ft\Do^2\Dt^{-2}\phit &
\Do^{-1}\Dtu\hsom\xi+\ft\Dou\Dt^{-1}\phi\\{}
\Do^{-1}\Dtu\hsom\xi+\ft\Dou\Dt^{-1}\phi &
\Do^{-2}\Dt^2\hsom\xit+\ft
\cmat
.
\eer
\ber\label{detdTb}
\det\dTb&=&-r^2\det\Tb,\\
\dTbinv&=&
\frac{\Do^{-1}\Dt^{-1}(\Do+\Dt\hsom\xi)^2}{rf\hsom(\phixi-1)^2}
\omat{cc}
2\phi & \Do^{-1}\Dtu\hsom[\phixi+1]\\
\Do^{-1}\Dtu\hsom[\phixi+1] &2\Do^{-2}\Dt^2\hsom\xi\cmat
.
\eer
From these relationships we conclude that 
\ber\label{detTbo}
\det\Tbo&=& 
\ft\Do^4(\Do+\Dt\hsom\xi)^{-4}(\phixi-1)^2(1-r^2)
=(1-r^2)\det\Tb
=\det\Tb+\det\dTb %% THIS IS NOT A MATTER OF COURSE
,\\\label{invTbo}
\Tboinv&=&(1-r^2)^{-1}
\left(\Tbinv-r^2\dTbinv
\right).
\eer

\subsection{The score~$\bgamma$ in the lithospheric parameters~$D$, $\ft$ and $r$}
\label{scorelapp}

The first derivative of the log-likelihood function~(\ref{firstlt}) is
given by the expression~(\ref{score}). The elements of the score
function~$\gammaL$ for a generic ``lithospheric'' parameter
$\tl\in\bthetaL=\bthetaLfullr$ are 
\be\label{gammaL}
\gammaL=
-\norml\left[
\frac{\pl\ln(\det\Tbo)}{\pl\tl}
+\hform{\left(\frac{\pl\hsum\Tboinv}{\pl\tl}\right)}\right]
=
-\norml
\left[2\mthL\ofk+\hform{\AbthL}\right]
.
\ee
We obtain these via eq.~(\ref{mthAL}), 
seeing that we will need the
derivatives of the (logarithm of the) 
determinant and the inverse of $\Tbo$. We compute these 
from their defining expressions
or via the identities for symmetric invertible
matrices~\cite[]{Strang91,Tegmark+97}  
\be\label{strang}
\frac{\pl\ln(\det\Ab)}{\pl\theta}=
\tr\left(\Abinv\frac{\pl\hsum\Ab}{\pl\theta}
\right)
\also
\frac{\pl\hsum\Abinv}{\pl\theta}=
-\Abinv\frac{\pl\hsum\Ab}{\pl\theta}\Abinv
.
\ee
We will thus also write that 
\be\label{dinvTdD}
\fracd{\pl\hsum\Tbinv}{\pl D}=
-\frac{2(\xi-1)^{-2}}{\ft D}
\omat{cc}
1+\ft\Do^2\Dt^{-2}+\ft\Dou\Dt^{-1}[\xi-1] &
\Do^{-1}\Dtu+ \phi/2-\half+\ft\Dou\Dt^{-1}+\half \ft[\xi-1]\\
\Do^{-1}\Dtu+ \phi/2-\half+\ft\Dou\Dt^{-1}+\half \ft[\xi-1] &
\ft+\Do^{-2}\Dt^2+\Do^{-1}\Dtu\hsom[\phi-1],
\cmat
\ee
\be\label{dinvTdft}
\fracd{\pl\hsum\Tbinv}{\pl \ft}=
-\fracd{\Do^{-2}(\Do+\Dt\hsom\xi)^2}{f^4(\phixi-1)^2}
\omat{cc}
1 & 
\Do^{-1}\Dtu\hsom\xi\\
\Do^{-1}\Dtu\hsom\xi & 
\Do^{-2}\Dt^2\hsom\xit
\cmat
=
-\fracd{1}{f^4}\Vb
,
\ee
\be
\fracd{\pl\hsum\dTbinv}{\pl D}=
-\frac{2(\xi-1)^{-2}}{rfD}
\omat{cc}
2 \Dou\Dt^{-1} +\xi-1
&
1+\phi/2+\xi/2\\
1+\phi/2+\xi/2
& 
2\Do^{-1}\Dtu +\phi-1
\cmat
.
\ee
From the above we then find that the expressions required by eq.~(\ref{mthAL})
to calculate the score in the lithospheric parameters are\\[0.25em]
\begin{minipage}[]{0.4\textwidth}{
\ber
m_D&=&%\frac{\pl\ln(\det\Tbo)}{\pl D}=
\frac{k^4(\Do^{-1}+\Dt^{-1})}{g(\phixi-1)}\label{md}
,\\
m_\ft&=&%=\frac{\pl\ln(\det\Tbo)}{\pl \ft}
\frac{1}{2\ft}\label{mft}
,\\
m_r&=&
\frac{-r}{1-r^2}\label{mr}
.
\eer}\end{minipage}\hspace{\fill}
\begin{minipage}[]{0.5\textwidth}{
\ber
\AbD&=&
(1-r^2)^{-1}
\left(
\frac{\pl\Tbinv}{\pl D}-r^2\frac{\pl\dTbinv}{\pl D}
\right)\label{AbD}
,\\
\Abft&=&
(1-r^2)^{-1}
\left(
\frac{\pl\Tbinv}{\pl\ft}+\frac{r^2}{2\ft}\dTbinv
\right)\label{Abft}
,\\
\Abr&=&
\frac{2r}{(1-r^2)^2}
\left(\Tbinv-\frac{1+r^2}{2}\dTbinv
\right)\label{Abr}
.
\eer
}\end{minipage}\\[1em]
Since the score vanishes at the estimate, in the uncorrelated case we
can solve eq.~(\ref{gammaL}) for the estimate $\hft$ directly. Using
eqs~(\ref{mft}) and~(\ref{Abft}) for the case where $r=0$, we can thus
write, with the help of the matrix~$\Vb$ defined in
eq.~(\ref{dinvTdft}), an expression for the estimate 
\be\label{direct1}
\hft=\norml\hform{\hsom\Vb}
.
\ee
In principle this would allow us to define a profile
likelihood~\cite[]{Pawitan2001}, but such a procedure and its properties
remain  outside of the scope of this text. 

\subsection{The score~$\bgamma$ in the spectral parameters~$\st$, $\nu$ and~$\rho$}
\label{scoresapp}

The elements of the  score function~$\gammaS$ for a generic ``spectral''
parameter $\ts\in\bthetaS=\bthetaSfull$ are 
\be\label{gammaS}
\gammaS=
-\norml
\left(\Soom\frac{\pl\Soo}{\pl\ts}\right)
\left(2-\hform{\hsom\Tboinv\hsomm}\right)
=
-\norml
\left[2\mthS\ofk+\hform{\AbthS}\right]
.
\ee
To compute these via eq.~(\ref{mthAS})
we need the derivatives of the Mat\'ern
form. Thus, directly from eq.~(\ref{materndef}), we 
obtain in particular,\\[0.25em]
\begin{minipage}[]{0.725\textwidth}{
\ber
m_\st&=&
\frac{1}{\st}\label{mst}
,\\
m_\nu&=&
\frac{\nu+1}{\nu}
+\ln\left(\frac{4\nu}{\pi^2\rho^2}\right)
-4\left(\frac{\nu+1}{\pi^2\rho^2}\right)
\left(\frac{4\nu}{\pi^2\rho^2}+k^2\right)^{-1}
\!\!-\ln
\left(\frac{4\nu}{\pi^2\rho^2}+k^2\right)
\label{mnu}
,\\
m_\rho&=&
-2\frac{\nu}{\rho}
+8\frac{\nu}{\rho}\left(\frac{\nu+1}{\pi^2\rho^2}\right)
\left({\frac{4\nu}{\pi^2\rho^2}+k^2}\right)^{-1}\label{mrho}
.
\eer}
\end{minipage}\hspace{\fill}
\begin{minipage}[]{0.25\textwidth}{
\ber
\label{Abst}\Ab_\st&=&-m_\st\Tboinv,\rule[-1.5em]{0em}{2em}\\
\label{Abnu}\Ab_\nu&=&-m_\nu\Tboinv,\rule[-0.5em]{0em}{1em}\\
\label{Abrho}\Ab_\rho&=&-m_\rho\Tboinv.\rule[-0.5em]{0em}{3em}
\eer}
\end{minipage}
As above in eq.~(\ref{direct1}), we pick up one direct solution,
namely 
\be\label{direct2}
\hst=\tnorml
\left(\frac{\Soo}{\st}\right)^{-1}\qform{\hsom\Tboinv}
,
\ee
where it is to be noted from eq.~(\ref{materndef}) that $(\Soo/\st)$
is indeed no longer dependent on~$\st$. With eq.~(\ref{direct1})
this would enable us to conduct a profile-likelihood estimation in a
reduced parameter space \cite[]{Pawitan2001}, but once again the
details are omitted here. 

\subsection{The Hessian~$\Fb$ and the Fisher matrix~$\mcF$}
\label{hessapp}

The Hessian or second derivative of the log-likelihood
function~(\ref{firstlt}), and its negative expectation or the Fisher
information matrix, are given by the expressions~(\ref{fththp})
and~(\ref{mcfththp}), respectively. Both of these contain the
terms~(\ref{md})--(\ref{Abr}) and~(\ref{mst})--(\ref{Abrho}) that we
have just derived, which renders them eminently calculable
analytically. In its raw form eq.~(\ref{mcfththp}) does not provide
much insight, but in Section~\ref{Fisher} we also introduced special
formulations for elements of the Fisher matrix that involve at least
one spectral variable, in which case the expressions~(\ref{diago2}),
(\ref{FLS}) and~(\ref{FSS}) for $\Fthsths$, $F_{\thetaL\hsumm\thetaS}$ and
$F_{\thetaSu\hsumm\thetapS}$, respectively, are of a common form. We do
not foresee needing the expressions for the Hessian: while 
optimization procedures  might benefit
from those, even in  eq.~(\ref{MLEalgorithm}) the Fisher matrix could
be substituted \cite[]{Cox+74}.  

We are thus left with determining the entries of the Fisher matrix
$F_{\thetaLu\!\thetapL}$  when only lithospheric variables are
present. The diagonal terms~$\Fthlthl$ are obtained via
eq.~(\ref{diago1}), which we repeat here specifically for this case as 
\be\label{diago1again}
\mcF_{\thetaL\hsumm\thetaL}=
\norml\big\{\left[\lambda\unthL\rmo\ofk\right]^2
+\left[\lambda\unthL\rmt\ofk\right]^2\big\},
\where
\lambda\unthL^{\pm}=
\mathrm{eig}\left(
\lform{\AbthL}
\right)
.
\ee
Only to obtain the cross terms involving different lithospheric
parameters do we need the full expression~(\ref{mcfththp}). Even this
case simplifies since, owing to eq.~(\ref{materndef}),
$\pl\unthL\Soo=0$, thereby yielding the expression
\be\label{Fthlthpl}\label{mcFthlthpl}
\mcF_{\thetaLu\hsumm\thetapL}=
\norml\left\{
2\frac{\pl m_{\thetaL'}\hsumm\ofk}{\pl\thetaL}+
\tr\left[\lform{\left(\frac{\pl\Ab_{\thetapL}}{\pl\thetaL}\right)}\right]
\right\}
, \ee 
where we recall from eq.~(\ref{mthAL}) that
$\pl_{\thetaLu}\Ab_{\thetapL}= \pl_{\thetaL}\pl_{\thetapL}\Tboinv$.
When $\thetaL\ne\thetaL'$, as is seen from eqs~(\ref{md})--(\ref{mr}),
the first term $\pl\unthL m_{\thetapL}=0$. When $\thetaL=\thetaL'$,
eqs~(\ref{diago1again})--(\ref{mcFthlthpl}) are exactly each others'
equivalent, and either expression can be used. We will not really need
the eigenvalues of the quadratic forms: their sums of squares (in
eq.~\ref{diago1again}) or sums (in eq.~\ref{mcFthlthpl}) suffice to
calculate the elements of the Fisher matrix. The specific eigenvalues
are only required if we should abandon the normal approximations and
develop an interest in calculating the distributions of
eq.~(\ref{SHAH}) exactly. 

Beginning with the flexural rigidity, we obtain
\ber\label{Fdd}
\lefteqn{
\Fdd=
\twnorml
\frac{k^8 \Do^{-2}\Dt^{-2}g^{-2}f^{-2}}{(1-r^2)(\phixi-1)^2}
\left(2f\Do\Dt[f-3r^2f-rf^2-r]\right.}\nnr\\
&&{}
\hspace{14em}+\left.\ft\Do^2[2+\ft-r^2+2rf]
+\Dt^2[1+2\ft- r^2\ft+2rf]
\right)
.
\eer
For the loading ratio, we obtain for the sum of squares
of the eigenvalues
\be\label{Fff}
\Fff=\frac{2-r^2}{2f^4(1-r^2)}
.
\ee
Finally, for the load-correlation coefficient we conclude that
\be\label{F}
\Frr=\frac{2(1+r^2)}{(1-r^2)^2}
.
\ee

For the cross terms that remain, we find, at last,
\be
\Fdf
=\norml
\frac{k^4\Do^{-1}\Dt^{-1}}{gf^3\hsom(1-r^2)(\phixi-1)}
\left(2f\Dt-r^2f[\Do+\Dt]-r\ft\Do+r\Dt\right)
,
\ee
\be
\Fdr
=\norml
\frac{2k^4\Do^{-1}\Dt^{-1}}{gf(1-r^2)(\phixi-1)}
\left(\ft\Do+\Dt-rf[\Do+\Dt]\right)
,
\ee
\be
\Ffr
=\frac{-r}{\ft(1-r^2)}
.
\ee

\subsection{Properties of admittance and coherence estimates --- and
  ``Cram\'er-Rao lite'' for the maximum-likelihood estimate}
\label{cramerrao}

Let us consider how the uncertainty on the parameters~$\hbt$
estimated via the maximum-likelihood method propagates to estimates of
the coherence and the admittance, $\widehat{\gto}$ and
$\widehat{\Qo}$, should we desire to construct those. Since
Section~\ref{properties} we have known that our 
estimate~$\hbt$, which is based on the likelihood~(\ref{firstlbar}) and thus
ultimately on the data $\Hbo\ofk$,  is centered on the truth~$\btruth$ as per
\be\label{allknown}
\hbt=\btruth+\mathbf{Y},
\also
\langle\hat{\theta}\rangle=\theta_0.
\ee
We know the distributional properties of~$\mathbf{Y}$ as having
a mean of zero and a variance that is proportional to the inverse of
the Fourier-domain sample size~$K$. Taking the Bouguer-topography
coherence as an example, we can again use the delta method to write
for its estimate
\be
\gto(\hbt)=\gto(\btruth)+\big[\bnabla\gto(\btruth)\big]\Trm\mathbf{Y},
\ee
from which easily follows that
\ber
\langle\gto(\hbt)\rangle&=&\gto(\btruth),\\
\var\{\gto(\hbt)\} &=& \big[\bnabla\gto(\btruth)\big]\Trm\var\{\mathbf{Y}\}\big[\bnabla\gto(\btruth)\big],\label{inform}
\eer
at identical wavenumbers~$k$, and a statement similar in form to
eq.~(\ref{inform}) for the covariance of the coherence estimate
between different wavenumbers~$k$ and $k'$.  With these we know the
relevant statistics of maximum-likelihood-based admittance and
coherence estimates.  

The ``traditional'' methods use estimates of coherence and admittance
to derive estimates of the parameters~$\btheta$. Regardless of how the
former are computed (via parameterized maximum-likelihood techniques
as in this paper, or non-parametrically using multitaper or other
spectral techniques), we know one important thing about their
statistics. No alternative estimate for the parameters that is
unbiased will beat the variance of our maximum-likelihood estimate. 

Let us imagine defining another unbiased estimator which would be
given by another function of the data, generically written
\be
\hat{\mathbf{t}},\where
\langle\hat{t}\rangle=\theta_0
,
\ee
and let us study the covariance of this hypothetical estimate with the
zero-mean score of the maximum-likelihood~(\ref{firstlbar}), 
defined in eq~(\ref{score}):
\ber\label{isone}
\cov\{\hat{t},\gamma_\theta\}=
\langle\hat{t}\,\gamma_\theta\rangle=
\frac{1}{K}\Big\langle\hat{t}\normu\gtk\Big\rangle
&=&\frac{1}{K}\underbrace{\int\!\!...\!\!\int}_{K}
\hat{t}
\left(
\normu\frac{1}{\pHbok}\frac{\pl\pHbok}{\pl\theta}
\right)
\left(
\prod_\kbp\pHbokp\,d\Hbokp
\right)\\
%\left(\int\!...\!\int\pHbok\hat{t}\,d\Hbok\right)=
%\norml\frac{\pl}{\pl \theta}\big(\langle\hat{t}\rangle\big)=
&=&\frac{1}{K}\underbrace{\int\!\!...\!\!\int}_{K}
\hat{t}
\frac{\pl}{\pl\theta}
\left(
\prod_\kb\pHbok\,d\Hbok
\right)
=
\frac{1}{K}
\frac{\pl}{\pl\theta}
\underbrace{\int\!\!...\!\!\int}_{K}
\hat{t}
\prod_\kb\pHbok\,d\Hbok
\\
&=&
\frac{1}{K}
\frac{\pl}{\pl\theta}
\langle
\hat{t}
\rangle
=\frac{1}{K}
\frac{\pl}{\pl\theta}
\theta
=\frac{1}{K}.
\eer
To obtain eq.~(\ref{isone}) we followed an argument as in
eqs~(\ref{logp})--(\ref{blabla}) while continuing to assume the
independence of the Fourier coefficients and using Leibniz' product rule of
differentiation. We now know from Cauchy-Schwartz that  
\be\label{cs}
\var\{\gamma_\theta\}\var\{\hat{t}\}\ge
\left(\cov\{\gamma_\theta,\hat{t}\}\right)^2=\frac{1}{K^2},
\ee
and thus, combining eq.~(\ref{cs}) with eqs~(\ref{kvar})
and~(\ref{Jdef}), we find that
\be
\var\{\hat{t}\}\ge\frac{1}{K^2}\frac{1}{\var\{\gamma_\theta\}}=
\frac{\Fthth^{-1}}{K}=\var\{\hat{\theta}\}
.
\ee
The maximum-likelihood estimate is asymptotically efficient: no other unbiased
estimate has a lower variance.

\subsection{Retrieval of spectral parameters}
\label{appretrieval}

Were we to observe a single random field $\mcH\ofx$, 
distributed as an isotropic Mat\'ern random field with the parameters
$\btheta=\bthetaS$, we would have 
\be\label{maternredux}
\langle d\mcH^{}\ofk \hsom d\mcH^*\!\ofkp \rangle=
\mcS\ofk\dtbk\,\dkkp=\mcS\ofsk \dbk=\frac{\st\nu^{\nu+1} 4^\nu}{\pi
  (\pi\rho)^{2\nu}} 
\left(\frac{4\nu}{\pi^2\rho^2}+k^2\right)^{-\nu-1}\!\!\dbk.
\ee
Its parameters could also be estimated using maximum-likelihood
estimation. Following the developments in Section~\ref{loglisec}
the blurred log-likelihood of observing the data under the
model~(\ref{maternredux}) would be written under the assumption of
independence as 
\be\label{secondlblurred}
\bar{\mcLS}\ofts=
\frac{1}{K}\left[\ln\prod_\kb\frac{\exp(-\Sbar^{-1}\hsomm\ofsk\hsom|H\ofk|^2)}
{\Sbar\ofsk}\right]
=-\norml
\left[
\ln \Sbar\ofsk
+\Sbar^{-1}\hsomm\ofsk\hsom|H\ofk|^2\right]
.
\ee
When the spectral blurring is being neglected, the likelihood becomes,
more simply,
\be\label{secondl}
\mcLS\ofts=
\frac{1}{K}\left[\ln\prod_\kb\frac{\exp(-\mcS^{-1}\hsomm\ofsk\hsom|H\ofk|^2)}
{\mcS\ofsk}\right]
=-\norml
\left[
\ln \mcS\ofsk
+\mcS^{-1}\hsomm\ofsk\hsom|H\ofk|^2\right]
.
\ee
The scores in this likelihood are then 
\be\label{scoreS}
\gSthS
=
-\norml
\mthS\ofsk\left[1- \mcS^{-1}\hsomm\ofsk\hsom|H\ofk|^2\right],\where
\mthS\ofsk=\mcS^{-1}\hsomm\ofsk\frac{\pl\mcS\ofsk}{\pl{\thetaS}}
,
\ee
which is only slightly different from the forms that they took in the
multivariable case, eqs~(\ref{score}) and~(\ref{mthAS}). In deriving
the variance of the score in the multivariate flexural case,
eq.~(\ref{sumint}), we neglected the complications of spectral
blurring, as we do here, and we also neglected the slight correlation
between wavenumbers, as we have here also. The simple form of
eq.~(\ref{scoreS}) allows us to re-examine the effect that wavenumber
correlations will have on the score by bypassing the development
outlined in eqs~(\ref{SHOH})--(\ref{SHAH}) and writing instead that
\be\label{covgammaS}
\cov\big\{\gSthSu,\gSthpS\big\}=\normltu\sumkp
\mthSu\hsomm\ofsk\hsom\mthSp\hsomm\ofsk 
\,\frac{\cov\{|H\ofk|^2,|H\ofkp|^2\}}
{\mcS(\kb)\hsom\mcS(\kb')}
.
\ee

Previously we wrote expressions for the covariance of the finite-length spectral
observation vector that took into account the blurring but not the
correlation, e.g. in approximating eq.~(\ref{specblur}) by
eq.~(\ref{covHH}), which we restate here for the univariate case as
\be
\cov\{H\ofk,H\ofkp\}=
\intnyq
W^{}_K(\kb-\kb'')W_K^*(\kb'-\kb'')\hsom\mcS(\kb'')\dbk''
\approx 
\bar\mcS\ofk\,\dkkp
.
\ee
We shall now  approximate this under slow variation of the spectrum,
relative to the decay of the window functions~$W_K$,  as
\be\label{covHHagain}
\cov\{H\ofk,H\ofkp\}
\approx \mcS(\kb)\intnyq W_K(\kb-\kb'')W_K^*\hsomm(\kb'-\kb'')\dbk''
=\mcS(\kb) \,c\hsom(\kb,\kb').
\ee
Using Isserlis' theorem \cite[]{Isserlis16,Percival+93,Walden+94}, we
then have for the covariance of the periodograms
\be
\cov\big\{|H\ofk|^2,|H\ofkp|^2\big\}=
\left|\cov\{H\ofk,H^*\!\ofkp\}\right|^2+
\left|\cov\{H\ofk,H\ofkp\}\right|^2
=\mcS^2(\kb) \hsom c^2\hsomm(\kb,\kb'),
\ee
since the first term, the pseudocovariance or relation
matrix vanishes in the half-plane for the complex-proper Gaussian
Fourier coefficients~\cite[]{Miller69,Thomson77a,Neeser+93} of real-valued
stationary variables. % Brillinger
We may thus conclude that
the covariance of the scores suffers mildly from wavenumber
correlation, 
\be
\cov\big\{\gSthSu,\gSthpS\big\}=\normltu\sumkp
\mthSu\hsomm\ofsk\hsom\mthSp\hsomm\ofsk 
\frac{\mcS^2(\kb) \hsom c^2(\kb,\kb')}{\mcS(\kb)\hsom\mcS(\kb')}
.
\ee
However, for very large observation windows or custom-designed tapering
procedures, we may write
\be
\cov\big\{\gSthSu,\gSthpS\big\}=\normlt \mthSu\hsomm\ofsk\hsom\mthSp\hsomm\ofsk
.
\ee
From eq.~(\ref{kvar}) we then also recover the entries of the Fisher
matrix for this problem as  exactly half the size of the multivariate
equivalent that we obtained in eq.~(\ref{FSS}), as expected,
\be\label{fisherS}
(\mcF\unS)_{\thetaSu\!\thetapS}=\norml  \mthSu\hsomm\ofsk\hsom\mthSp\hsomm\ofsk
,
\ee
which are to be used in the construction of confidence intervals for
the parameters $\st$, $\rho$ and $\nu$ of the isotropic Mat\'ern
distribution as determined by this procedure. The expressions
for~$\mthS$ were listed in Appendix~\ref{scoresapp}. Refer again also to
Table~\ref{tablespectral}, which we have only now completed filling. 

\subsection{Testing correlation via the likelihood-ratio test}\label{appratio}

We seek to evaluate the null and alternative hypotheses 
\be\label{hypotest}
\mathscr{H}_0:\quad r=0 \quad \mbox{versus} \quad \mathscr{H}_1:\quad
r\neq0.
\ee
Our definition of the log-likelihood $\mcL\oft$ in
eq.~(\ref{firstl}) included the correlation coefficient~$r$ between
initial-loading topographies as a parameter to be 
estimated from the data. In contrast, the log-likelihood
$\mcLw\oftw=\mcL\oftwz$ of eq.~(\ref{firstlw}) did not. 
The Hessian of~$\mcL$ is~$\Fb$ and that of~$\mcLw$ is~$\Fbw$, and from
eq.~(\ref{inprob}) we know that $\Fb$ converges in probability to the
negative Fisher matrix $-\bmcF$ and, similarly, $\Fbw$ converges to
the constant~$-\bmcFw$. This gives us the elements to evaluate the
different scenarios. 

Should we evaluate ``uncorrelated data'' using a ``correlated model'',
we need a significance test for the addition of the correlation
parameter. Since the hypotheses~(\ref{hypotest})  refer to nested
models, $\bthetaw$ containing some of the same entries 
as~$\btheta$, see eqs~(\ref{btheta})--(\ref{bthetawi}), otherwise put  
\be
\btheta=[\bthetawt \,\,\, r]\Tit
,\ee
standard likelihood-ratio theory \cite[]{Cox+74} applies. Let the
truth under $\mathscr{H}_0$ be given by the parameter vector
\be
\btruth=[\btruthwt\,\,\,0]\Tit,
\ee
and let us consider having found two maximum-likelihood
estimates,  
\ber\label{twodiff1}
\hbt&=&\arg\max \mcL\oft\quad=\quad[\hbtwt\,\,\,\hat{r}]\Tit
,\\\label{twodiff2}
\hbtw_u&=&\arg\max \mcLw\oftw\quad\ne\quad\hspace{0.285em}\hbtw.
\eer
Note that $\mcL(\hbt)\le\mcLw(\hbtw)$ and
and that the estimates of
`everything-but-the-correlation-coefficient' are different from the
full estimates depending 
on whether the correlation coefficient is included as a parameter to
be estimated or not. We now define the maximum-log-likelihood
ratio statistic from the evaluated likelihoods
\be\label{X}
X=
2K\big[\mcL\ofth-\mcLw\oftwhc\big]=
2K\big[\mcL\ofth-\mcLw\oftwhc+
\mcLw\oftrw-\mcL\oftr\big]=X_1-X_2,
\ee
whereby we have used that, evaluated at the truth under
$\mathscr{H}_0$, the likelihood values $\mcLw\oftrw=\mcL\oftr$, and defined the
auxiliary quantities  
\be
X_1=2K\big[\mcL\ofth-\mcL\oftr\big],\also
X_2=2K\big[\tilde\mcL\oftwhc-\tilde\mcL\oftrw\big].
\ee
By Taylor expansion of the log-likelihoods around the
truth, to second order and with the first-order
derivatives vanishing, we then have
\be\label{restate}
X_1\inlaw-\sqrt{K}\big[\hbt-\btruth\big]\Tit
\bmcF(\btruth)\big[\hbt-\btruth\big]\sqrt{K}
\also
\label{restate2}
X_2\inlaw-\sqrt{K}\big[\hbtwc-\btruthw\big]\Tit
\bmcFw(\btruthw)\big[\hbtwc-\btruthw\big]\sqrt{K}
,
\ee
where we have used the limiting behavior~(\ref{inprob}). For more
generality, we consider maximum-likelihood problems with a 
partitioned parameter vector  
\be\label{thetax}
\btheta=[\bthetawt\,\,\,\bthetaxt]\Tit
,\ee
whereby $\bthetax$ may contain any number of extra parameters,
$\bthetax=[r]$ being the case under consideration. Introducing
notation as we go along, the Fisher matrix for such
problems partitions into four blocks \cite[see also][]{Kennett+98} such
that we can write,
\be\label{Fparts}
\bmcF=
\omat{cc}
\bmcFw & \bmcFx\\
\bmcFxt & \bmcFo
\cmat
.
\ee
The submatrices $\bmcFx$ and $\bmcFo$ contain the negative
expectations of the second derivatives of the likelihood~$\mcL$, with
respect to at least one of the `extra'
parameters~$\thetax\in\bthetax$, suitably arranged with the mnemonic
subscripts $\times$ and $\circ$. The corner matrices~$\Fbw$
and~$\bmcFw$ contain the second derivatives of the likelihood~$\mcLw$
in only the `simpler' subset of parameters~$\thetaw\in\bthetaw$. The
inverse of the Fisher matrix is given by
\be\label{Finvdef}
\bmcF^{-1}
=
\omat{cc}
\bmcFxo^{-1} & -\bmcFxo^{-1}\bmcFx^{}\bmcFo^{-1}\\
-\bmcFox^{-1}\bmcFxt\bmcFw^{-1} &\bmcFox^{-1}
\cmat
,
\ee
thereby defining the auxiliary matrices, and, via the Woodbury
identity, their inverses, as
\ber
\label{Woodbury1}\label{Woodbury}
\bmcFxo&=&\bmcFw-\bmcFx^{}\bmcFo^{-1}\bmcFxt
\also
\bmcFxo^{-1}=\bmcFw^{-1}+\bmcFw^{-1}\bmcFx^{}
\bmcFox^{-1}\bmcFxt\bmcFw^{-1},
\\
\label{Woodbury2}
\bmcFox&=&\bmcFo-\bmcFxt\bmcFw^{-1}\bmcFx
\also
\bmcFox^{-1}=\bmcFo^{-1}+\bmcFo^{-1}\bmcFxt
\bmcFxo^{-1}\bmcFx^{}\bmcFo^{-1}
.
\eer
This yields the variances of the vectors partitions. Recalling from
eq.~(\ref{summary2}) that  
\be\label{summary3}
\sqrt{K}(\hbt-\btruth) \sim \mcN(\bzero,\bmcF^{-1}\oftr)
,
\ee
we may use eqs~(\ref{Finvdef})--(\ref{Woodbury}) to express the
marginal distribution of the partition~$\hbtx$ under the null hypothesis,   
\be\label{summary4}
\sqrt{K}\hsom\hbtx \sim \mcN(\bzero,\bmcFox^{-1}\oftr)
.
\ee
In this general framework we rewrite likelihood-ratio
statistic~(\ref{X}) with the help of
eqs~(\ref{restate})--(\ref{thetax}) as  
\be\label{X1minX2}
X=X_1-X_2\approx
-\sqrt{K}
\left[
\omat{c}\hbtw\\\hbtx\cmat-
\omat{c}\hbtwc\\\bzero\cmat
\right]\Tit
\omat{cc}
\bmcFw & \bmcFx\\
\bmcFxt & \bmcFo
\cmat
\left[\omat{c}\hbtw\\\hbtx\cmat-
\omat{c}\hbtwc\\\bzero\cmat\right]
\sqrt{K}.
\ee
In order to figure out the properties of the likelihood-ratio test we
now need to understand the properties of the difference between the
`correlated' and `uncorrelated' estimates $\hbtw-\hbtwc$ of
eqs~(\ref{twodiff1})--(\ref{twodiff2}). We may note directly from
\cite{Cox+74} that  
\be
\hbtwc=\hbtw+
\bmcFw^{-1}\!\bmcFx\,\hbtx
.
\ee
Inserting this relation into eq.~(\ref{X1minX2}) the limiting behavior
of the likelihood-ratio test statistics becomes
\be\label{X1minX2x}
X\approx%\inlaw
-\sqrt{K}\,\hbtxt
\left( -\bmcFxt\bmcFw^{-1}\bmcFx + \bmcFo\right)
\hbtx
\sqrt{K}=
-\sqrt{K}\hsom\hbtxt
\bmcFox\hsom
\hbtx\sqrt{K}
,
\ee
where we have used eq.~(\ref{Woodbury2}). From eq.~(\ref{summary4})
then follows that the distribution of~$X$ is the sum of squared
zero-mean Gaussian variates divided by their variance, i.e.,
chi-squared with as many degrees of freedom as the difference in
number of parameters between the alternative models described
by~$\btheta$ and~$\bthetaw$, a conclusion first reached
by~\cite{Wilks38}. For a derivation rooted in the geometry of contours
of the likelihood surface, see \cite{Fan+2000}. 

In our particular case, the only complementary variable is the
correlation~$r$ between the two initial-loading terms, and the
likelihood-ratio test statistic of eq.~(\ref{X}) becomes
\be\label{xsimchi}
X=2K\big[\mcL\ofth-\mcLw\oftwhc\big]\sim\chi^2_1
,
\ee
which is how we may test the alternative hypotheses of initial-load
correlation and absence thereof.

\subsection{\textit{A posteriori} justification for the behavior of the
  synthetic tests}
\label{numex}

We owe the reader a short theoretical justification of why using the
unblurred likelihoods~$\mcL$ of eq.~(\ref{firstlt}) for the variance
calculations (the black curves in Figs.~\ref{syms1}--\ref{syms4})
accurately predicts the outcome of experiments (the grey-shaded
histograms) conducted on the basis of the blurred likelihoods~$\Lbar$
of eq.~(\ref{firstlbar}). The blurring enters through the spectral
term, which is~$\bbmcSo$ instead of~$\bmcSo$ as we recall from
eq.~(\ref{Sdefinebar}), and it affects the likelihood~(\ref{firstlbar})
through its determinant and inverse. Instead of the purely numerical
evaluation of the convolutions of the type~(\ref{simublur}) and
conducting all subsequent operations on the result, which is how we
construct~$\Lbar$ in the numerical experiments, in principle, in the
notation suggested by eqs~(\ref{specdensij})--(\ref{specdensoij}), we
could attempt to explicitly evaluate, though this would be cumbersome,
\be\label{detbarS0}
\det\bbmcSo\ofk=
\intnyqt
\left|W(\kb-\kb')\right|^2
\left|W(\kb-\kb'')\right|^2
\left[\mcSo_{11}\ofkp\mcSo_{22}\ofkpp-\mcSo_{12}\ofkp\mcSo_{21}\ofkpp\right]
\dbk'\dbk''
,
\ee
for the determinant. For the inverse
(see eq.~\ref{Cayley}), we might calculate
\be\label{invbarS0}
\bbmcSoinv\ofk =
\frac{1}{\det\bbmcSo\ofk}
\intnyq
\left|W(\kb-\kb')\right|^2\, 
\omats{cc}
\hsem\mcSo_{22}\ofkp  & -\mcSo_{12}\ofkp \\
-\mcSo_{21}\ofkp  & \hsem\mcSo_{11}\ofkp 
\cmats
\dbk',
\ee
and construct derivatives of the kind
\be\label{dthinvbarS0}
\frac{\pl\bbmcSoinv\ofk}{\pl\theta} =
-\frac{1}{\det\bbmcSo\ofk}
\left(
\frac{\pl\det\bbmcSo\ofk}{\pl\theta}\,\bbmcSoinv\ofk+
\intnyq
\left|W(\kb-\kb')\right|^2\, 
\omats{cc}
\hsem\plth\mcSo_{22}\ofkp  & -\plth\mcSo_{12}\ofkp \\
-\plth\mcSo_{21}\ofkp  & \hsem\plth\mcSo_{11}\ofkp 
\cmats
\dbk'
\right)
.
\ee
Of course, should the spectral windows be delta functions,
eqs~(\ref{detbarS0})--(\ref{invbarS0}) would reduce to $\cSoo^2\det\Tbo$
and $\cSoom\Tboinv$ (see eqs~\ref{detTbo}--\ref{invTbo}), as expected
on the basis of eq.~(\ref{split}). With these expressions, we could
proceed to forming the first and second derivatives of the blurred
likelihood~(see eqs~\ref{gammaL}--\ref{strang}). For example, for the score in the
blurred likelihood we would then have
\be
\label{scoreagain}
\frac{\pl\Lbar}{\pl\theta}
=-\norml
\left[
\frac{\pl\ln(\det\bbmcSo)}{\pl\theta}
+\qform{\left(\frac{\pl\bbmcSoinv}{\pl\theta}\right)}\right]
=-\norml
\left[
\tr\left(\bbmcSoinv\frac{\pl\hsum\bbmcSo}{\pl\theta}\right)
+\qform{\left(-\bbmcSoinv\frac{\pl\hsum\bbmcSo}{\pl\theta}\bbmcSoinv\right)}\right]
,
\ee
and then the derivatives of eq.~(\ref{scoreagain}) would be needed to
determine the variance of the maximum-blurred-likelihood estimate in a
manner analogous to eqs~(\ref{kvar}) and~(\ref{Jdef}).

In short, a full analytical treatment would be very involved, and a
purely numerical solution would not give us very much insight. How
then can we understand that we can approximate the variance of our
maximum-blurred-likelihood estimator by replacing the second
derivatives of the blurred likelihood with those of its unblurred
form? We can follow \cite{Percival+93} and regard the blurring as
introducing a bias given by, to second order in the Taylor expansion,
\ber
\label{Sdefinebaragain}
\bbmcSo\ofk-\bmcSo\ofk 
&=&
\intnyq
\left|W_K(\kb-\kb')\right|^2
\left[
\bmcSo\ofkp-
\bmcSo\ofk
\right]
\dbk'
=
\intnyq
\left|W_K(\kb')\right|^2
\left[
\bmcSo(\kb+\kb')-
\bmcSo\ofk
\right]
\dbk'\\
&=&
\frac{1}{2}\intnyq
\left|W_K(\kb')\right|^2
\left[
\kbp\Trm\hsomm
%\Big\{\bnabla_{\kbp}\bnabla_{\kbp}\bmcSo\big|_\kb\Big\}
\Big\{\bnabla\bnabla\Trm\hsomm\bmcSo\big|_\kb\Big\}
\hsom
\kbp
\right]
\dbk'
=\tr
\left\{\frac{1}{2}
\Big[\bnabla\bnabla\Trm\hsomm\bmcSo\big|_\kb\Big]
\intnyq
\left|W_K(\kb')\right|^2\hsomm
\Big[\kbp\kbp\Trm\Big]
\dbk'
%\Big\{\bnabla_{\kbp}\bnabla_{\kbp}\bmcSo\big|_\kb\Big\}
\right\}
\\
&=&
\tr
\left\{\frac{1}{2}
\intnyq
\left|W_K(\kb')\right|^2\hsomm
\Big[\kbp\kbp\Trm\Big]
\dbk'
%\Big\{\bnabla_{\kbp}\bnabla_{\kbp}\bmcSo\big|_\kb\Big\}
\right\}
\tr
\Big[\bnabla\bnabla\Trm\hsomm\bmcSo\big|_\kb\Big]
,
\label{factor}
\eer
where we have used the hermiticity and periodicity of both
the spectral density~$\bmcSo$ and the spectral
window~$\left|W_K\right|^2$, and the evenness and energy normalization 
of the latter. For more general (e.g. non-radially symmetric or
non-separable) windows the equations will change, but not the conclusions.
 The first factor in eq.~(\ref{factor}) is a measure of the
bandwidth of the spectral window, which we shall call~$\beta^2(W)$,
and the second is a measure of the spectral variability via the
curvature of the spectral matrix. Thus the blurred spectral matrix is
the sum of the unblurred spectral matrix and a second term which
decays much faster with wavenumber than the first: 
\be\label{bls}
\bbmcSo\ofk=\bmcSo\ofk +\beta^2(W)\bnabla\bnabla\Trm\hsomm\bmcSo\ofk
.
\ee
The matter that concerns us here is how the blurring affects the
derivatives of the blurred spectrum and thus the derivatives of the
blurred likelihood. What transpires is that the differentiation with
respect to the parameters~$\theta$ does not change the relative order
of the terms in eq.~(\ref{bls}), in the sense that the correction
terms are only important at low values of the wavenumber~$k$.

Since the mean score is zero, by virtue of eq.~(\ref{blabla}), the
correction term becomes important, which leads to a bias of the
estimate. But since the variance of the score is not zero, see
eq.~(\ref{kvar}), the correction term is dwarfed by the contribution 
from the unblurred term. Hence we should, as we have, use the blurred
likelihood~(\ref{firstlbar}) to conduct numerical maximum-likelihood
experiments on finite data patches, but we can, as we have shown,
predict the variance of the resulting estimators using the analytical
expressions based on the unblurred likelihood~(\ref{firstl}).

\end{document}